\documentclass[11pt]{article}              

\usepackage[margin=25mm]{geometry}
\usepackage{graphicx}
\usepackage{colortbl}
\usepackage{subfigure}
 
\usepackage{fancyhdr}
\usepackage{amssymb,amsfonts,amsmath,amsthm}
\usepackage{natbib}
\usepackage{bstnotations}

\usepackage[english]{babel} 
\usepackage{graphicx}
\usepackage{float}
\usepackage{color}
\usepackage{soul}
\usepackage{geometry}
\usepackage{natbib}
\usepackage{setspace}
\usepackage{caption}
\usepackage{multirow}

\usepackage{authblk}
	
\graphicspath{{./},{./Figures/}}

\newcommand{\res}{{\mathcal R}}

\newcommand{\hsp}{\hspace{0.3mm}}

\newcommand{\nval}{n_{\rm{val}}}
\newcommand{\vli}{v_l^{(i)}}

\newcommand{\vri}{v_r^{(i)}}
\newcommand{\zkli}{z_{k,l}^{(i)}}
\newcommand{\zkri}{z_{k,r}^{(i)}}
\newcommand{\Pki}{P_k^{(i)}}
\newcommand{\Pkj}{P_k^{(j)}}
\newcommand{\Imax}{I_{\rm max}}
\newcommand{\Derrmin}{\Delta\widehat{err}_{\rm min}}

\newcommand{\errCV}{\widehat{err}_{\rm CV3}}
\newcommand{\errLOO}{\widehat{err}_{\rm LOO}^*}
\newcommand{\errG}{\widehat{err}_G}

\defcitealias{Chevreuil2013}{Chevreuil, Rai et~al., 2013}
\defcitealias{Chevreuil2013least}{Chevreuil, Lebrun et~al., 2013}
\defcitealias{Nouy2010b}{Nouy, 2010}
\defcitealias{Khoromskij2011}{Khoromskij and Schwab, 2011}
\defcitealias{Doostan2013}{Doostan et~al., 2013}
\defcitealias{Hadigol2014}{Hadigol et~al., 2014}
\defcitealias{RaiThesis}{Rai, 2014}
\defcitealias{Validi2014}{Validi, 2014}
\defcitealias{Konakli2015UNCECOMP}{Konakli and Sudret, 2014}

\begin{document}
\title{Reliability analysis of high-dimensional models using low-rank tensor approximations} 

\author[1]{K. Konakli} \author[1]{B. Sudret} 

\affil[1]{Chair of Risk, Safety and Uncertainty Quantification,
  
  ETH Zurich, Stefano-Franscini-Platz 5, 8093 Zurich, Switzerland}

\date{}
\maketitle

\abstract{Engineering and applied sciences use models of increasing complexity to simulate the behavior of manufactured and physical systems. Propagation of uncertainties from the input to a response quantity of interest through such models may become intractable in cases when a single simulation is time demanding. Particularly challenging is the reliability analysis of systems represented by computationally costly models, because of the large number of model evaluations that are typically required to estimate small probabilities of failure. In this paper, we demonstrate the potential of a newly emerged meta-modeling technique known as low-rank tensor approximations to address this limitation. This technique is especially promising for high-dimensional problems because: (i) the number of unknowns in the generic functional form of the meta-model grows only linearly with the input dimension and (ii) such approximations can be constructed by relying on a series of minimization problems of small size independent of the input dimension. {  In example applications involving finite-element models pertinent to structural mechanics and heat conduction, low-rank tensor approximations built with polynomial bases are found to outperform the popular sparse polynomial chaos expansions in the estimation of tail probabilities when small experimental designs are used.} It should be emphasized that contrary to methods particularly targeted to reliability analysis, the meta-modeling approach also provides a full probabilistic description of the model response, which can be used to estimate any statistical measure of interest. 
\\[1em] 

{\bf Keywords}: uncertainty propagation -- reliability analysis --
meta-models -- low-rank approximations -- polynomial chaos expansions }

\maketitle

\section{INTRODUCTION}

Analysis of the response of engineered and physical systems under uncertainties is of key importance in risk assessment and decision making in a wide range of fields. To this end, it is typical to use a computer model to represent the behavior of a system and perform repeated simulations to propagate uncertainties from the input to a response quantity of interest. However, because of the growing complexity of the computer models used across engineering and sciences, in many practical situations, a single simulation is time consuming, thus rendering uncertainty propagation non-affordable. Such situations are often encountered in reliability analysis due to the large number of model evaluations required to compute small failure probabilities.  As a result, meta-modeling techniques are gaining increasing popularity. The key idea thereof is to substitute a computationally expensive model with a statistically equivalent one, so-called meta-model, which can be easily evaluated. Using the meta-model, the analyst can perform statistical analysis of a response quantity of interest at low cost.

Of interest herein is non-intrusive meta-modeling, in which the original model is treated as a ``black box''. Building a meta-model in an non-intrusive manner relies on the evaluation of the original model at a set of points in the input space, called experimental design. The efficiency of a meta-modeling technique depends on its ability to provide sufficiently accurate representations of the exact model response over the entire input range by using relatively small experimental designs. This can be particularly challenging in cases when determining the tails of the response distribution with high accuracy is important, as in the estimation of small exceedence probabilities required in reliability analysis.

In this paper, we demonstrate the potential of the newly-emerged approach called \emph{low-rank tensor approximations} to provide meta-models appropriate for reliability analysis. Although different types of tensor decompositions may be used (see \eg \cite{Kolda2009, Hackbusch2012, Grasedyck2013}), we confine our attention to canonical tensor formats. In this context, low-rank approximations express the model response as a sum of a small number or rank-one tensors, where a rank-one tensor is a product of univariate functions in each of the input parameters. The idea of such decompositions originates in the work of Hitchcock \cite{Hitchcock1927} in the first half of the 20th century and has been employed within the last 50 years in a wide range of fields, including - but not limited to - psychometrics \cite{Carroll1970a, Harshman1970foundations}, chemometrics \cite{Appellof1981strategies, Bro1997parafac}, neuroscience \cite{Mocks1988topographic, Andersen2004structure}, fluid mechanics \cite{Felippa1990mixed, Ammar2006new}, signal processing \cite{Sidiropoulos2000parallel,DeLathauwer2007tensor}, image analysis \cite{Shashua2001linear, Furukawa2002appearance} and data mining \cite{Acar2006collective, Beylkin2009multivariate}. This technique is now attracting an expanding interest from the more recently established community of uncertainty quantification \cite{Nouy2010b, Khoromskij2011, Chevreuil2013, Chevreuil2013least, Doostan2013, Hadigol2014, RaiThesis, Validi2014, Konakli2015UNCECOMP}.

The focus of the present study is on low-rank tensor approximations that are made of polynomial functions due to the simplicity and versatility these offer. The considered meta-models therefore constitute an alternative to the widely used polynomial chaos expansions. In the latter, the number of unknown coefficients grows exponentially with the input dimension, requiring prohibitively large experimental designs when high-dimensional models are considered. Conversely, in low-rank tensor approximations, the number of unknown coefficients grows only linearly with the input dimension, which makes the approach particularly promising for dealing with high dimensionality. Such approximations can be constructed through a series of minimization problems of small size that is independent of the input dimension.

The paper is organized as follows: In Section~2, we describe the mathematical setup of non-intrusive meta-modeling and review basic concepts of reliability analysis with meta-models. In Section~3, we present the formulation of canonical low-rank approximations with polynomial bases and detail an algorithm for their construction; in the same section, we present, in a comparative way, the polynomial chaos expansions approach. In Section~4, we confront canonical low-rank approximations to state-of-art polynomial chaos expansions in the estimation of response probability density functions and of failure probabilities in reliability analysis. The paper concludes with a summary of the main findings and respective outlooks.

\section{META-MODELS FOR UNCERTAINTY PROPAGATION AND RELIABILITY ANALYSIS}

\subsection{Non-intrusive meta-modeling and error estimation}

We consider a computational model $\cm$ that represents the behavior of a physical or engineering system of interest. We denote by $\ve X=\{X_1 \enum X_M\}$ the $M$-dimensional input vector of $\cm$ and by $\ve Y=\{Y_1 \enum Y_N\}$ the $N$-dimensional response vector. To account for the uncertainty in the input and the resulting uncertainty in the response, the elements of $\ve X$ and $\ve Y$ are described by random variables. For the sake of simplicity, the case of a scalar model response ($N=1$) is considered hereafter. Therefore, $\cm$ is considered as the mapping:
\begin{equation}
\label{eq:model}
\ve X \in \cd_{\ve X} \subset \Rr^M\longmapsto Y=\cm (\ve X) \in \Rr,
\end{equation}
where $\cd_{\ve X}$ denotes the support of $\ve X$. Note that the case of a vector model response can be addressed by separately treating each element of $\ve Y$ as in the case of a scalar response. 

In general, the map in Eq.~(\ref{eq:model}) is not known in a closed analytical form and may represent a computationally intensive process. A meta-model $\widehat{\cm}$ is an analytical function that mimics the behavior of $\cm$; in other words, $\widehat{\cm}(\ve X)$ possesses similar statistical properties with $\cm(\ve X)$, while maintaining an easy-to-evaluate form. Replacing a complex computational model with a meta-model allows efficient uncertainty propagation from the random input to a response quantity of interest in cases when this is practically infeasible by using the original model due to the high computational cost.

In non-intrusive meta-modeling, which is of interest herein, the original computational model is treated as a ``black box''. Thus, in order to develop a meta-model in a non-intrusive manner, one only needs a set of $N$ realizations of the input vector $\ce=\{\ve \chi^{(1)} \enum \ve \chi^{(N)}\}$, \emph{called experimental design} (ED), and the corresponding set of model responses $\cy=\{\cm(\ve \chi^{(1)}) \enum \cm(\ve \chi^{(N)})\}$. We underline that non-intrusive approaches do not require any further knowledge of the original model, which is used therein without any modification.

To define measures of accuracy of the meta-model response $\widehat{Y}=\widehat{\cm}(\ve X)$, we first introduce the discrete $L_2$ semi-norm:
\begin{equation}
\label{eq:norm}
\norme a {\cx}=\left(\frac{1}{n}\sum_{i=1}^n a^2(\ve x_i)\right)^{1/2},
\end{equation}
where $a$ represents a function: $\ve x~\in~\cd_{\ve X}~\longmapsto~a(\ve x)~\in~\Rr$ and $\cx=\{\ve x_1 \enum \ve x_n\}\subset \cd_{\ve X}$ denotes a set of $n$ realizations of $\ve X$. A good measure of accuracy of the meta-model response is the \emph{generalization error} $Err_G$, which is defined as the mean-square error of the difference $(Y-\widehat Y)$ and can be estimated by:
\begin{equation}
\label{eq:hatErrG}
\widehat{Err}_G =\left\|\cm-\widehat {\cm}\right\|_{\cx_{\rm{val}}}^2,
\end{equation}
where $\cx_{\rm{val}}=\{{\ve x}_1 \enum {\ve x}_{n_{\rm{val}}}\}$ is a sufficiently large set of realizations of $\ve X$, called \emph{validation set}. The estimate of the relative generalization error $\widehat{err}_G$, is obtained by normalizing $\widehat{Err}_G$ with the empirical variance of $\cy_{\rm{val}}=\{\cm({\ve x}_1)\enum \cm({\ve x}_{n_{\rm{val}}})\}$, which denotes the set of model responses at the validation set. Unfortunately, a validation set is not available in typical meta-modeling applications, where a large number of model evaluations is not affordable.  An alternative estimate that relies solely on the ED is the \emph{empirical error} $\widehat{Err}_E$, which is given by:
\begin{equation}
\label{eq:ErrE}
\widehat{Err}_E =\left\|\cm-\widehat {\cm}\right\|_{\ce}^2.
\end{equation}
In the above equation, the subscript ${\ce}$ emphasizes that the semi-norm is evaluated at the points of the ED. The relative empirical error $\widehat{err}_E$, is obtained by normalizing $\widehat{Err}_E$ with the empirical variance of $\cy=\{\cm({\ve \chi}^{(1)})\enum \cm({\ve \chi}^{(N)})\}$, which denotes the set of model responses at the ED. Although the empirical error reuses the points of the ED, it has the strong drawback that it tends to underestimate the actual generalization error, which might be severe in cases of overfitting.

{  By using the information contained in the ED only, one can obtain fair approximations of the generalization error by means of cross-validation (CV) techniques (see \eg \cite{Viana2009,Arlot2010}). In brief, the basic idea of $k$-fold CV is to randomly partition the ED into $k$ sets of approximately equal size, build the meta-model by relying on all but one of the partitions, and use the excluded set to evaluate the generalization error. By alternating through the $k$ sets, one obtains $k$ meta-models of which the average generalization error serves as the error estimate of the meta-model built with the full ED.}

\subsection{Reliability analysis}
\label{sec:rel}

{  Reliability analysis aims at computing the probability that the system under consideration fails to satisfy prescribed criteria. A failure criterion is mathematically represented by the so-called \emph{limit-state function} $g(\ve Z)$; in a general case, $\ve Z=\{\ve X,\hsp \cm(\ve X),\hsp \ve X'\}$ may depend on the input parameters of the model describing the system, response quantities obtained from the model and additional random parameters gathered in $\ve X'$. Conventionally, the limit-state function is formulated so that failure corresponds to $g(\ve z)\leq 0$; the set of points that satisfy this condition comprise the \emph{failure domain} $\cd_f$ with respect to $g$, \ie $\cd_f=\{\ve z:g(\ve z)\leq 0\}$. The associated probability of failure is therefore given by:
\begin{equation}
\label{eq:Pf}
P_f=\int_{\cd_f} f_{\ve Z}(\ve z) \hsp \di{\ve z},
\end{equation}
where $f_{\ve Z}$ denotes the probability density function (PDF) of $\ve Z$. Note that in a general case, failure of a system may be defined in terms of multiple limit-state functions representing different failure criteria.

A universal method for computing the integral in Eq.~(\ref{eq:Pf}) is Monte Carlo simulation (MCS). The MCS approach involves generating a sufficiently large sample of realizations of $\ve Z$, say $\{\ve z_1 \enum \ve z_n\}$, and then, estimating $P_f$ as the empirical mean:
\begin{equation}
\label{eq:Pf_MCS}
\widehat{P}_{f,\hsp \rm MCS}=\dfrac{1}{n} \sum_{k=1}^{n}I_{\cd_f}(\ve z_k),
\end{equation}
where $I_{\cd_f}$ denotes the indicator function of the failure domain. Obviously, the MCS-based estimator in Eq.~(\ref{eq:Pf_MCS}) is unbiased, which means $\Esp{\widehat{P}_{f,\hsp \rm MCS}}=P_f$.
Typically, of interest in reliability analysis are failure events with small probabilities of occurrence, \ie $P_f<<1$, leading to the following approximation of the coefficient of variation (CoV) of $\widehat{P}_{f,\hsp \rm MCS}$:
\begin{equation}
\label{eq:CoV_Pf}
\delta_{\widehat{P}_{f,\hsp \rm MCS}}\approx 1/\sqrt{n \hsp P_f}.
\end{equation}

Eq.~(\ref{eq:CoV_Pf}) indicates that estimation of a failure probability with magnitude of the order of $10^{-k}$ with CoV $<10\%$ requires that a number of samples larger than $10^{k+2}$ is used. Clearly, MCS is impractical for computing small failure probabilities in cases when a single evaluation of the model response is computationally costly. This limitation is overcome when the original model  $\cm$ is substituted by a meta-model  $\widehat{\cm}$. Accordingly, the actual failure domain $D_f$ is approximated by ${\widehat{\cd}_f}=\{\ve z:\widehat{g}(\ve z)\leq 0\}$, where $\widehat{g}(\ve Z)=g(\ve X,\hsp \widehat{\cm}(\ve X),\hsp \ve X')$. Once the meta-model is available, evaluation of $\widehat{P}_{f,\hsp \rm MCS}$ in Eq.~(\ref{eq:Pf_MCS}) by using large Monte Carlo samples becomes essentially costless.

Various techniques have been devised with the purpose of computing efficiently the small failure probabilities that are of interest in reliability analysis; a thorough review listing the advantages and drawbacks of different methods can be found in \cite{Morio2014}. We herein briefly describe three widely-used methods, which are considered in the application section of the present study:
\begin{itemize}
\item \emph{First-order reliability method} (FORM) \cite{Hasofer1974, Rackwitz1978}: FORM relies on determining the design point $P^*$, \ie the point of the failure domain that is closest to the origin in the standard normal space. The failure domain is then approximated by the half space defined as the hyperplane that is tangent to the limit-state surface $\{\ve z:g(\ve z)=0\}$ at $P^*$, leading to the first-order approximation of $P_f$.
\item \emph{Second-order reliability method} (SORM) \cite{Breitung1989, DerKiureghian1991}: SORM provides a correction to the FORM solution by approximating the limit-state surface at the design point by a second-order surface.
\item \emph{Importance sampling} (IS) \cite{Melchers1989, Au2003b}: The basic idea in IS is to recast the definition of $P_f$ in Eq.~(\ref{eq:Pf}) by means of an auxiliary PDF that is more efficient in generating samples within the failure domain; appropriate weights are introduced in the computation of the integral in order to account for the change in the PDF.  
\end{itemize}
We emphasize that the aforementioned methods are particularly targeted to reliability analysis; on the other hand, meta-modeling comprises a more general tool for uncertainty propagation, which may be used to conduct any type of statistical analysis of the model response, \eg PDF estimation, evaluation of statistical moments and confidence intervals, analysis of variance, and so forth (see \eg \cite{Xiu2003, Acharjee2006, SudretHDR, Najm2009, Jones2013, DemanKonakli2016} among a vast literature). However, in cases when the analyst is only interested in the computation of failure probabilities, any reliability-analysis technique can be used in conjunction with an appropriate meta-model (see \eg \cite{Li2012bayesian, Balesdent2013, Dubourg2013}). It is underlined that the accuracy of reliability analysis based on a meta-model approximation relies on the ability of the latter to accurately represent the response of the original model at the tails of its distribution.}

\section{LOW-RANK TENSOR APPROXIMATIONS}

\subsection{Formulation using polynomial bases}
\label{LRA_form}

We consider the map in Eq.~(\ref{eq:model}) assuming that the components of $\ve X$ are independent, with the marginal PDF of $X_i$ denoted by $f_{X_i}$ for $\{i=1 \enum M\}$. Let $\cm^{\rm LRA}$ denote a meta-model of $\cm$ belonging to the class of low-rank approximations (LRA); as mentioned in the Introduction, the term ``rank" herein refers to the so-called ``canonical rank". The corresponding approximation of $Y=\cm (\ve X)$ has the general form:
\begin{equation}
\label{eq:LRA}
Y^{\rm LRA} =\cm^{\rm LRA}(\ve X)=\sum_{l=1}^R b_l \hsp w_l(\ve X),
\end{equation}
in which $b_l$ is a normalizing constant and $w_l$ is a \emph{rank-one function} of $\ve X$. The rank-one function $w_l$ is a product of univariate functions of the components of $\ve X$:
\begin{equation}
\label{eq:rank1}
w_l(\ve X)= \prod_{i=1}^M {v_l^{(i)}(X_i)},
\end{equation}
where $\vli$ denotes a univariate function of $X_i$. Accordingly, $R$ in Eq.~(\ref{eq:LRA}) represents the number of rank-one components retained in the approximation. Naturally, representations with a small number of rank-one components are of interest, thus named \emph{low-rank}.

In order to obtain a representation of $Y=\cm (\ve X)$ in terms of polynomial functions, we expand $\vli$ onto a polynomial basis $\{\Pki,~k \in \Nn\}$ that is orthonormal with respect to $f_{X_i}$, \ie satisfies:
\begin{equation}
\label{eq:orthonorm_cond}
\innprod {P_j^{(i)}} {P_k^{(i)}} =\int_{\cd_{X_i}}{P_j^{(i)}(x_i) \hsp P_k^{(i)}(x_i) \hsp f_{X_i}(x_i) \hsp \di{x_i}}=\delta_{jk},
\end{equation}
where $\cd_{X_i}$ denotes the support of $X_i$ and $\delta_{jk}$ is the Kronecker delta symbol, equal to one if $j=k$ and zero otherwise.  Accordingly, the univariate function of $X_i$ takes the form:
\begin{equation}
\label{eq:vli}
\vli(X_i)=\sum_{k=0}^{p_i} \zkli \hsp \Pki (X_i),
\end{equation}
where $\Pki$ is the $k$-th degree univariate polynomial in the $i$-th input variable of maximum degree $p_i$ and $\zkli$ is the coefficient of $\Pki$ in the $l$-th rank-one term. By substituting Eq.~(\ref{eq:vli}) into Eq.~(\ref{eq:LRA}), we obtain:
\begin{equation}
\label{eq:LRA_pol}
Y^{\rm LRA} =\cm^{\rm LRA}(\ve X) = \sum_{l=1}^R b_l \left(\prod_{i=1}^M\left(\sum_{k=0}^{p_i} \zkli \hsp \Pki (X_i)\right)\right).
\end{equation}
Disregarding the redundant parameterization arising from the normalizing constants, the number of unknowns in Eq.~(\ref{eq:LRA_pol}) is $R \cdot \sum_{i=1}^M (p_i+1)$, which grows \emph{only linearly} with the input dimension $M$. We will see later that this is a key factor for the higher efficiency of LRA as compared to polynomial chaos expansions when dealing with high-dimensional problems. 

Classical algebra allows one to build a family of polynomials satisfying Eq.~(\ref{eq:orthonorm_cond}) \cite{Abramowitz}. For standard distributions, the associated families of orthonormal polynomials are well known; for instance, a uniform variable with support $[-1,1]$ is associated with the family of Legendre polynomials, whereas a standard normal variable is associated with the family of Hermite polynomials \cite{Xiu2002wiener}. However, it is common in practical situations that the input variables do not follow standard distributions. In such cases, the random vector $\ve X$ is first transformed into a basic random vector $\ve U$ (\eg a standard normal or standard uniform random vector) through an isoprobabilistic transformation $\ve X=T^{-1}(\ve U)$ and then, the model response $\cm (T^{-1}(\ve U))$ is expanded onto the polynomial basis associated with $\ve U$. Cases with mutually dependent input variables can also be treated through an isoprobabilistic transformation into a vector of independent variables, \eg the Nataf transformation in the case of a joint PDF with Gaussian copula \cite{Lebrun2009a, Lebrun2009b}. We underline that although the focus of the present work is on LRA developed with polynomial functions, the use of such functions is not a constraint in a general case.

\subsection{Construction with greedy approaches}
\label{sec:LRA_alg}

Different non-intrusive algorithms have been proposed recently for developing LRA in the form of Eq.~(\ref{eq:LRA_pol}); see \eg \cite{Chevreuil2013, Chevreuil2013least, Doostan2013, RaiThesis, Validi2014}. A common attribute of these algorithms is that the computation of the polynomial coefficients relies on an alternated least-squares (ALS) minimization approach. The ALS technique consists in solving a series of small-size least-squares minimization problems, where each minimization is performed along \emph{a single dimension}. Chevreuil et~al.~\cite{Chevreuil2013} proposed to construct LRA in a greedy manner by successively adding rank-one components and updating the entire set of normalizing constants following each increase of the rank. Aspects of this algorithm were further investigated by Konakli and Sudret~\cite{Konakli2015arxiv}. This greedy approach is employed in the present study and described analytically below.

Let $Y_r^{\rm LRA} = \cm_r^{\rm LRA}(\ve X)$ denote the rank-$r$ approximation of $Y=\cm(\ve X)$:
\begin{equation}
\label{eq:Y_r}
Y_r^{\rm LRA} = \cm_r^{\rm LRA}(\ve X)=\sum_{l=1}^r b_l \hsp w_l (\ve X),
\end{equation}
where:
\begin{equation}
\label{eq:w_l}
w_l(\ve X)=\prod_{i=1}^M\left(\sum_{k=0}^{p_i} \zkli \hsp \Pki (X_i)\right).
\end{equation}
The employed algorithm comprises a sequence of pairs of a \textit{correction step} and an \textit{updating step}, so that the $r$-th correction step yields the rank-one component $w_r$ and the $r$-th updating step yields the set of coefficients $\{b_1 \enum b_r\}$. Details on these steps are given next.

\textbf{Correction step}: Let $\res_r(\ve X)$ denote the residual after the completion of the  $r$-th iteration:
\begin{equation}
\label{eq:res}
\res_r(\ve X)=\cm(\ve X)-\cm^{\rm LRA}_r(\ve X).
\end{equation}
The sequence is initiated by setting $\cm^{\rm LRA}_0(\ve X)=0$ leading to $\res_0(\ve X)=\cm(\ve X)$. In the $r$-th correction step, the new rank-one tensor $w_r$ is determined by minimizing the empirical error with respect to the current residual:
\begin{equation}
\label{eq:solve_wr}
w_r=\mathrm{arg} \underset{\omega \in \cw}{\hsp\mathrm{\min}} \left\|\res_{r-1}-\omega\right\|_{\ce}^2,
\end{equation}
where $\cw$ represents the space of rank-one tensors. Eq.~(\ref{eq:solve_wr}) is solved by means of an ALS scheme that involves successive minimizations along the dimensions $\{1\enum M\}$. In the minimization along dimension $j$, the polynomial coefficients in all other dimensions are ``frozen'' at their current values; the coefficients ${\ve z}_r^{(j)}=\{z_{1,r}^{(j)} \ldots z_{p_j,r}^{(j)}\}$ are therefore obtained as:
\begin{equation}
\label{eq:solve_zr}
\ve z_r^{(j)}= \mathrm{arg}\underset{\ve \zeta \in\Rr^{p_j+1}}{\mathrm{min}}\left\|\res_{r-1}-C^{(j)}\cdot\left(\sum_{k=0}^{p_j} \zeta_k \hsp \Pkj\right)\right\|_{\ce}^2,
\end{equation}
where $C^{(j)}$ represents the ``frozen'' component:
\begin{equation}
\label{eq:Cj}
C^{(j)}(X_1 \enum X_{j-1},X_{j+1} \enum X_M)
=\prod_{i\neq j}\vri(X_i)
=\prod_{i\neq j}\left(\sum_{k=0}^{p_i} \zkri \hsp \Pki(X_i)\right).
\end{equation}
Because Eq.~(\ref{eq:solve_zr}) involves only $(p_j+1)$ unknowns ($p_j<20$ in typical applications), it can be easily solved using the ordinary least squares (OLS) method.

The correction step is initiated by assigning arbitrary values to the unknowns and may involve several iterations over the set of dimensions. Note that assigning initial values to the functions $\vri$ (Eq.~(\ref{eq:vli})) is sufficient; for instance, unity values may be used. Konakli and Sudret~\cite{Konakli2015arxiv} investigated the effect of the number of iterations performed in a correction step on the accuracy of LRA. They proposed a stopping criterion combining the number of iterations $I_r$ with the decrease in the relative empirical error $\Delta\widehat{err}_r$ in two successive iterations. The relative empirical error $\widehat{err}_r$ is obtained by normalizing the error measure:
\begin{equation}
\label{eq:Err_r}
\widehat{Err}_r=\left\|\res_{r-1}-w_r\right\|_{\ce}^2
\end{equation}
with the empirical variance of $\cy=\{\cm({\ve \chi}^{(1)})\enum \cm({\ve \chi}^{(N)})\}$, the latter denoting the set of model responses at the ED. Accordingly, the algorithm exits the $r$-th correction step if either $I_r$ reaches a maximum allowable value $\Imax$ or $\Delta \widehat{err}_r$ becomes smaller than a prescribed threshold $\Derrmin$. Based on numerical investigations in different case studies, Konakli and Sudret~\cite{Konakli2015arxiv} proposed to use $\Imax=50$ and $\Derrmin = 10^{-6}$. 

\textbf{Updating step}: After the completion of a correction step, the algorithm moves to an updating step, in which the set of coefficients $\ve b=\{b_1 \ldots b_r\}$ is obtained by minimizing the empirical error with respect to the response of the original model:
\begin{equation}
\label{eq:solve_b}
\ve b =\mathrm{arg} \underset{\ve \beta \in\Rr^{r}}{\mathrm{min}}\left\|\cm-\sum_{l=1}^r \beta_l \hsp w_l\right\|_{\ce}^2.
\end{equation}
Note that in each updating step, the size of vector $\ve b$ is increased by one. In the $r$-th updating step, the value of the new element $b_r$ is determined for the first time, whereas the values of the existing elements $\{b_1 \enum b_{r-1}\}$ are updated (recomputed). Because Eq.~(\ref{eq:solve_b}) involves only $r$ unknowns (recall that small ranks are of interest in LRA), it can be easily solved using OLS.

Construction of a rank-$R$ representation in the form of Eq.~(\ref{eq:LRA_pol}) requires repeating pairs of a correction and an updating step for $r=1\enum R$. The algorithm is summarized below.\vspace{2mm}

\textbf{Algorithm 1}: Construction of a rank-$R$ representation of $Y=\cm(\ve X)$ with polynomial bases, using a set of input samples $\ce=\{\ve \chi^{(1)} \enum \ve \chi^{(N)}\}$ and the corresponding model responses $\cy=\{\cm(\ve \chi^{(1)}) \enum \cm(\ve \chi^{(N)})\}$:
\begin{enumerate}
\item Set $\res_0(\ve{\chi}^{(q)})=\cm(\ve{\chi}^{(q)})$, $q =1\enum N$.
\item For $r=1\enum R$, repeat steps (a)-(f):
\begin{enumerate}
\item Assign initial values to $\vri$, $i=1 \enum M$ (\eg unity values).
\item Set $I_r=0$ and  $\Delta \widehat{err}_r=\epsilon>\Derrmin$.
\item While $\Delta \widehat{err}_r>\Derrmin$ and $I_r<I_{\rm max}$, repeat steps i-iv:
\begin{enumerate}
\item Set $I_r \leftarrow I_r+1$.
\item Determine ${\ve z}_r^{(i)}=\{z_{0,r}^{(i)} \ldots z_{p_i,r}^{(i)}\}$, $i=1 \enum M$, using Eq.~(\ref{eq:solve_zr}).
\item Use the current values of ${\ve z}_r^{(i)}$, $i=1 \enum M$, to update $w_r$.
\item Compute $\widehat{Err}_r$ using Eq.~(\ref{eq:Err_r}) and update $\Delta \widehat{err}_r$.
\end{enumerate}
\item Determine $\ve b=\{b_1 \ldots b_r\}$ using Eq.~(\ref{eq:solve_b}).
\item Evaluate $\cm^{\rm LRA}_r(\ve{\chi}^{(q)})$, $q =1\enum N$, using Eq.~(\ref{eq:Y_r}).
\item Evaluate $\res_r(\ve{\chi}^{(q)})$, $q =1\enum N$, using Eq.~(\ref{eq:res}).
\end{enumerate}
\end{enumerate}

Algorithm~1 describes the construction of LRA for a given rank $R$. However, in a typical application, the optimal rank is not known \textit{a priori}. Because Algorithm~1 yields a set of LRA of progressively increasing rank $r=1 \enum R$, the optimal among those can be selected using error-based criteria. In the present study, the optimal LRA is identified by means of 3-fold CV, as proposed by Chevreuil et al. \cite{Chevreuil2013least} (see Section 2 for details on $k$-fold CV). Thus, we set $r=1 \enum r_{\rm max}$ in Step 2 of Algorithm 1, where $r_{\rm max}$ is a maximum allowable candidate rank, and at the end, select the  optimal rank $R\in\{1 \enum r_{\rm max}\}$ as the one yielding the minimum 3-fold CV error estimate. Konakli and Sudret~\cite{Konakli2015arxiv} investigated the accuracy of rank selection based on 3-fold CV in different case studies and found that it leads to optimal or nearly optimal LRA in terms of the relative generalization errors, with the latter estimated using large validation sets. We note that the 3-fold CV error estimate may also be used to select the optimal polynomial degrees (see \cite{Konakli2015arxiv} for an investigation of the accuracy of this approach).

\subsection{Comparison to polynomial chaos expansions}
\label{sec:PCE}

A popular method for developing meta-models with polynomial bases is the use of polynomial chaos expansions (PCE). In this section, we provide a brief description of the PCE technique, noting its similarities with LRA.

{  We consider again the map in Eq.~(\ref{eq:model}) assuming that the components of $\ve X$ are independent with a joint PDF $f_{\ve X}$. Analogously to LRA, the case of dependent input variables can be herein treated with an appropriate isoprobabilistic transformation (see Section~\ref{LRA_form}). A PCE approximation of $Y=\cm (\ve X)$ has the form \cite{Xiu2002wiener,Soize2004}:}
\begin{equation}
\label{eq:PCE}
Y^{\rm PCE}=\cm^{\rm{PCE}}(\ve X)=\sum_{\ua \in \ca}{y_{\ua}} \hsp \Psi_{\ua}(\ve {X}),
\end{equation}
where $\ca$ is a set of multi-indices $\ua=(\alpha_1 \enum \alpha_M)$, $\{\Psi_{\ua},~\ua \in \ca\}$ is a set of multivariate polynomials that are orthonormal with respect to $f_{\ve X}$ and $\{y_{\ua},~\ua \in \ca\}$ is the set of polynomial coefficients. The orthonormal polynomial bases in Eq.~(\ref{eq:PCE}) can be obtained by tensorization of univariate polynomials that are orthonormal with respect to the marginals $f_{X_i}$:
\begin{equation}
\label{eq:mult_pol}
\Psi_{\ua}(\ve X)=\prod_{i=1}^M P^{(i)}_{\alpha_i}(X_i),
\end{equation}
where $P^{(i)}_{\alpha_i}$ is a univariate polynomial of degree ${\alpha_i}$ in the $i$-th input variable belonging to an appropriate family. Obviously, the families of the univariate polynomials used to formulate the multivariate PCE basis are the same as the families of polynomials that form the bases of the univariate functions in LRA (see Section~\ref{LRA_form}). However, as seen in Eq.~(\ref{eq:LRA_pol}), LRA retain the tensor-product form of Eq.~(\ref{eq:mult_pol}), whereas the expanded form is considered in PCE. Thus, LRA with polynomial bases can be seen as equivalent \emph{compressed} representations of PCE.

Different truncation schemes may be employed to determine the set of multi-indices $\{\ua \in \ca\}$ in Eq.~(\ref{eq:PCE}).  When the maximum degree of $P^{(i)}_{\alpha_i}$ is set to $p_i$, \ie $\ca=\{\ua \in \Nn^M: \alpha_i\leq p_i,~i=1 \enum M\}$, the expansion in Eq.~(\ref{eq:PCE}) relies on exactly the same polynomial functions with those used in Eq.~(\ref{eq:LRA_pol}). For this case, let us compare the number of unknowns in LRA and PCE considering a common maximum polynomial degree in all dimensions,  \ie $p_i=p$ for $i=1 \enum M$. One has $(p+1)^M$ unknowns in the PCE representation versus $(p+1)\cdot M \cdot R$ in LRA when redundant parameters are disregarded. Note that the number of unknowns grows \emph{exponentially} with $M$ in PCE, but \emph{only linearly} in LRA. For a typical engineering problem with dimensionality $M=10$, considering polynomials of low degree $p=3$ and an example low rank $R=10$, the aforementioned formulas yield $1,048,576$ PCE coefficients  versus a mere $400$ unknowns in LRA.

A more efficient truncation scheme is the \emph{hyperbolic} scheme proposed by Blatman and Sudret \cite{BlatmanPEM2010}. This is defined by the condition that the $q$-norm of any multi-index does not exceed a value $p^t$, \ie $\ca=\{\ua \in \Nn^M: \|\ua\|_q \leq p^t\}$ with:
\begin{equation}
\label{eq:qnorm} 
\|\ua\|_q=\left(\sum_{i=1}^M{\alpha_i}^q\right)^{1/q}, \hspace{2mm} 0<q\leq1.
\end{equation}
When $q=1$, multivariate polynomials of maximum \emph{total} degree $p^t$ are retained in the expansion. The corresponding number of terms in the truncated series is:
\begin{equation}
\label{eq:card} 
{\rm{card}}\hsp\ca={M+p^t \choose p^t} =\frac{(M+p^t)!}{M!p^t!},
\end{equation}
which grows \emph{polynomially} with $M$. Smaller values of $q$ impose limitations to the number of terms that include interactions between two or more input variables. Optimal values of $p^t$ and $q$ in the hyperbolic truncation scheme can be determined by means of error-based criteria (\eg the leave-one-out error described later).

Once the basis has been specified, the set of coefficients $\ve y=\{y_{\ua},~\ua \in \ca\}$ may be computed by minimizing the empirical error of the approximation:
\begin{equation}
\label{eq:PCE_OLS}
\ve y= \mathrm{arg} \underset{\ve {\upsilon}\in\Rr^{\mathrm{card}\hsp\ca}}{\mathrm{min}}\left\|\cm-\sum_{\ua \in \ca}\upsilon_{\ua}\hsp\Psi_{\ua}\right\|_{\ce}^2.
\end{equation}
Even by employing a  hyperbolic truncation scheme, the number of unknowns in Eq.~(\ref{eq:PCE_OLS}) can be very large in high-dimensional problems, requiring EDs of non-affordable size. Note that contrary to LRA, where the computation of the polynomial coefficients in each dimension is performed separately, the entire set of PCE coefficients is determined from a single minimization problem. To improve efficiency in the latter, one may substitute Eq.~(\ref{eq:PCE_OLS}) with a respective regularized problem. By penalizing the $L_1$ norm of $\ve y$, insignificant terms are disregarded from the set of predictors, leading to \emph{sparse} PCE. An efficient method to solve $L_1$-regularized problems is the least angle regression (LAR) method \cite{Efron2004}. A variation proposed by Blatman and Sudret~\cite{BlatmanJCP2011} under the name \emph{hybrid} LAR consists in using the LAR method to determine the best set of predictors and then, computing the PCE coefficients with OLS.

{  The PCE accuracy can be assessed by means of the leave-one-out error $\widehat{Err}_{\rm LOO}$, corresponding to the CV error for the extreme case $k=N$ \cite{Allen1971}. Using algebraic manipulations, this error can be computed based on a \emph{single} PCE that is built with the full ED (see \cite{BlatmanJCP2011} for details). The corresponding relative error, denoted by $\widehat{err}_{\rm LOO}$, is obtained after normalizing $\widehat{Err}_{\rm LOO}$ with the empirical variance of $\cy=\{\cm(\ve \chi^{(1)}) \enum \cm(\ve \chi^{(N)})\}$. Because $\widehat{err}_{\rm LOO}$ can be too optimistic, Blatman and Sudret~\cite{BlatmanJCP2011} proposed the use of the \emph{corrected} leave-one-out error, which includes a multiplication factor derived by Chapelle et al.~\cite{Chapelle2002}.}

\section{EXAMPLE APPLICATIONS}

{  In this section, we confront LRA to sparse PCE in uncertainty propagation through four models with different characteristics and dimensionality. In the first example, we consider a structural-mechanics model of dimension $M=5$ having an analytical rank-one structure. The following three examples involve finite-element models; in particular, we consider a truss model with independent input of dimension $M=10$, a heat-conduction model with thermal conductivity described by a random field, which is approximated by a series expansion of dimension $M=53$, and a frame model with correlated input of dimension $M=21$. For the aforementioned models, we investigate the comparative accuracy of LRA and PCE in the estimation of small failure probabilities $\widehat{P}_f$ and of the corresponding \emph{reliability indices} $\widehat{\beta}=-\Phi^{-1}(\widehat{P}_f)$, where $\Phi$ denotes the standard normal cumulative distribution function (CDF).

In all applications, the EDs used to build the LRA and PCE meta-models are obtained using Sobol pseudo-random sequences \cite{Niederreiter1992}. The LRA meta-models are built by implementing Algorithm~1 in Section~\ref{sec:LRA_alg}. A common maximum polynomial degree $p_1=\ldots=p_M=p$ is considered in all dimensions, with its optimal value selected by means of 3-fold CV. The involved minimization problems are solved using the OLS method. In building the PCE meta-models, a candidate basis is first determined by employing a hyperbolic truncation scheme and then, a sparse expansion is obtained by evaluating the PCE coefficients with the hybrid LAR method, as described in Section~\ref{sec:PCE}. The optimal combination of the maximum total polynomial degree $p^t$ and the parameter $q$ controlling the truncation, where $q\in\{0.25,0.50,0.75,1.0\}$, is selected as the one leading to the minimum corrected leave-one-out error (see Section~\ref{sec:PCE}). The PCE meta-models are built using the UQLab software \cite{MarelliICVRAM2014, UQdoc_09_104}; an implementation of the algorithm for developing LRA in the same software in currently underway.}

\subsection{Beam deflection}
In the first example, we perform reliability analysis of a simply supported beam subjected to a concentrated load at the midspan. The beam has a rectangular cross-section of width $b$ and height $h$, length $L$ and material Young's modulus $E$. The magnitude of the concentrated load is denoted by $P$. The aforementioned parameters are modeled as independent random variables following the distributions listed in Table~\ref{tab:beam_input}. The response quantity of interest is the midspan deflection, which is obtained through basic structural mechanics as:
\begin{equation}
\label{eq:beam_u}
U=\dfrac{P \hsp L^3}{4 E \hsp b \hsp h^3}.
\end{equation}
Because $U$ is a product of lognormal random variables, the response PDF can be herein obtained analytically. In particular, $U$ follows a lognormal distribution with parameters (mean and standard deviation of the corresponding normal variable $\log{U}$) given by: 
\begin{equation}
\lambda_U=-\log(4)+\lambda_P+3\lambda_L-\lambda_E-\lambda_b-3\lambda_h
\end{equation} 
and
\begin{equation}
\zeta_U=\left(\zeta_P^2+9\zeta_L^2+\zeta_E^2+\zeta_b^2+9\zeta_h^2\right)^{1/2},
\end{equation} 
where $\lambda_{X_i}$ and $\zeta_{X_i}$ respectively denote the mean and standard deviation of $\log X_i$.

\begin{table} [!ht]
\centering
\caption{Beam-deflection problem: Distributions of input variables.}
\label{tab:beam_input}
\begin{tabular}{c c c c}
\hline Variable & Distribution & mean & CoV \\
\hline $b$ [m] & Lognormal & 0.15 & 0.05 \\
$h$ [m] & Lognormal & 0.3  & 0.05 \\
$L$ [m] & Lognormal & 5    & 0.01 \\
$E$ [MPa] & Lognormal & 30,000  & 0.15 \\
$P$ [KN] & Lognormal & 10 & 0.20 \\
\hline       
\end{tabular}
\end{table}

We develop LRA and sparse PCE meta-models of $U=\cm(P,\hsp L,\hsp E,\hsp b,\hsp h)$ using two EDs of size $N=30$ and $N=50$. For both types of meta-models, we use Hermite polynomials to build the basis functions, after an isoprobabilistic transformation of the input variables to standard normal variables. In the LRA algorithm, we define the stopping criterion in the correction step by setting $\Imax=50$ and $\Derrmin= 10^{-8}$ (it was shown in \cite{Konakli2015arxiv} that in the considered problem, selecting a small value for $\Derrmin$ is critical for the LRA accuracy).  Parameters and error estimates of the LRA and PCE meta-models are listed in Tables~{\ref{tab:beam_LRA}} and~{\ref{tab:beam_PCE}},respectively. In particular, Table~{\ref{tab:beam_LRA}} lists the rank $R$ and polynomial degree $p$ of the LRA meta-model, the 3-fold CV error estimate $\errCV$ and the generalization error $\errG$. Table~{\ref{tab:beam_PCE}} lists the parameter $q$ controlling the truncation scheme and the total polynomial degree $p^t$ of the PCE meta-model, the corrected leave-one-out error $\errLOO$ and the generalization error $\errG$. The generalization errors are estimated using a validations set of size $\nval=10^6$ sampled with MCS. The ED-based error estimates $\errCV$ and $\errLOO$ are fairly close to the corresponding generalization errors except for the LRA meta-model when $N=50$. In the latter case, $\errCV$ underestimates $\errG$ by approximately one order of magnitude; however, $\errG$ is sufficiently small. Note that for each ED, the generalization error of LRA is 2-3 orders of magnitude smaller than that of sparse PCE, which can be justified by the rank-one structure of the herein considered model.

\begin{table} [!ht]
\centering
\caption{Beam-deflection problem: Parameters and error estimates of LRA meta-models.}
\label{tab:beam_LRA}
\begin{tabular}{c c c c c}
\hline
$N$ & $R$ & $p$ & $\errCV$ & $\errG$ \\
\hline
30 & 1 & 2  &  $1.21\cdot10^{-4}$ & $2.32\cdot10^{-4}$ \\
50 & 1 & 3  &  $ 3.14\cdot10^{-7}$ & $ 2.63\cdot10^{-6}$ \\
\hline       
\end{tabular}
\end{table}

\begin{table} [!ht]
\centering
\caption{Beam-deflection problem: Parameters and error estimates of PCE meta-models.}
\label{tab:beam_PCE}
\begin{tabular}{c c c c c}
\hline
$N$ & $q$ & $p^t$ & $\errLOO$ & $\errG$ \\
\hline
30 & 1 & 2  &  $3.13\cdot10^{-2}$ & $1.47\cdot10^{-2}$ \\
50 & 1 & 2  &  $1.56\cdot10^{-3}$ & $1.81\cdot10^{-3}$ \\
\hline       
\end{tabular}
\end{table}

\begin{figure}[!ht]
\centering
\includegraphics[width=0.45\textwidth] {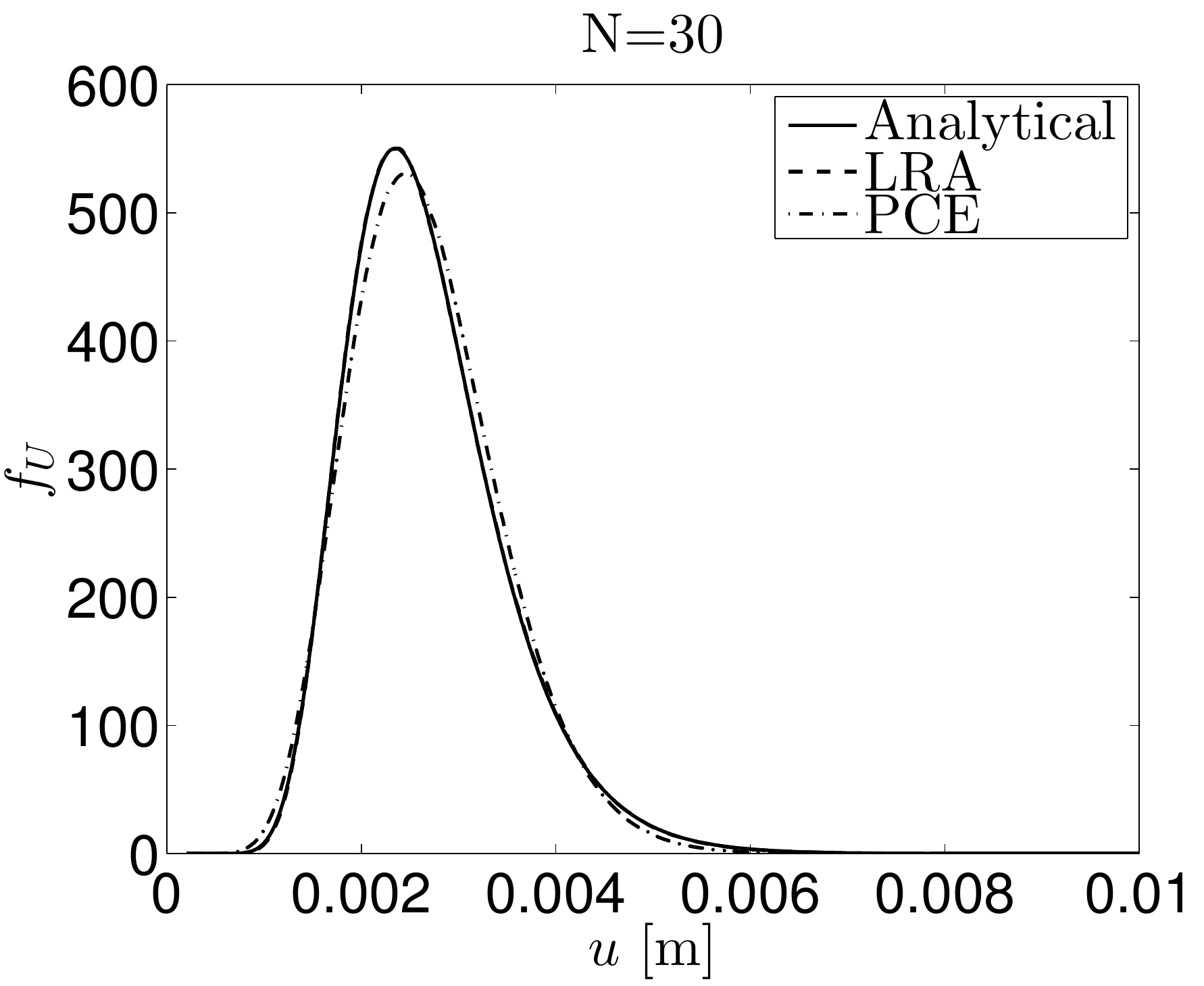}
\includegraphics[width=0.45\textwidth] {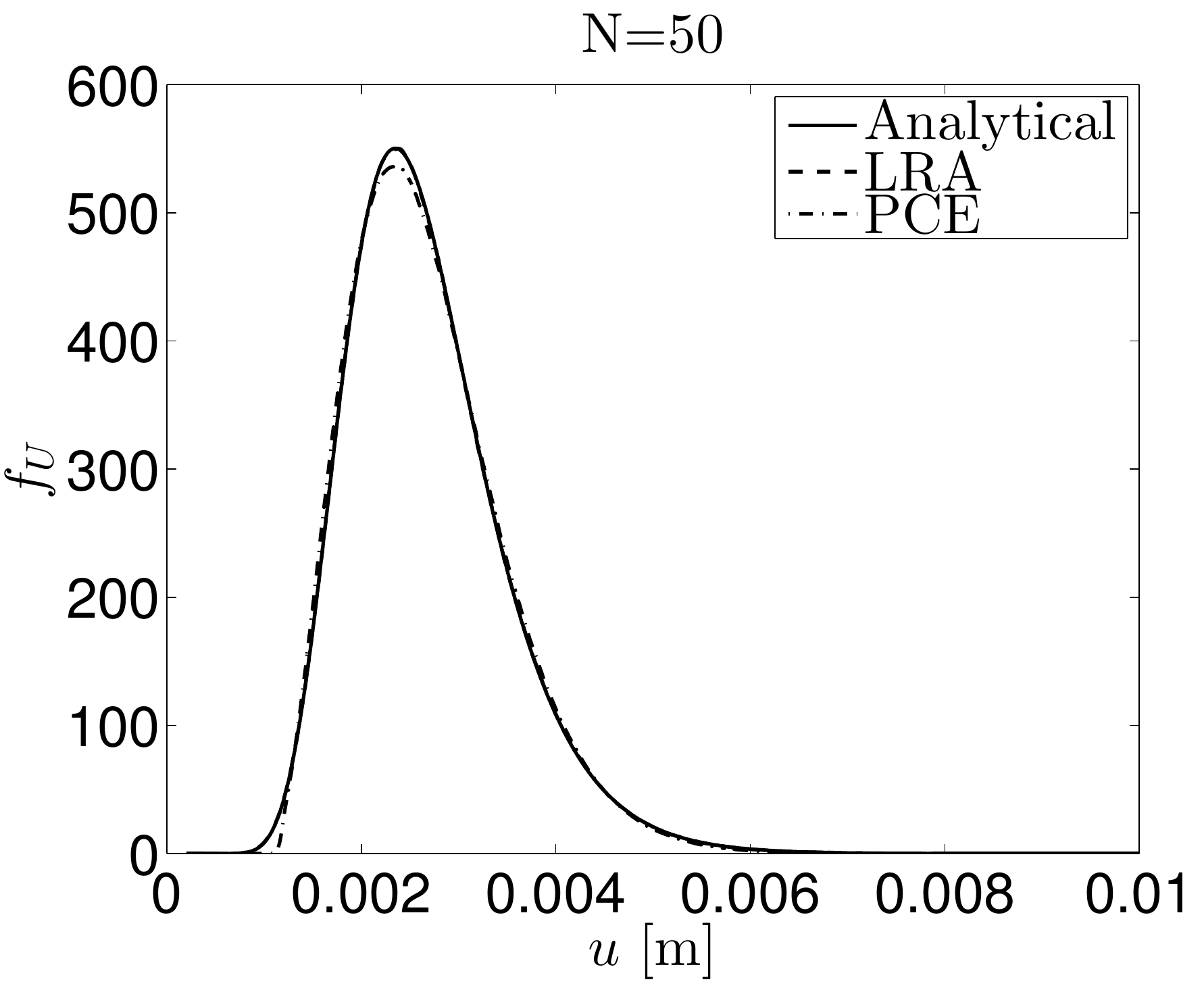}
\caption{Beam-deflection problem: Probability density function of the response (normal scale).}
\label{fig:beam_KDE}
\end{figure}

\begin{figure}[!ht]
\centering
\includegraphics[width=0.45\textwidth] {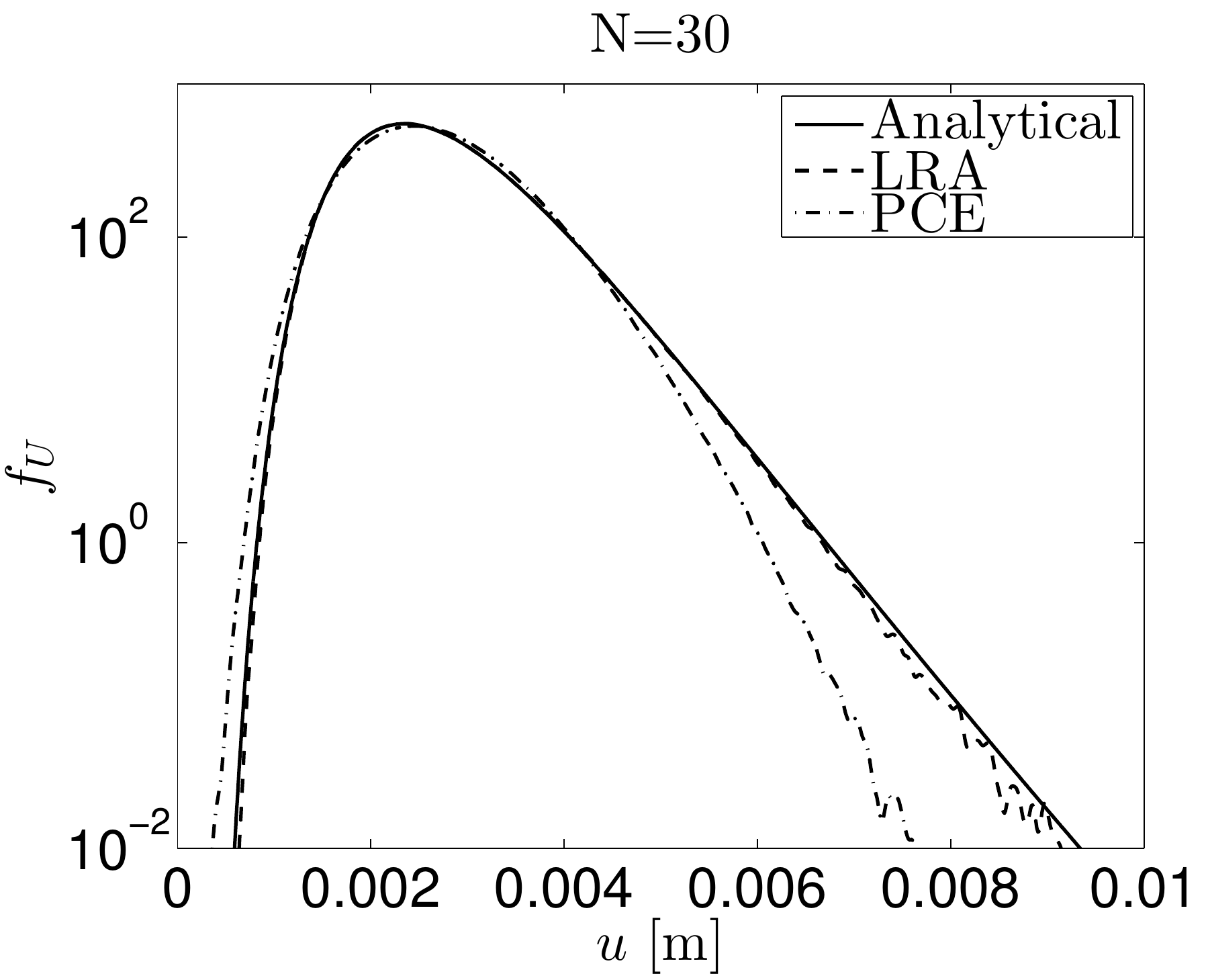}
\includegraphics[width=0.45\textwidth] {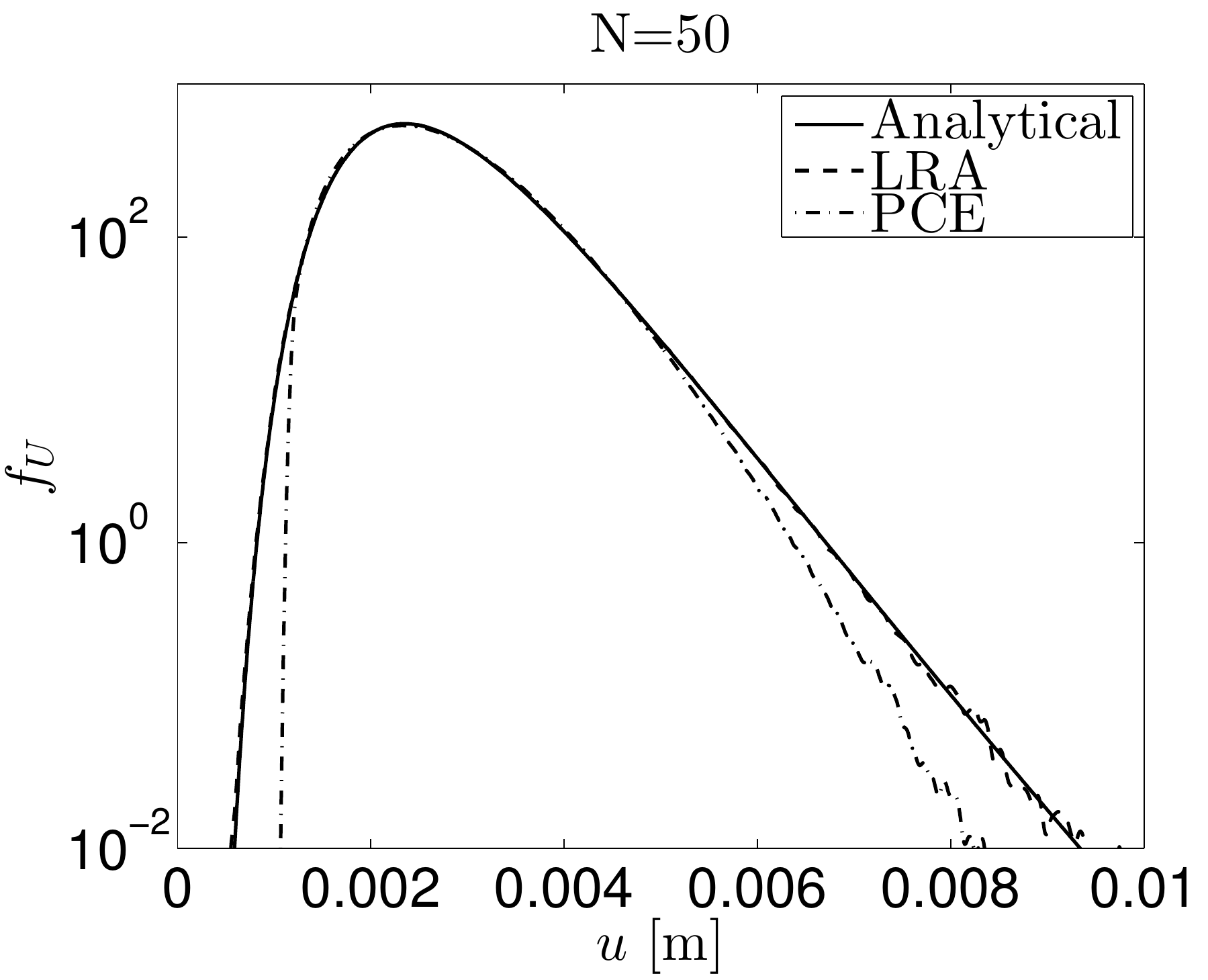}
\caption{Beam-deflection problem: Probability density function of the response (log-scale).}
\label{fig:beam_KDElog}
\end{figure}

In Figure~\ref{fig:beam_KDE}, we compare the analytical response PDF $f_U$ to the respective kernel density estimates (KDEs) obtained with the LRA and sparse PCE meta-models. The KDEs are based on the meta-model responses at a set of $n=10^7$ points in the input space sampled with MCS. In Figure~\ref{fig:beam_KDElog}, we show a similar comparison but using a logarithmic scale in the vertical axis in order to highlight the behavior at the tails of the PDF. Clearly, for the considered EDs of relatively small size, LRA yield superior estimates of the PDF as compared to PCE. It is remarkable that with the LRA approach, an ED of size as small as $N=30$ is sufficient to obtain an excellent approximation of $f_U$ in the normal scale and a fairly good approximation of the tails.

\begin{figure}[!ht]
\centering
\includegraphics[width=0.45\textwidth] {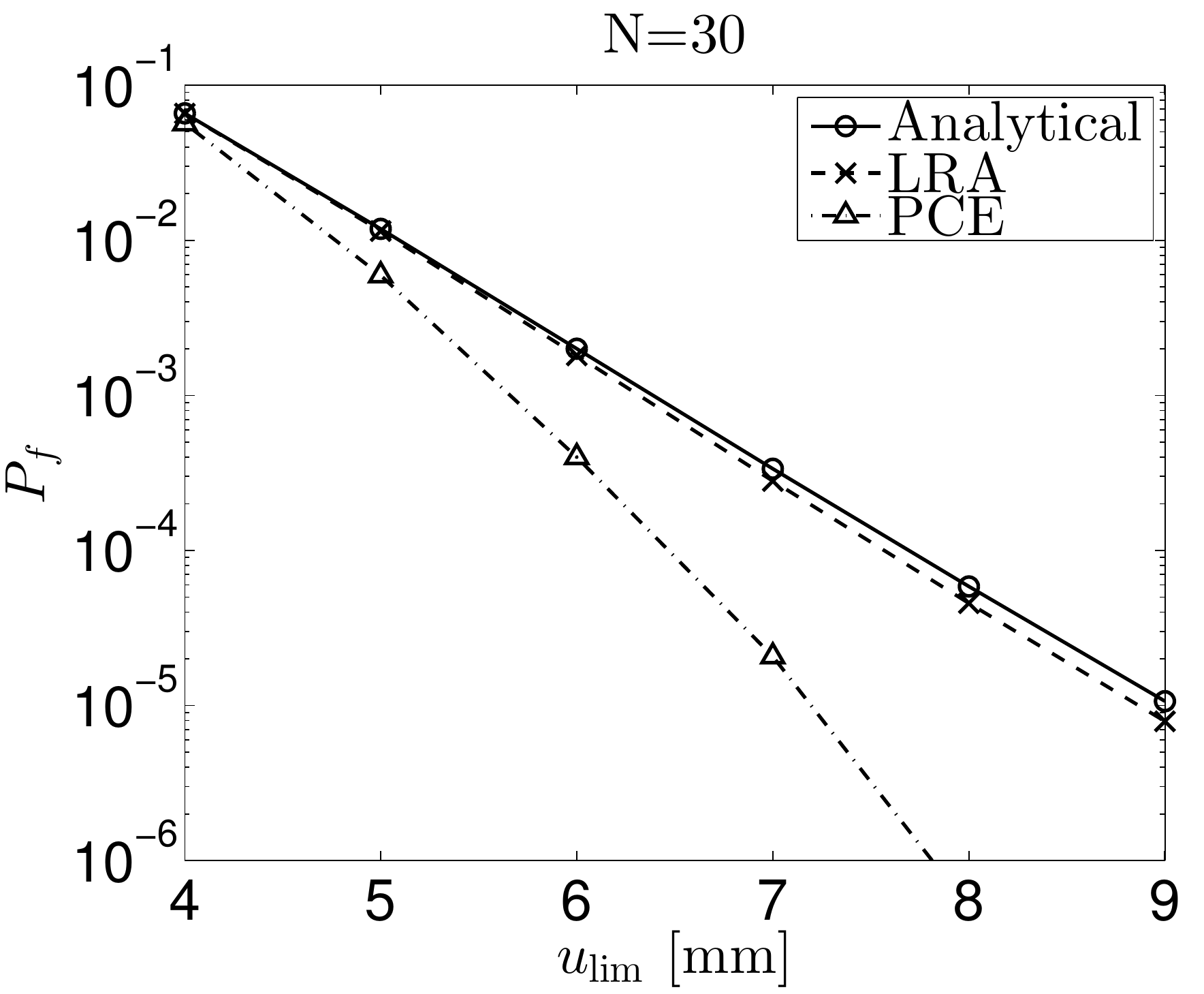}
\includegraphics[width=0.45\textwidth] {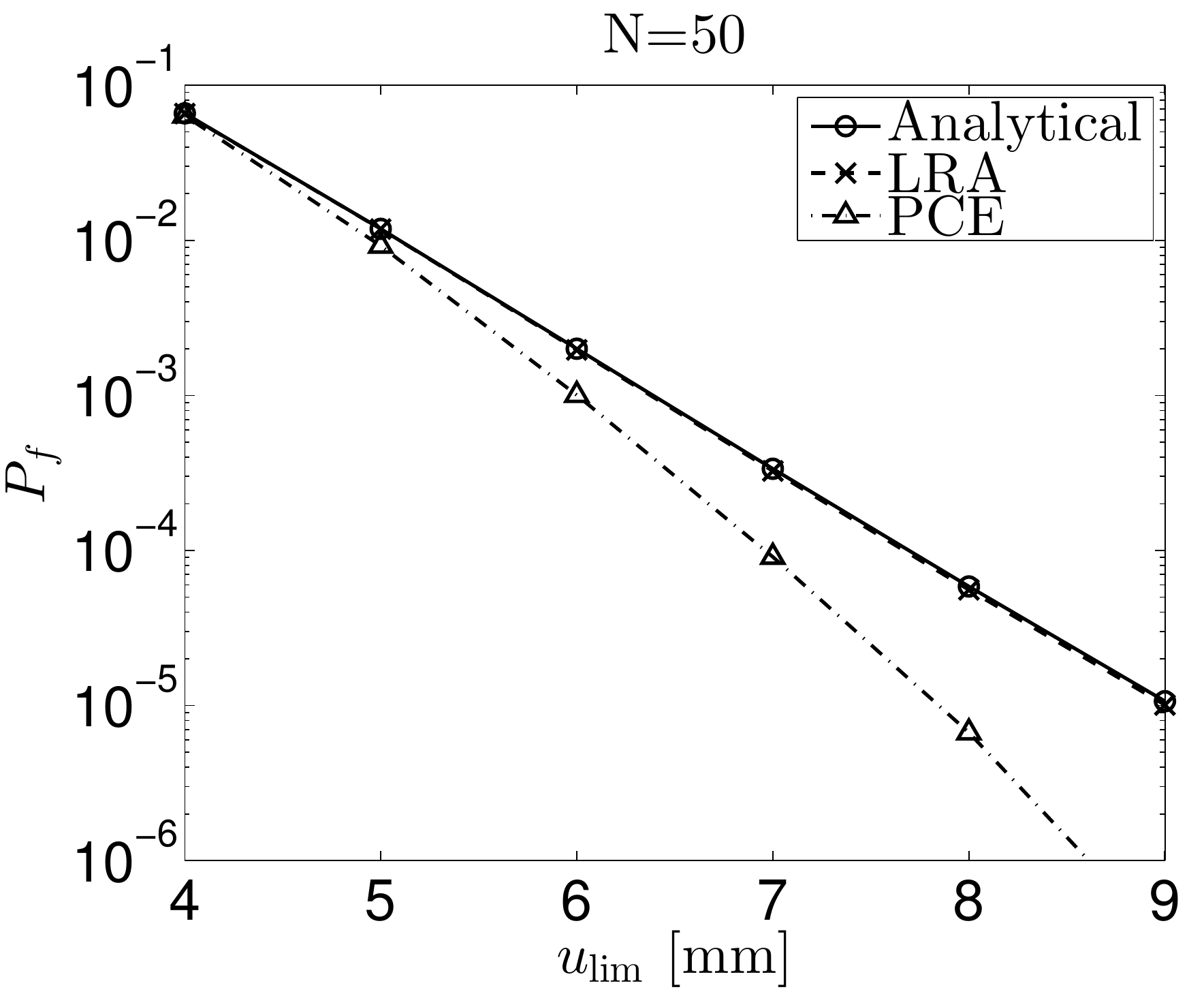}
\caption{Beam-deflection problem: Failure probabilities.}
\label{fig:beam_Pf}
\end{figure}

\begin{figure}[!ht]
\centering
\includegraphics[width=0.45\textwidth] {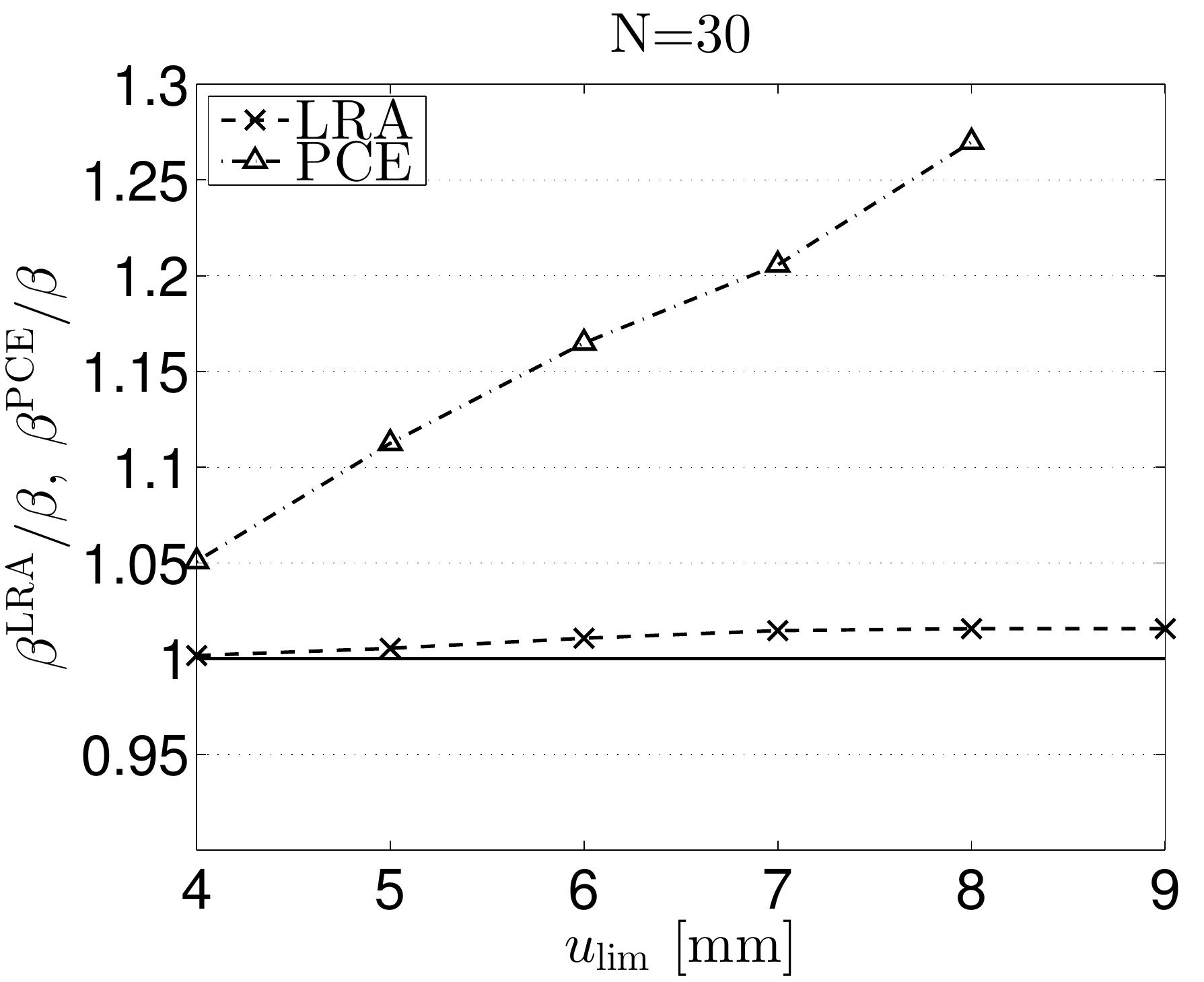}
\includegraphics[width=0.45\textwidth] {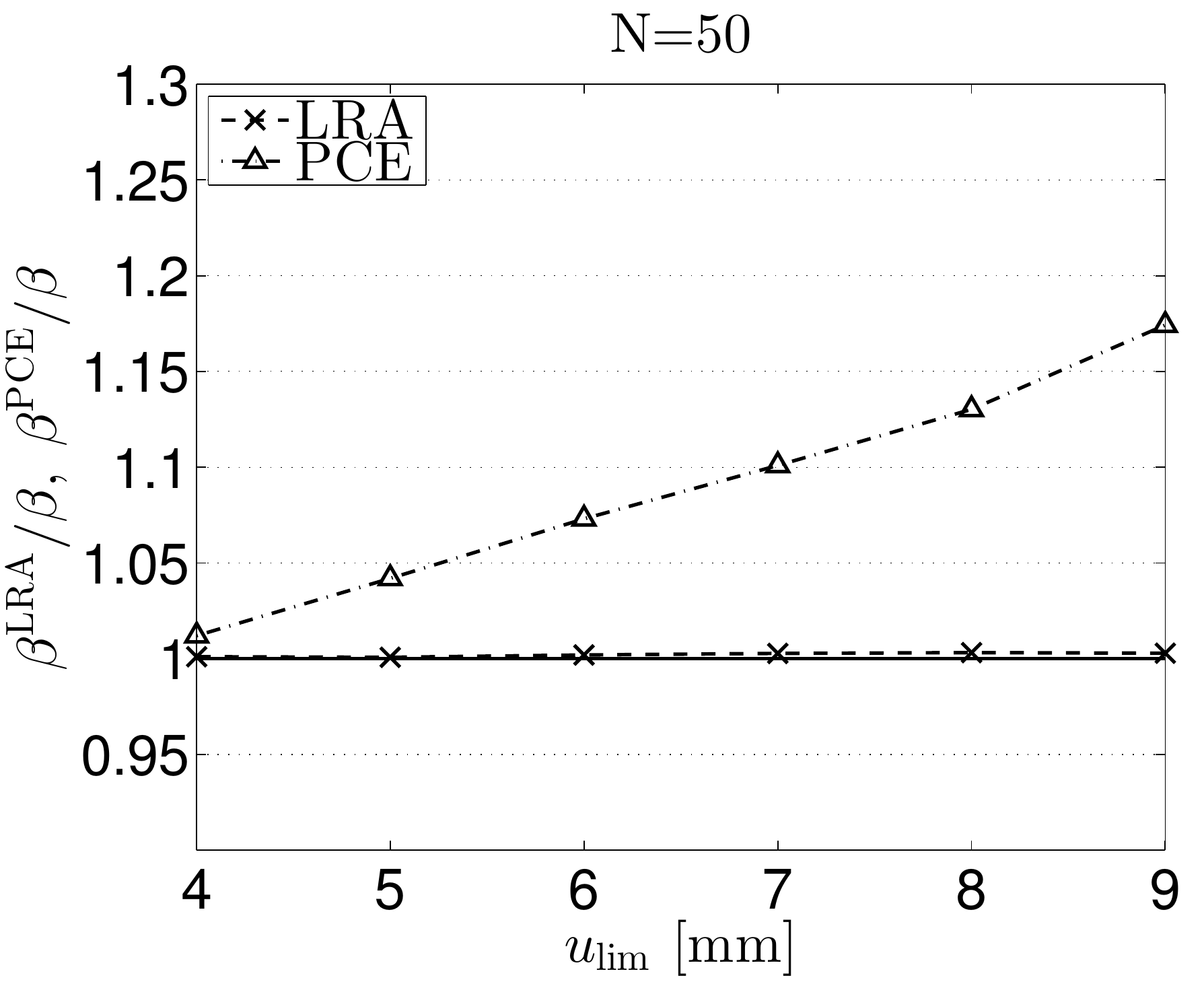}
\caption{Beam-deflection problem: Ratios of meta-model-based to reference reliability indices.}
\label{fig:beam_beta}
\end{figure}

In the sequel, we use the LRA and sparse PCE meta-models to estimate the failure probability  $P_f=\Pp(u_{\rm lim}-U\leq0)=\Pp(U\geq u_{\rm lim})$, \ie the probability that the beam deflection $U$ exceeds a prescribed threshold $u_{\rm lim}$. The estimates of the failure probability obtained with the meta-models are compared to the analytical solution. By varying the deflection threshold $u_{\rm lim}$ in the range $[4, 9]$~mm, the analytical failure probability varies in the range $[6.60 \cdot 10^{-2}, 1.07\cdot 10^{-5}]$. The LRA- and PCE-based estimates of the failure probabilities are obtained using a MCS approach with an input sample of size $n=10^7$. According to Eq.~(\ref{eq:CoV_Pf}), this sample size would be sufficient to estimate the smallest failure probability with a CoV$<0.10$ if the meta-models were exact representations of the actual model in Eq.~(\ref{eq:beam_u}). It is worth mentioning that the current implementation of PCE and LRA allows one to sample $10^7$ values in a matter of a few seconds using a standard desktop. Figure~\ref{fig:beam_Pf}  depicts the estimates of the failure probability for the two EDs of size $N=30$ and $N=50$ together with the respective analytical solutions. For $N=30$, the LRA-based estimates are close to the analytical solutions, especially for the smaller deflection thresholds, whereas the PCE-based estimates are overall highly inaccurate. For $N=50$, the LRA-based estimates are excellent in the entire range examined, whereas the PCE-based estimates remain poor for the larger deflection thresholds. Figure~\ref{fig:beam_beta} shows the corresponding ratios of the reliability index estimates based on the LRA and sparse PCE meta-models, respectively denoted by $\beta^{\rm LRA}$ and $\beta^{\rm PCE}$, to the reliability index based on the analytical solution, denoted by $\beta$. For the considered deflection thresholds, $\beta$ varies in the range $[1.51, 4.25]$. Note that with the LRA approach, we estimate the largest $\beta$ with a relative error smaller than $2\%$ using only $N=30$ evaluations of the actual model. The values of the failure probabilities and corresponding reliability indices depicted in Figures~\ref{fig:beam_Pf} and \ref{fig:beam_beta} are listed in the Appendix.

\subsection{Truss deflection}

\begin{figure}[!ht]
\centering
\includegraphics[trim = 15mm 70mm 15mm 70mm, width=0.6\textwidth]{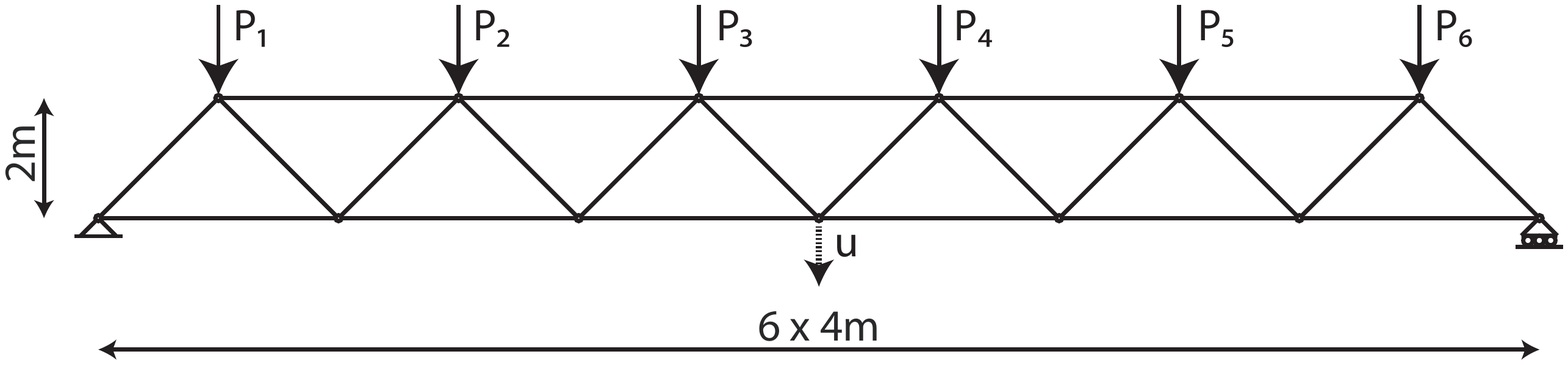}
\caption{Truss structure.}
\label{fig:006}
\end{figure}

In the second example, we conduct reliability analysis of the truss structure shown in Figure \ref{fig:006} with respect to the midspan deflection $u$. The random input comprises the six vertical loads, denoted by $P_1 \enum P_6$, the cross-sectional area and Young's modulus of the horizontal bars, respectively denoted by $A_1$ and $E_1$, and the cross-sectional area and Young's modulus of the vertical bars, respectively denoted by $A_2$ and $E_2$. The distributions of the input random variables are listed in Table \ref{tab:truss_input}. The deflection is computed with an in-house finite-element analysis code developed in the Matlab environment.

\begin{table} [!ht]
\centering
\caption{Truss-deflection problem: Distributions of input random variables.}
\label{tab:truss_input}
\begin{tabular}{c c c c}
\hline Variable & Distribution & mean & CoV \\
\hline
$A_1$ [m] & Lognormal & 0.002 & 0.10 \\
$A_2$ [m] & Lognormal & 0.001  & 0.10 \\
$E_1, E_2$ [MPa] & Lognormal & 210,000   & 0.10\\
$P_1\enum P_6$ [KN] & Gumbel & 50  & 0.15 \\
\hline       
\end{tabular}
\end{table}

We develop LRA and sparse PCE meta-models of $U=\cm(A_1, A_2, E_1, E_2, P_1\enum P_6)$ using two EDs of size $N=50$ and $N=100$. For both types of meta-models, we use Hermite polynomials to build the basis functions, after an isoprobabilistic transformation of the input variables to standard normal variables. In the LRA algorithm, we define the stopping criterion in the correction step by setting $\Imax=50$ and $\Derrmin= 10^{-6}$.  Parameters and error estimates of the LRA and PCE meta-models are listed in Tables~{\ref{tab:truss_LRA}} and {\ref{tab:truss_PCE}}, respectively. The generalization errors $\errG$ are estimated using a validation set of size $\nval=10^6$ sampled with MCS. For $N=50$, the generalization error of LRA is nearly an order of magnitude smaller than that of sparse PCE. For $N=100$, the generalization errors of the two types of meta-models are fairly close.

\begin{figure}[!ht]
\centering
\includegraphics[width=0.45\textwidth] {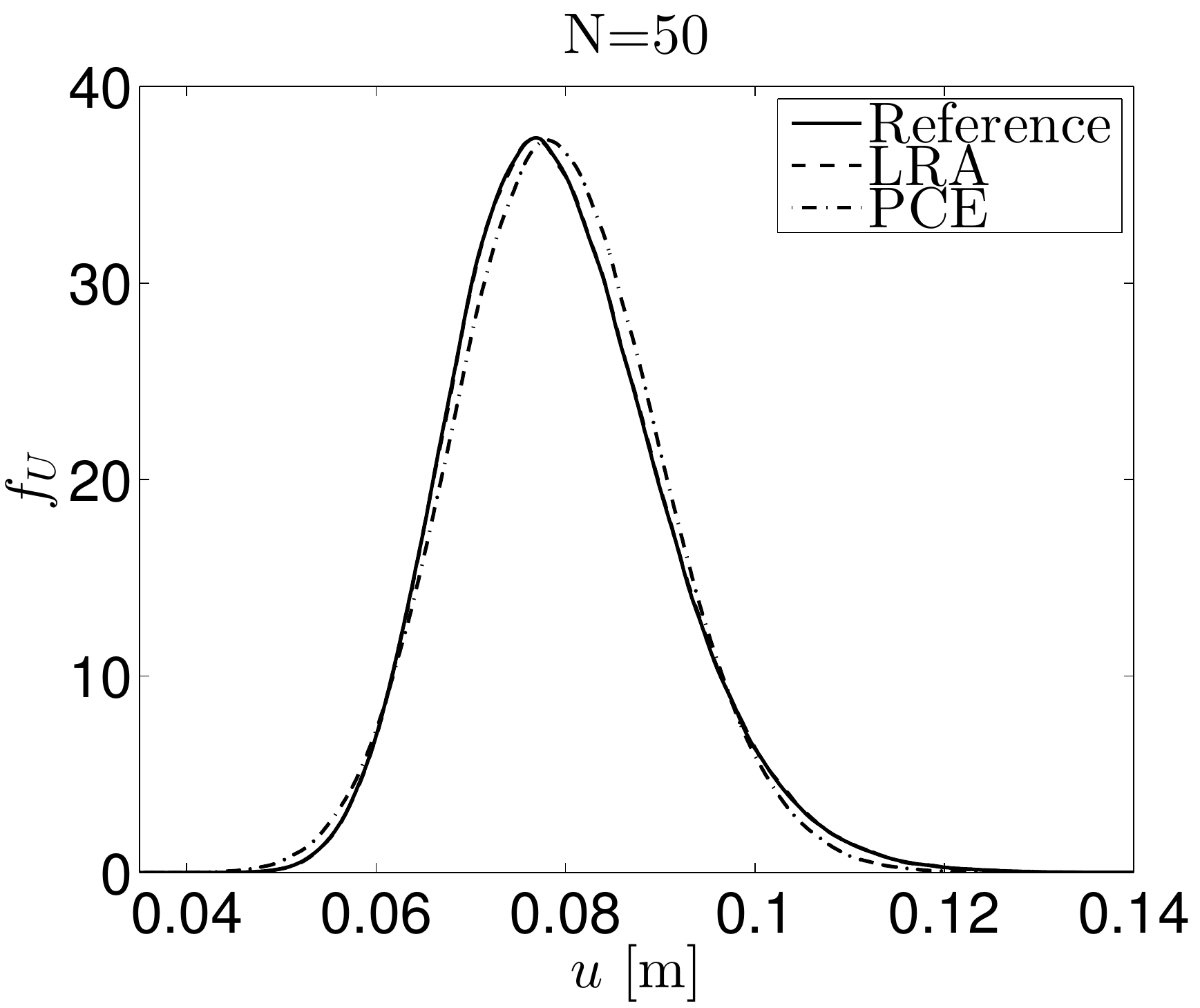}
\includegraphics[width=0.45\textwidth] {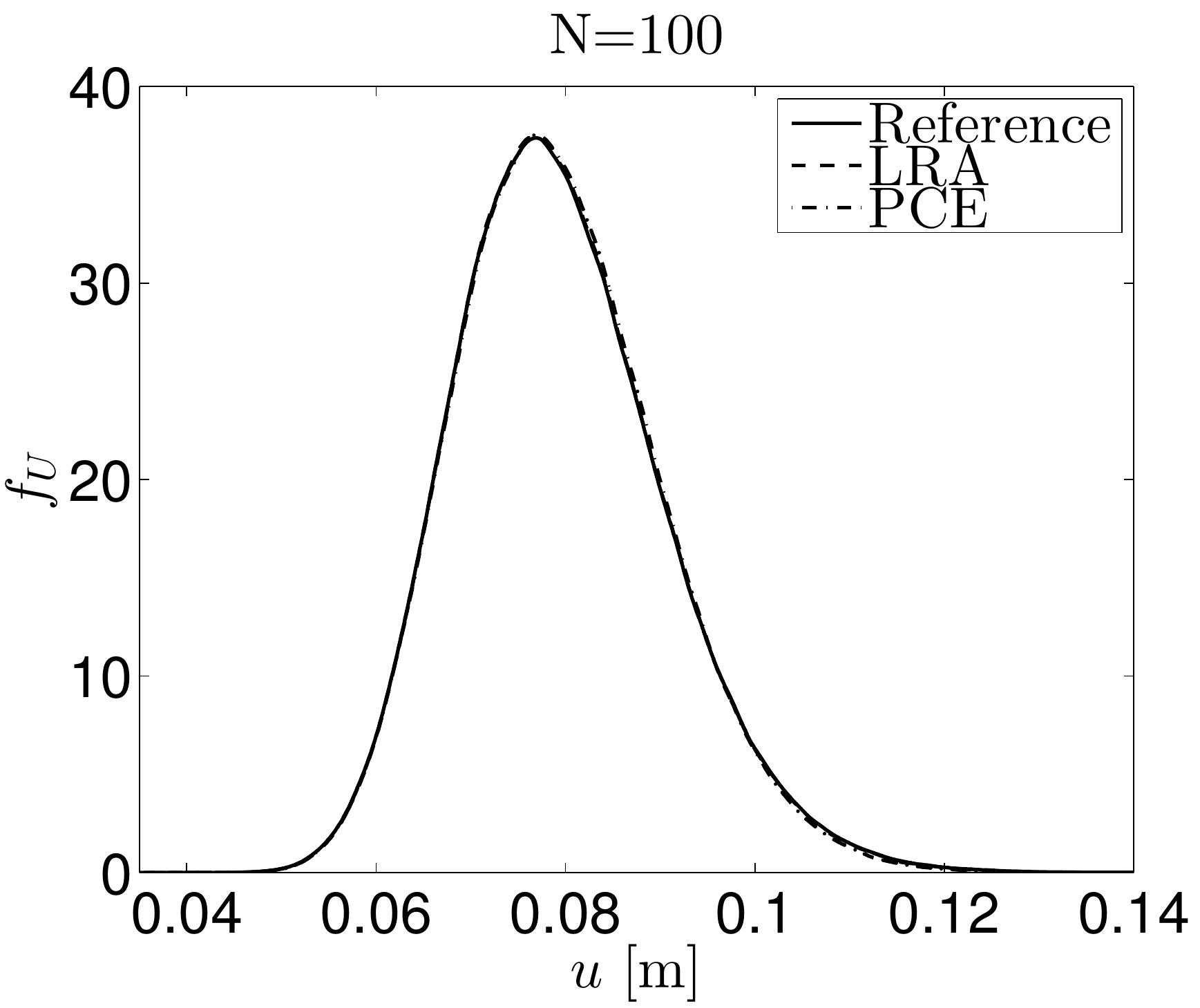}
\caption{Truss-deflection problem: Probability density function of the response (normal scale).}
\label{fig:truss_KDE}
\end{figure}

\begin{figure}[!ht]
\centering
\includegraphics[width=0.45\textwidth] {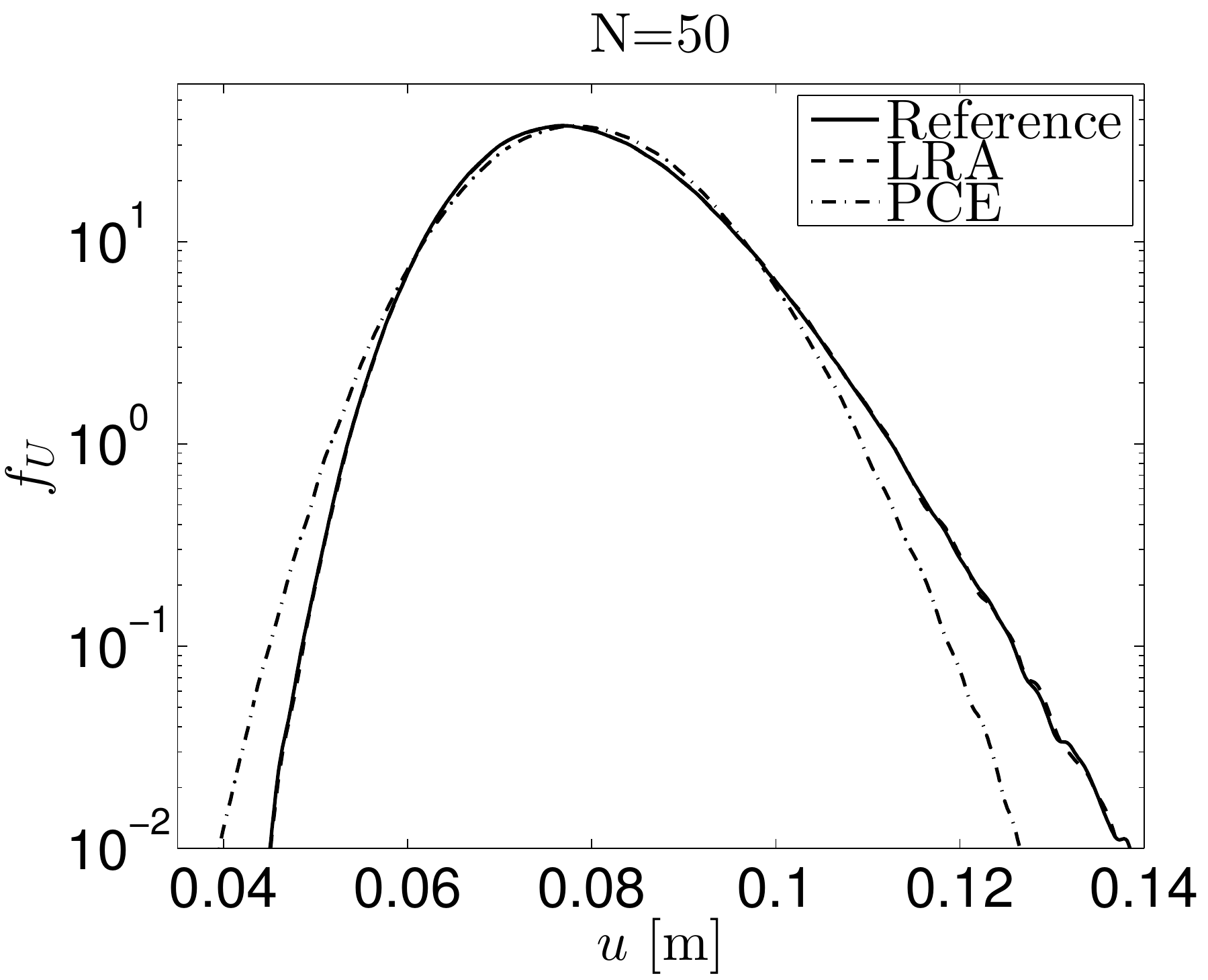}
\includegraphics[width=0.45\textwidth] {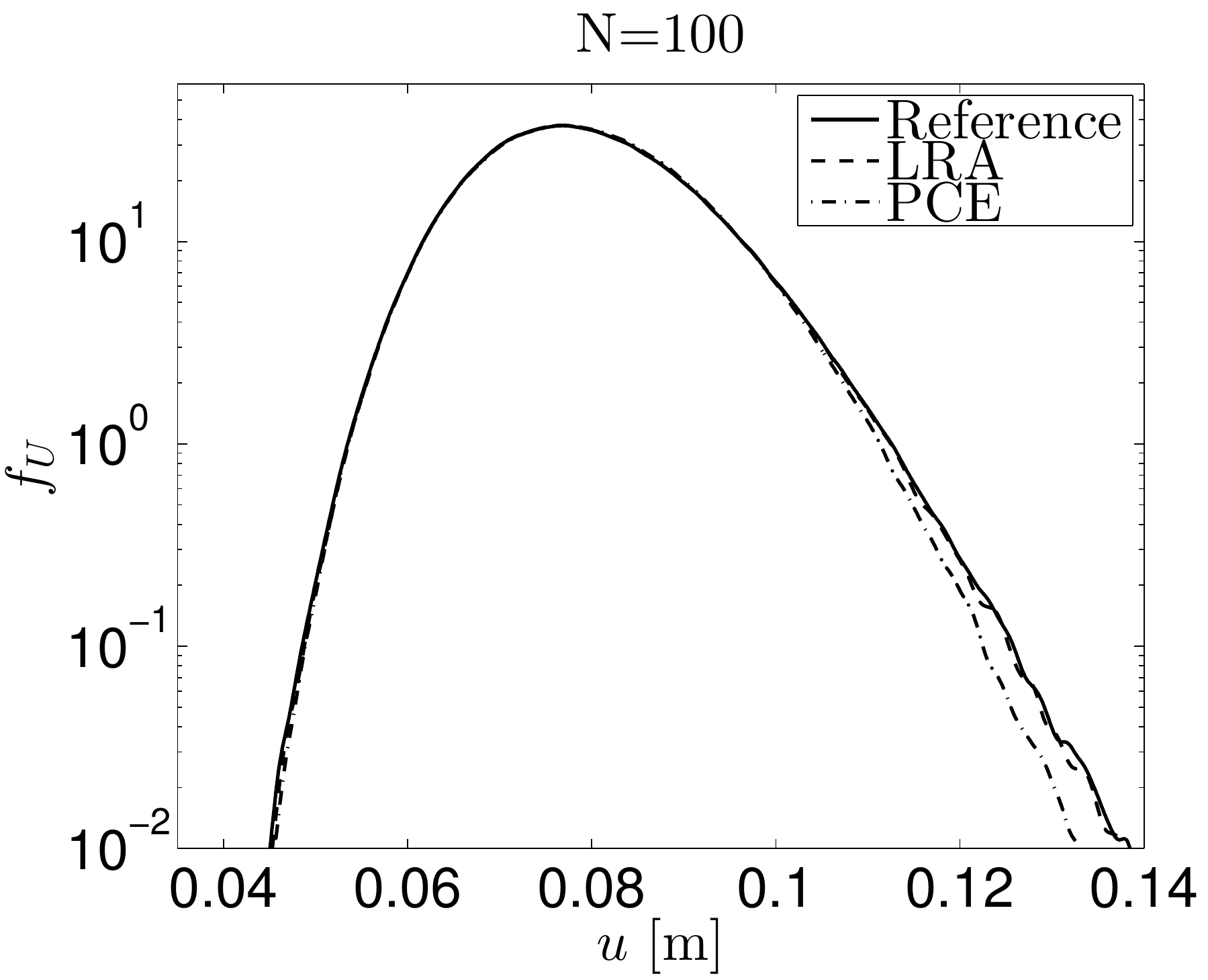}
\caption{Truss-deflection problem: Probability density function of the response (log-scale).}
\label{fig:truss_KDElog}
\end{figure}

In Figure \ref{fig:truss_KDE}, we compare the KDEs of the response PDF $f_U$ obtained with the LRA and sparse PCE meta-models with that obtained with the actual model, which is considered the reference solution for $f_U$. All aforementioned KDEs are based on the evaluation of the different models at a MCS sample of $n=10^6$ points in the input space. In Figure \ref{fig:truss_KDElog}, the same KDEs are shown in the logarithmic scale in order to emphasize the behavior at the tails. When the LRA approach is employed, the ED of size $N=50$ is sufficient to approximate the response PDF with high accuracy in its entire range including the tails. The PCE solution converges more slowly to the reference solution. For $N=50$, the discrepancy between the PCE-based KDE from the reference one is obvious even in the normal scale; for $N=100$, the PCE-based KDE remains inaccurate for $u>0.11$~m ($99.1$-th percentile).

\begin{table} [!ht]
\centering
\caption{Truss-deflection problem: Parameters and error estimates of LRA meta-models.}
\label{tab:truss_LRA}
\begin{tabular}{c c c c c}
\hline
$N$ & $R$ & $p$ & $\widehat{err}_{\rm CV3}$ & $\widehat{err}_G$ \\
\hline
50 & 1 & 2  &  $6.26\cdot10^{-3}$ & $2.85\cdot10^{-3}$ \\
100 & 1 & 2  &  $3.33\cdot10^{-3}$ & $  2.10\cdot10^{-3}$ \\
\hline       
\end{tabular}
\end{table}

\begin{table} [!ht]
\centering
\caption{Truss-deflection problem: Parameters and error estimates of PCE meta-models.}
\label{tab:truss_PCE}
\begin{tabular}{c c c c c}
\hline
$N$ & $q$ & $p^t$ & $\widehat{err}_{\rm LOO}^*$ & $\widehat{err}_G$ \\
\hline
50 & 0.25 & 2  &  $4.25\cdot10^{-2}$ & $1.24\cdot10^{-2}$ \\
100 & 1 & 2  &  $3.50\cdot10^{-3}$ & $2.56\cdot10^{-3}$ \\
\hline       
\end{tabular}
\end{table}

\begin{figure}[!ht]
\centering
\includegraphics[width=0.45\textwidth] {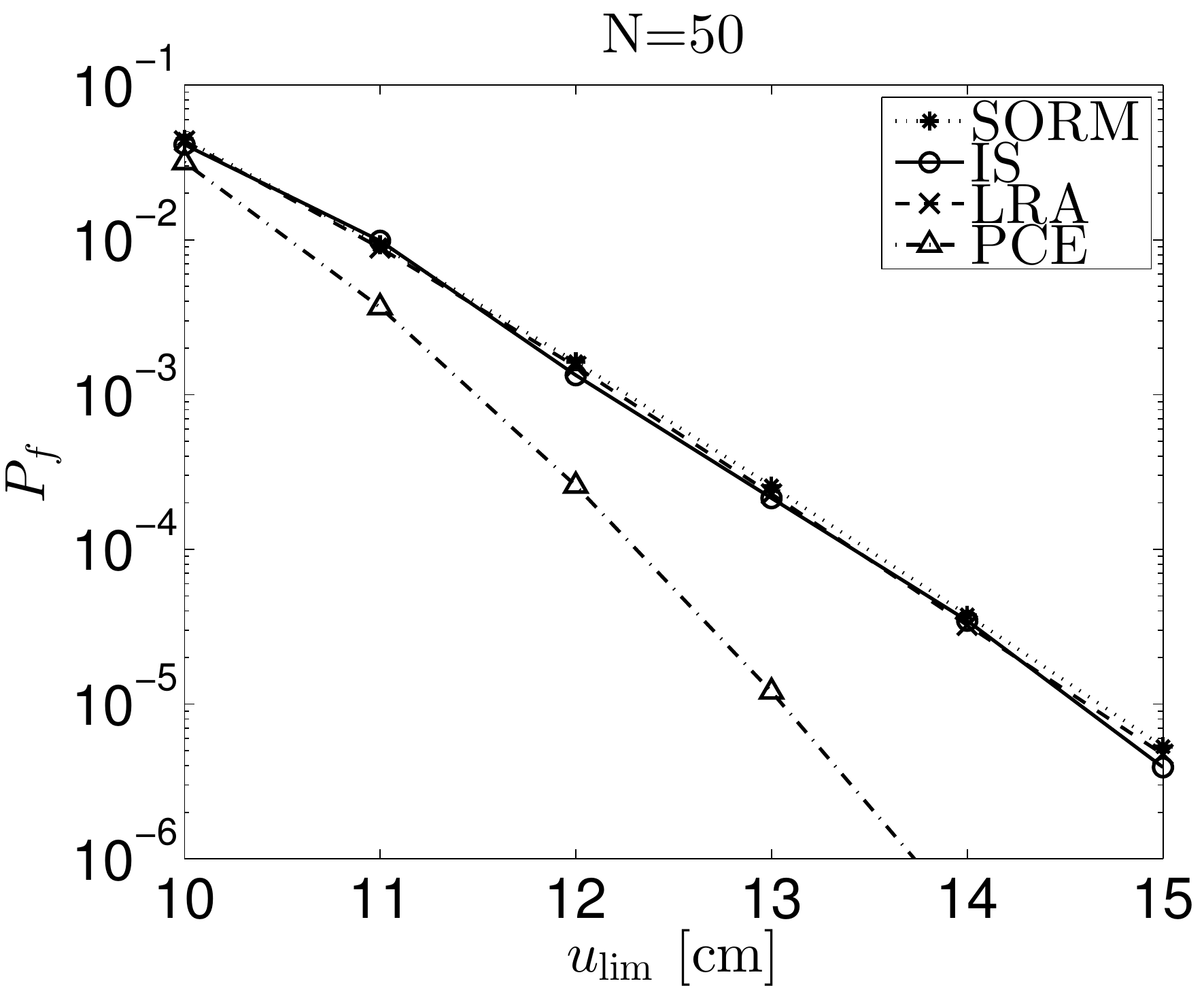}
\includegraphics[width=0.45\textwidth] {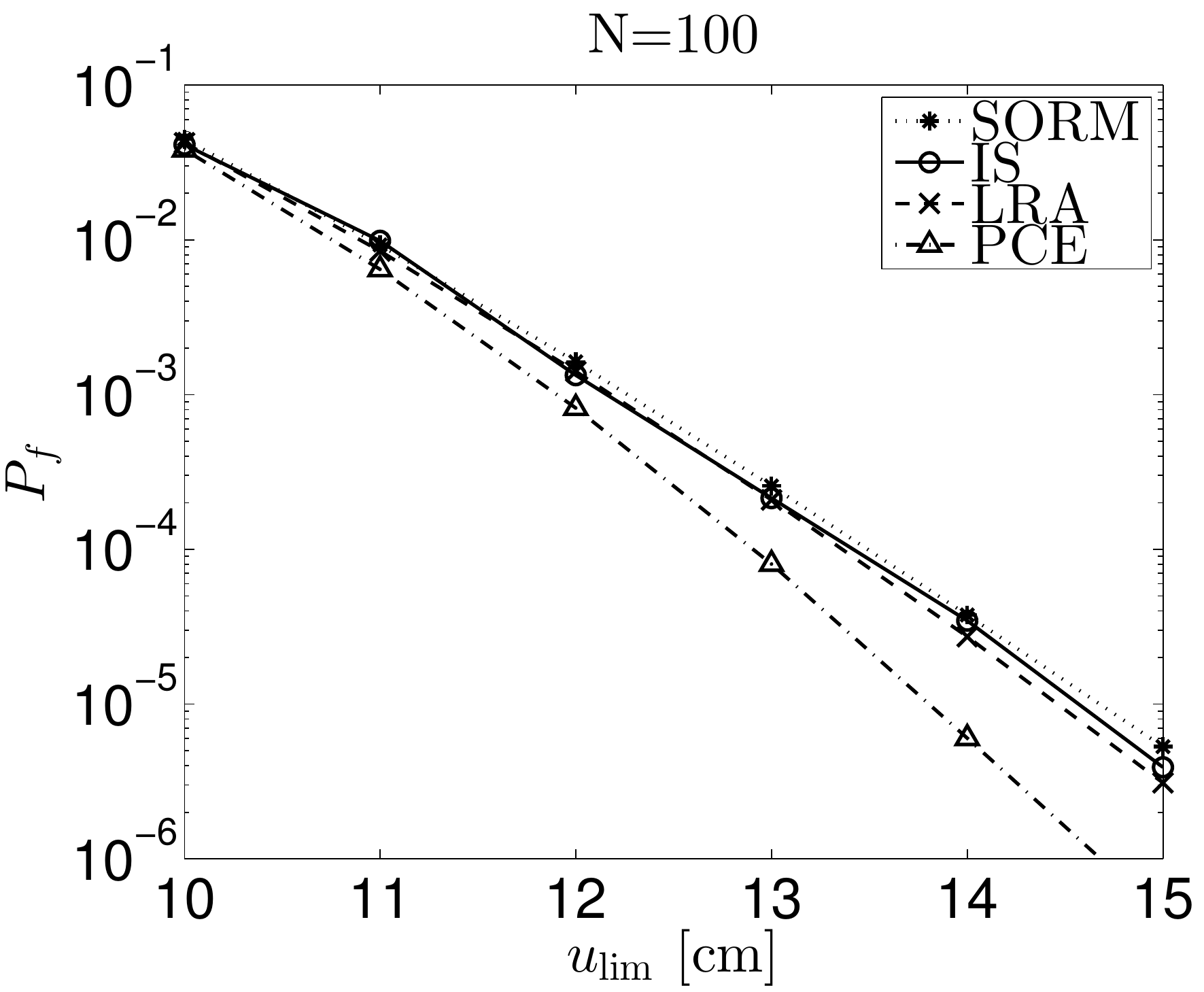}
\caption{Truss-deflection problem: Failure probabilities.}
\label{fig:truss_Pf}
\end{figure}

\begin{figure}[!ht]
\centering
\includegraphics[width=0.45\textwidth] {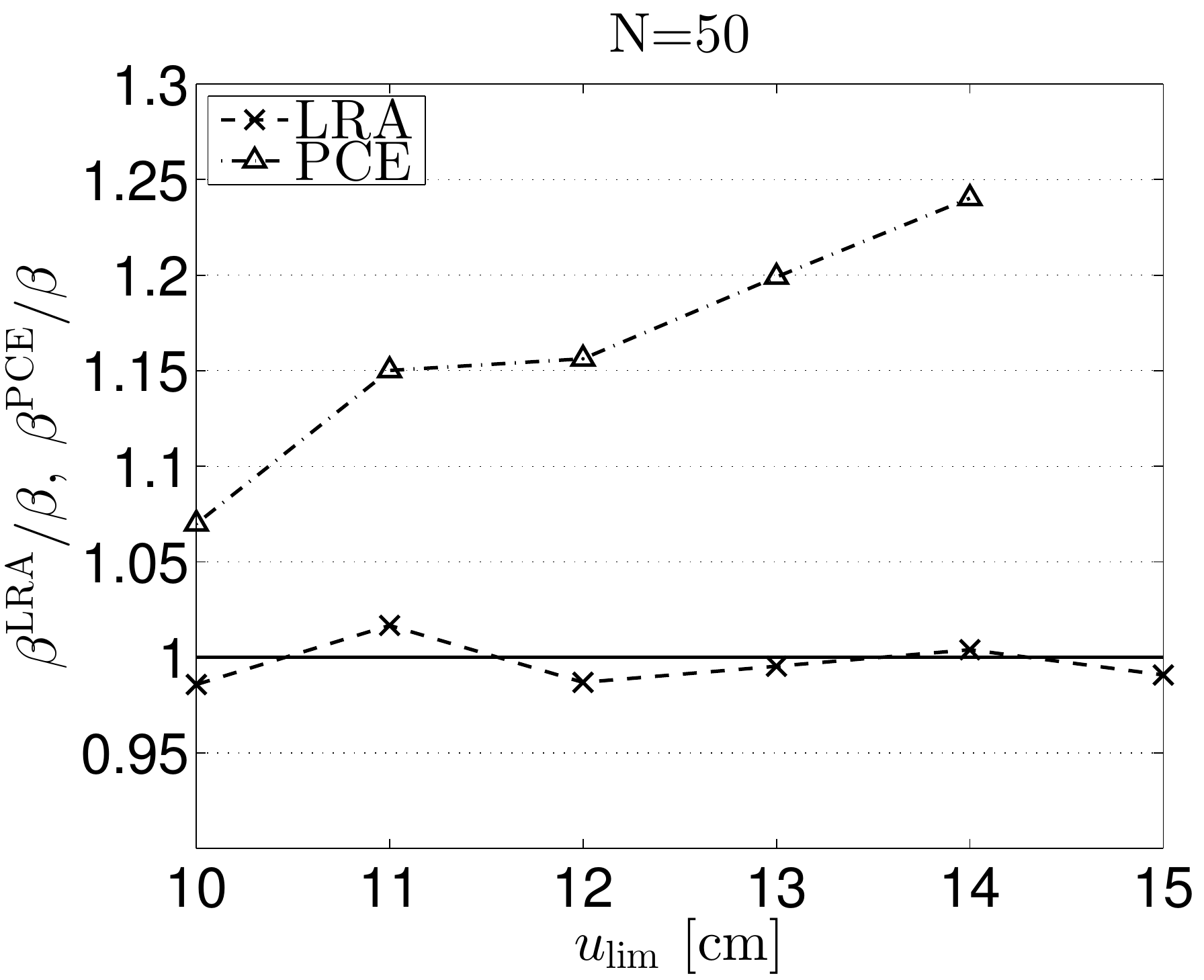}
\includegraphics[width=0.45\textwidth] {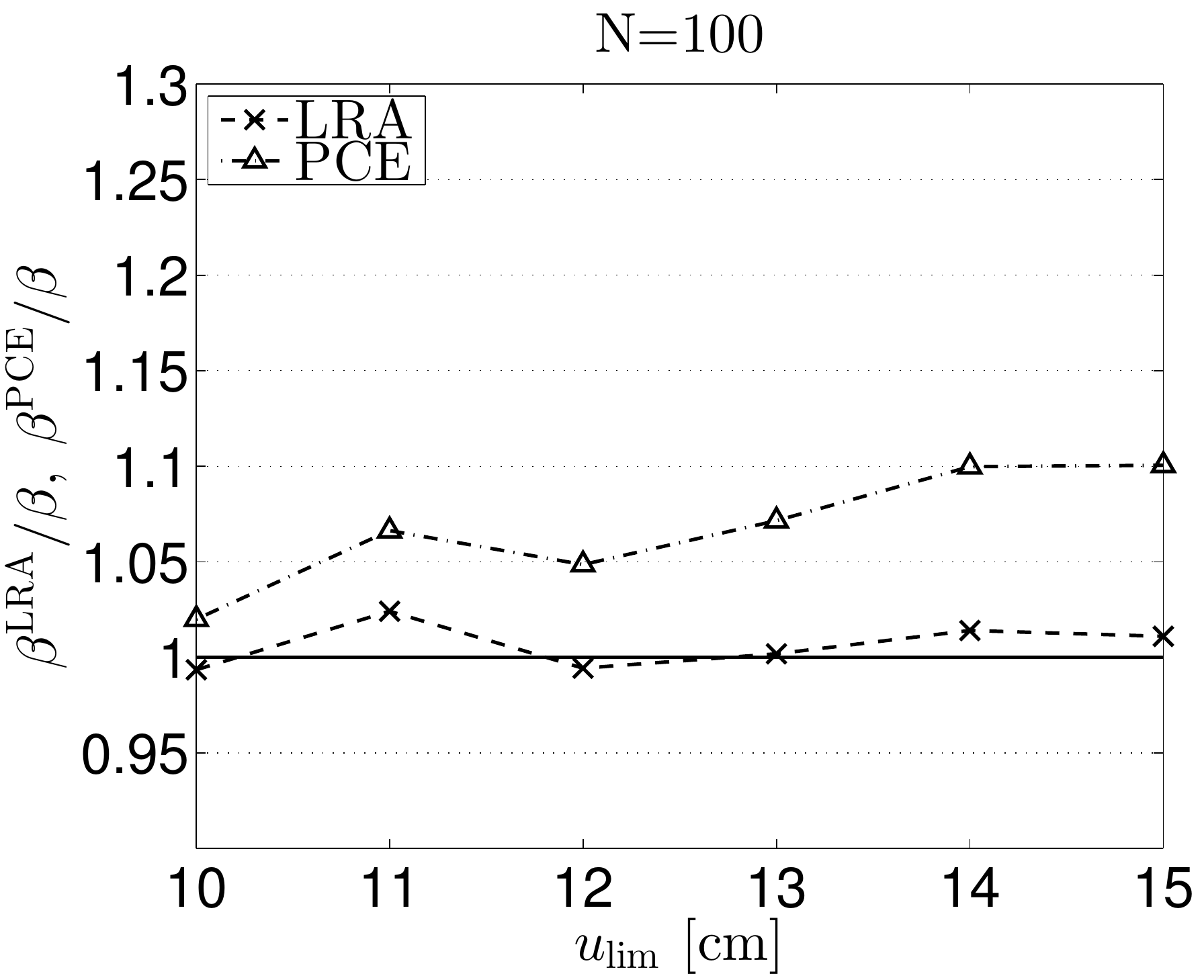}
\caption{Truss-deflection problem: Ratios of meta-model-based to reference reliability indices.}
\label{fig:truss_beta}
\end{figure}

Next, we assess the comparative accuracy of LRA and sparse PCE in estimating the failure probabilities $P_f=\Pp(u_{\rm lim}-U\leq0)=\Pp(U\geq u_{\rm lim})$, with the deflection threshold $u_{\rm lim}$ varying in $[10, 15]$~cm. The LRA- and PCE-based estimates are compared with respective reference values obtained with (i) SORM and (ii) IS (see Section \ref{sec:rel} for a brief description of these methods). In the IS approach, we utilize the results of a previous analysis with FORM and sequentially add samples of size $N_{\rm IS}=100$ until the coefficient of variation of the estimated failure probability becomes smaller than $0.10$. The SORM- and IS-based failure probabilities are computed with the software UQLab \cite{MarelliICVRAM2014, UQdoc_09_107}. The IS-based estimates, considered the reference solution, vary in the range $[4.13 \cdot 10^{-2}, 3.90 \cdot 10^{-6}]$. The LRA- and PCE-based estimates are obtained using a MCS approach with an input sample of size $n=3\cdot10^7$, which is sufficient to estimate the smallest failure probability with $\rm CoV<0.10$, \ie with a coefficient of variation similar to that of the reference IS solution. The results are shown in Figure \ref{fig:truss_Pf}. It is remarkable that with the LRA approach, an ED of size as small as $N=50$ proves sufficient to evaluate failure probabilities of the order of $10^{-6}$. The PCE-based estimates converge to the reference solution with increasing $N$, but at a slower rate than LRA. Figure~\ref{fig:truss_beta} shows the corresponding ratios of the LRA- and PCE-based reliability indices, $\beta^{\rm LRA}$ and $\beta^{\rm PCE}$ respectively, to the reference reliability indices $\beta$, evaluated from the IS estimates of the failure probabilities. The latter varies in the range $[1.74, 4.47]$. The relative difference between the LRA-based and the reference reliability indices in all cases do not exceed $3\%$. The values of the failure probabilities and corresponding reliability indices depicted in Figures~\ref{fig:truss_Pf} and \ref{fig:truss_beta} are listed in the Appendix.

\begin{figure}[!ht]
\centering
\includegraphics[width=0.45\textwidth] {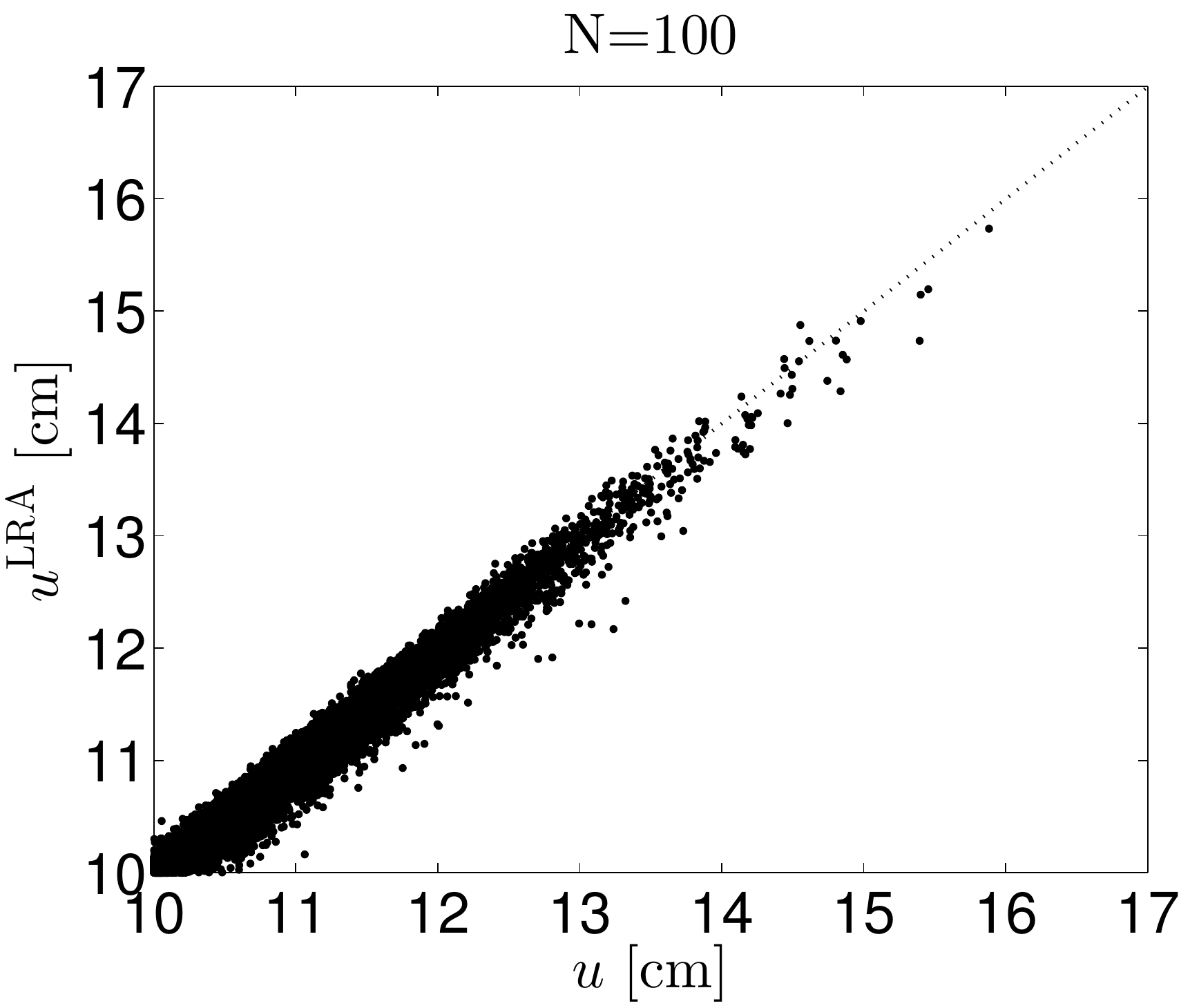}
\includegraphics[width=0.45\textwidth] {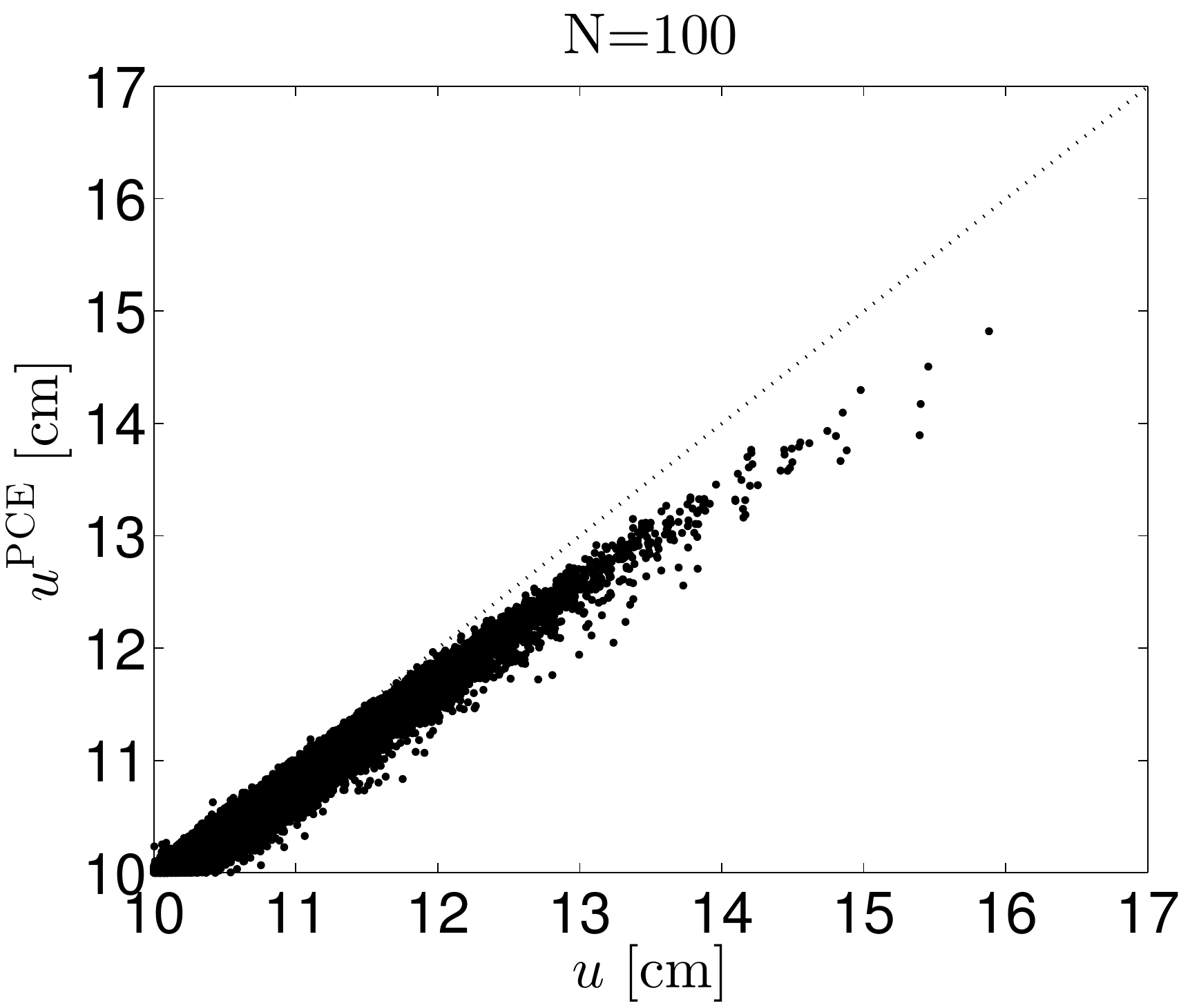}
\caption{Truss-deflection problem: Meta-model versus actual model responses at a subset of the validation set corresponding to the upper tail of the response distribution.}
\label{fig:truss_u_uhat}
\end{figure}

It is worth noting that although the two types of meta-models obtained with $N=100$ are characterized by similar generalization errors, use of LRA leads to significantly superior estimates of the tail probabilities. This can be explained by examining the behavior of the meta-models at the upper tail of the response distribution. In Figure~\ref{fig:truss_u_uhat}, we plot the responses of the LRA and sparse PCE meta-models, denoted by $u^{\rm LRA}$ and $u^{\rm PCE}$ respectively, versus the actual model responses, denoted by $u$, at the points $\ve x$ of the validation set satisfying the condition $u=\cm(\ve x)\geq10~\rm cm$. The figure shows that LRA clearly outperform PCE at the upper tail of the response distribution, with the latter yielding obviously biased values. Because the considered points of the validation set belong to the upper $5$-th percentile of the response distribution, they have a rather small contribution to the generalization error. The above observations become more pronounced by considering smaller percentiles at the upper tail of the response distribution. In order to capture the meta-model performance in particular regions of interest, we introduce the \emph{conditional generalization error}:
\begin{equation}
\label{eq:hatErrGcond}
\widehat{Err}_G^{\rm C} =\left\|\cm-\widehat {\cm}\right\|_{\cx_{\rm val}^{\rm C}}^2.
\end{equation}
The conditional generalization error is computed similarly to the generalization error in Eq.~(\ref{eq:hatErrG}), but by considering only a subset $\cx_{\rm val}^{\rm C}$ of the validation set $\cx_{\rm val}$, defined by an appropriate condition. The corresponding relative error is obtained after normalization with the empirical variance of $\cy_{\rm val}^{\rm C}$, which denotes the set of model responses at $\cx_{\rm val}^{\rm C}$. In reliability analysis, we are interested in conditional errors evaluated at subsets of the validation set defined as:
\begin{equation}
\label{eq:hatErrGcond2}
\cx_{\rm val}^{\rm C}=\{\ve x \in \cx_{\rm val}: u=\cm(\ve x)\geq u_{\rm lim}\}.
\end{equation}
In Table~\ref{tab:truss_errcond}, we list the relative conditional generalization errors of the LRA and sparse PCE meta-models obtained with $N=100$, considering the same values of $u_{\rm lim}$ as in the above reliability analysis. These errors are significantly smaller for LRA than for PCE (about one order of magnitude smaller for the larger response thresholds), which is consistent with Figure~\ref{fig:truss_u_uhat} and with the results of the reliability analysis.
 
\begin{table} [!ht]
\centering
\caption{Truss-deflection problem: Relative conditional generalization errors of the LRA and PCE meta-models obtained with $N=100$ for different response thresholds.}
\label{tab:truss_errcond}
\begin{tabular}{c c c}
\hline
$u_{\rm lim}$ [cm] & LRA & PCE \\
\hline
10 & $2.89\cdot10^{-2}$ & $8.68\cdot10^{-2}$ \\
11 & $5.70\cdot10^{-2}$ & $2.58\cdot10^{-1}$ \\
12 & $1.04\cdot10^{-1}$ & $6.45\cdot10^{-1}$ \\
13 & $2.00\cdot10^{-1}$ & $1.45\cdot10^{0~}$ \\
14 & $3.94\cdot10^{-1}$ & $3.55\cdot10^{0~}$ \\
15 & $2.67\cdot10^{0~}$ & $2.62\cdot10^{1~}$ \\
\hline   
\end{tabular}
\end{table}

To further highlight the efficiency of the LRA meta-modeling approach, in Table~\ref{tab:truss_eval}, we list the number of evaluations of the actual (finite-element) model required by SORM and IS to compute the failure probability for each threshold. The given number for IS includes the model evaluations required to obtain the FORM estimate. To limit the number of model evaluations in the sequential analyses performed for the increasing thresholds, we start a new FORM analysis from the previous design point. Note that by using the LRA meta-model, we can obtain $P_f$ values similar to those computed with SORM and IS, while relying on a much smaller number of model evaluations. We underline that once a meta-model is built, the failure probability for any threshold can be estimated without any additional model evaluations, whereas a new set of model evaluations for each threshold is required by the FORM, SORM and IS techniques.

\begin{table} [!ht]
\centering
\caption{Truss-deflection problem: Number of model evaluations required in the computation of failure probabilities with SORM and IS.}
\label{tab:truss_eval}
\begin{tabular}{c c c}
\hline $u_{\rm lim}$ [cm] & SORM & IS \\
\hline
10 & 387 & 475 \\         
11 & 285 & 473 \\         
12 & 297 & 585 \\         
13 & 309 & 597 \\         
14 & 321 & 709 \\         
15 & 333 & 921 \\
\hline       
\end{tabular}
\end{table}

\subsection{Heat conduction with spatially varying diffusion coefficient}

\begin{figure}[!ht]
\centering
\includegraphics[trim = 0mm 15mm 0mm 10mm, width=0.49\textwidth] {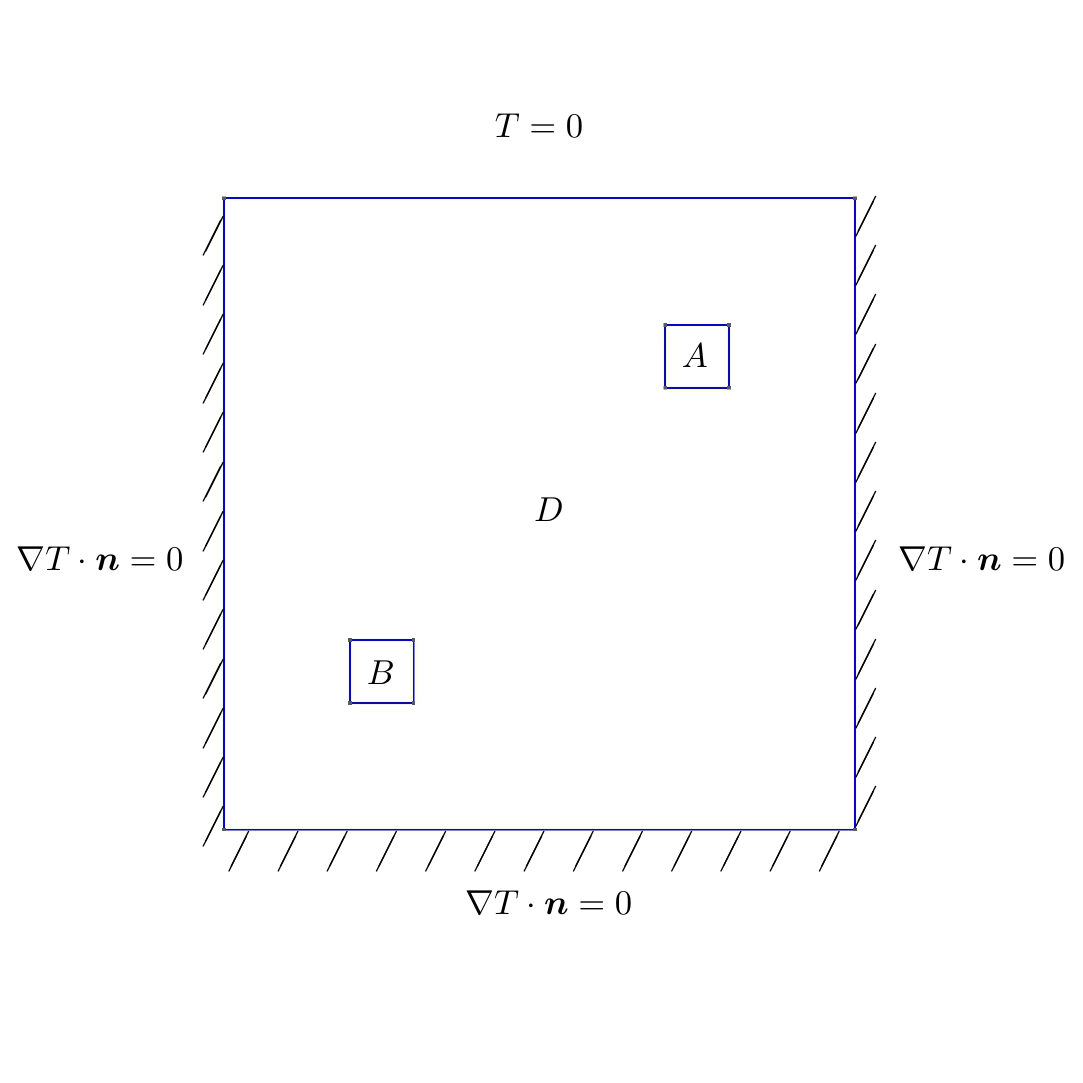}
\includegraphics[trim = 0mm 15mm 0mm 10mm,width=0.49\textwidth] {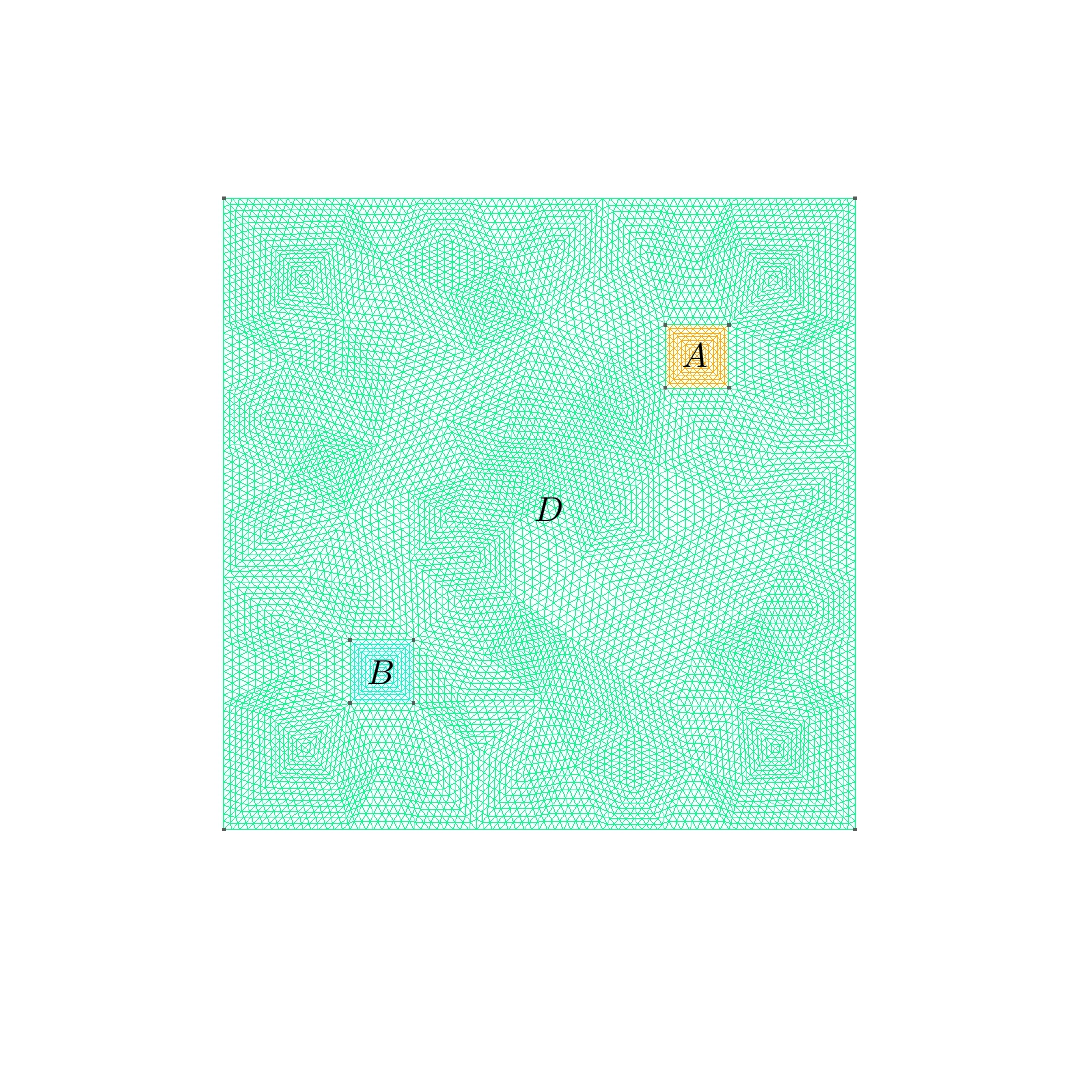}
\caption{Heat-conduction problem: Domain and boundary conditions (left); finite-element mesh (right).}
\label{fig:RF_domain}
\end{figure}

The present example, inspired by \cite{Nouy2010b}, concerns two-dimensional stationary heat-conduction defined on the square domain $D=(-0.5,0.5)~\rm m\times(0.5,0.5)~\rm m$ shown in Figure~\ref{fig:RF_domain}. The temperature field $T(\ve z)$, $\ve z \in D$, is described by the partial differential equation:
\begin{equation}
\label{eq:diffusion_eq}
-\nabla(\kappa(\ve z)\hsp \nabla T(\ve z)) =Q \hsp I_A(\ve z),
\end{equation}
with boundary conditions  $T=0$ on the top boundary and $\nabla T \cdot \ve n=0$ on the left, right and bottom boundaries, where $\ve n$ denotes the vector normal to the boundary. In Eq.~(\ref{eq:diffusion_eq}), $Q=2\cdot 10^3~\rm W/m^3$, $A=(0.2,0.3)~\rm m\times(0.2,0.3)~\rm m$ is a square domain within $D$ (see Figure~\ref{fig:RF_domain}) and $I_A$ is the indicator function equal to $1$ if $\ve z\in A$ and $0$ otherwise. The diffusion coefficient $\kappa(\ve z)$ is a lognormal random field defined as:
\begin{equation}
\label{eq:diffusion_coef}
\kappa(\ve z)=\exp[a_{\kappa}+b_{\kappa} \hsp g(\ve z)],
\end{equation}
where $g(\ve z)$ denotes a standard Gaussian random field with autocorrelation function:
\begin{equation}
\label{eq:autocorr}
\rho(\ve z,\ve z')=\exp{(-\|\ve z-\ve z'\|^2/\ell^2)}.
\end{equation}
In Eq.~(\ref{eq:diffusion_coef}), the parameters $a_{\kappa}$ and $b_{\kappa}$ are such that the mean and standard deviation of $\kappa$ are $\mu_{\kappa}=1~\rm W/°C\cdot m$ and $\sigma_{\kappa}=0.3~\rm W/°C\cdot m$, respectively, while in Eq.~(\ref{eq:autocorr}), $\ell=0.2 ~\rm m$.

\begin{figure}[!ht]
\centering
\includegraphics[trim = 25mm 25mm 25mm 25mm, width=0.60\textwidth] {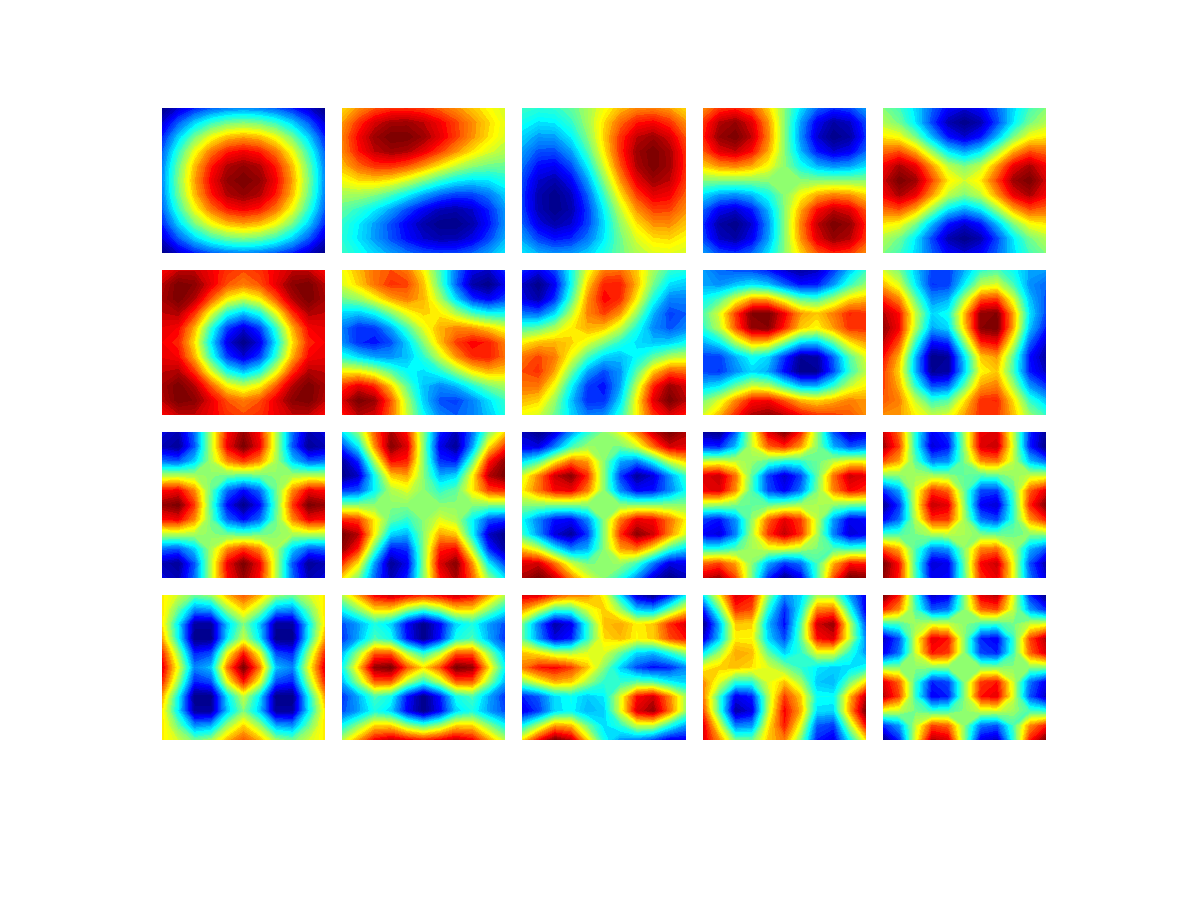}
\caption{Heat-conduction problem: Shapes of the first 20 basis functions in the EOLE discretization (from left-top to bottom-right row-wise).}
\label{fig:RF_modes}
\end{figure}

To solve Eq.~(\ref{eq:diffusion_eq}), the Gaussian random field $g(\ve z)$ in Eq.~(\ref{eq:diffusion_coef}) is first discretized using the expansion optimal linear estimation (EOLE) method \cite{Li1993optimal}. Let $\{\ve{\zeta}_1 \enum \ve{\zeta}_n\}$ denote the points of an appropriately defined grid in $D$. By retaining the first $M$ terms in the EOLE series, $g(\ve z)$ is approximated by:
\begin{equation}
\label{eq:EOLE}
\widehat{g}(\ve z) = \sum_{i=1}^M \frac{\xi_i}{\sqrt{l_i}}\ve {\phi}_i^{\rm{T}} \ve C_{\ve z \ve \zeta}(\ve z),
\end{equation}
where $\{\xi_1 \enum \xi_M\}$ are independent standard normal variables; $\ve C_{\ve z \ve \zeta}$ is a vector with elements $\ve C_{\ve z \ve \zeta}^{(k)}=\rho(\ve z,\ve \zeta_k)$ for $k=1 \enum n$; and $(l_i,\ve{\phi}_i)$ are the eigenvalues and eigenvectors of the correlation matrix $\ve C_{\ve \zeta \ve \zeta}$ with elements $\ve C_{\ve \zeta \ve \zeta}^{(k,l)}=\rho(\ve{\zeta}_k,\ve{\zeta}_l)$ for $k,l=1 \enum n$. In \cite{Sudret2000stochastic}, it is recommended that for a square-exponential autocorrelation function, the size of the element in the EOLE grid must be $1/2-1/3$ of $\ell$. Accordingly, in the present numerical application, we use a square grid with element size $0.01~\rm m$, thus comprising $n=121$ points. The number of terms in the EOLE series is determined according to the rule:
\begin{equation}
\label{eq:EOLE_M}
\sum_{i=1}^{M}l_i/\sum_{i=1}^{n} l_i \geq 0.99,
\end{equation}
herein leading to $M=53$. The shapes of the first 20 basis functions $\{{\phi}_i^{\rm{T}} \ve C_{\ve z \ve \zeta}(\ve z), \hsp i=1 \enum 20\}$ are shown in Figure~\ref{fig:RF_modes}.

\begin{figure}[!ht]
\centering
\includegraphics[width=0.49\textwidth] {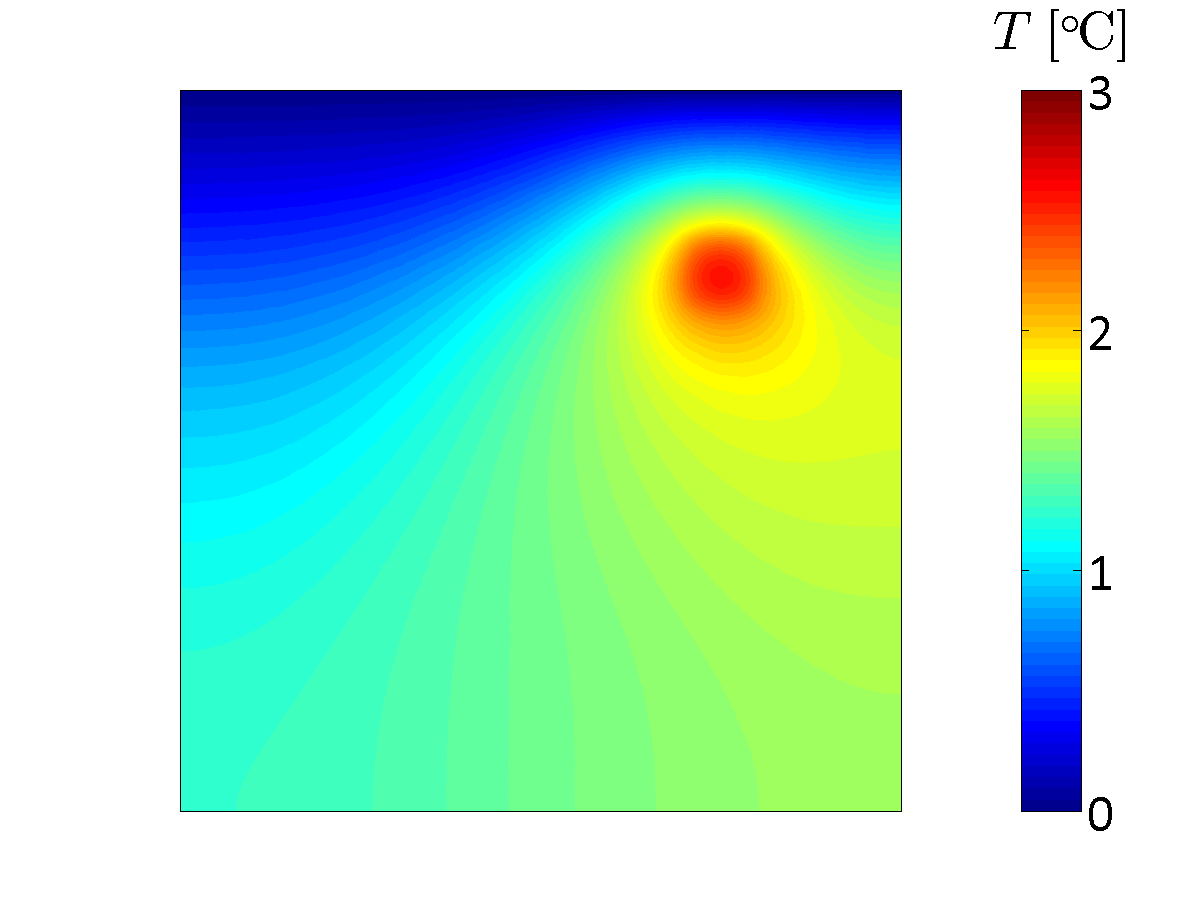}
\includegraphics[width=0.49\textwidth] {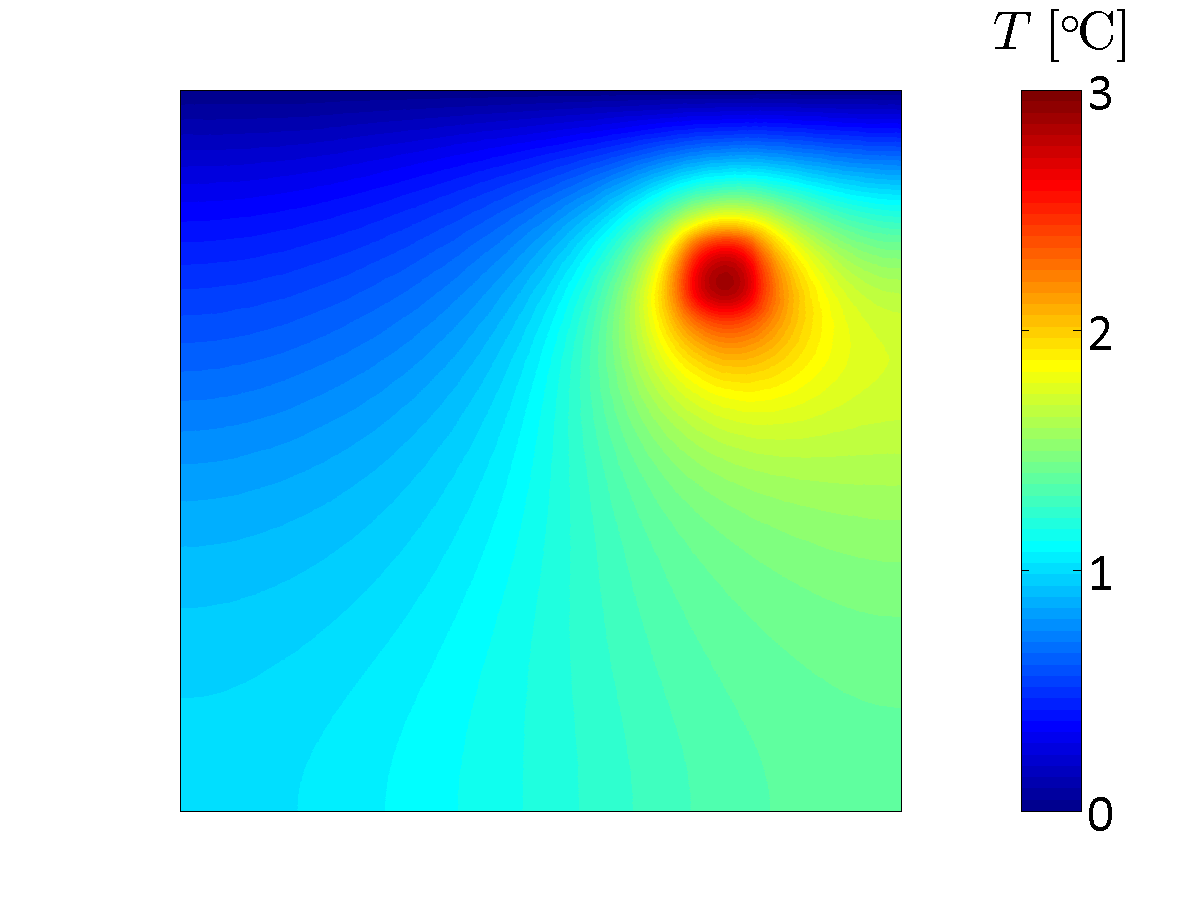}
\caption{Heat-conduction problem: Example realizations of the temperature field.}
\label{fig:RF_maps}
\end{figure}

The response quantity of interest is the average temperature in the square domain $B=(-0.3,-0.2)~\rm m\times(-0.3,-0.2)~\rm m$ (see Figure~\ref{fig:RF_domain}), denoted by $\widetilde{T}$:
\begin{equation}
\label{eq:diffusion_Y}
\widetilde{T}=\frac{1}{|B|}\int_{\ve z \in B}T(\ve z) \hsp d\ve z.
\end{equation}
For a given realization of $\{\xi_1 \enum \xi_M\}$, the ``exact'' model response is obtained with an in-house finite-element analysis code developed in the Matlab environment. The employed finite-element discretization in 16,000 triangular T3 elements is depicted in Figure~\ref{fig:RF_domain}; this discretization is obtained using software \emph{Gmsh} \cite{Geuzaine2009gmsh}. Figure~\ref{fig:RF_maps} shows the temperature field $T(\ve z)$ for two example realizations of the conductivity random field.

\vspace{8mm}

We develop LRA and sparse PCE meta-models of $\widetilde{T}=\cm(\xi_1,\enum,\xi_{53})$ using EDs of size varying from $50$ to $2,000$. Because the random input herein comprises standard normal variables, we build the basis functions by relying on the associated family of Hermite polynomials. In the LRA algorithm, we define the stopping criterion in the correction step by setting $\Imax=50$ and $\Derrmin= 10^{-6}$.  Parameters and error estimates of the resulting LRA and PCE meta-models are listed in Tables~{\ref{tab:RF_LRA}} and {\ref{tab:RF_PCE}}, respectively. The generalization errors $\errG$ are estimated using a validation set of size $\nval=10^4$ sampled with MCS. The two types of meta-models are characterized by generalization errors of the same order of magnitude when a certain ED is considered, except for $N=2,000$. Note that the LRA meta-models exhibit smaller $\errG$ than sparse PCE only when $N\leq200$. For both types of meta-models, the ED-based error estimates are well approximated by the respective generalization errors, particularly for the larger EDs.

\begin{table} [!ht]
\centering
\caption{Heat-conduction problem: Parameters and error estimates of LRA meta-models.}
\label{tab:RF_LRA}
\begin{tabular}{c c c c c}
\hline
$N$ & $R$ & $p$ & $\widehat{err}_{\rm CV3}$ & $\widehat{err}_G$ \\
\hline
50    &  3  &  1  &  $6.51\cdot 10^{-2}$  &  $1.08\cdot 10^{-1}$ \\
100   &  1  &  1  &  $3.56\cdot 10^{-2}$  &  $2.46\cdot 10^{-2}$ \\
200   &  1  &  1  &  $1.79\cdot 10^{-2}$  &  $1.38\cdot 10^{-2}$ \\
500   &  1  &  2  &  $1.13\cdot 10^{-2}$  &  $9.68\cdot 10^{-3}$ \\
1,000 &  1  &  2  &  $8.40\cdot 10^{-3}$  &  $8.19\cdot 10^{-3}$ \\
2,000 &  1  &  2  &  $7.81\cdot 10^{-3}$  &  $7.72\cdot 10^{-3}$ \\
\hline       
\end{tabular}
\end{table}

\begin{table} [!ht]
\centering
\caption{Heat-conduction problem: Parameters and error estimates of PCE meta-models.}
\label{tab:RF_PCE}
\begin{tabular}{c c c c c}
\hline
$N$ & $q$ & $p^t$ & $\widehat{err}_{\rm LOO}^*$ & $\widehat{err}_G$ \\
\hline
50    &  0.25  &  2  &  $1.71\cdot 10^{-1}$  &  $2.53\cdot 10^{-1}$ \\
100   &  0.25  &  1  &  $3.69\cdot 10^{-2}$  &  $3.89\cdot 10^{-2}$ \\
200   &  0.25  &  3  &  $4.08\cdot 10^{-2}$  &  $2.58\cdot 10^{-2}$ \\
500   &  0.50  &  5  &  $8.36\cdot 10^{-3}$  &  $9.39\cdot 10^{-3}$ \\
1,000 &  0.75  &  3  &  $2.65\cdot 10^{-3}$  &  $2.19\cdot 10^{-3}$ \\
2,000 &  0.75  &  3  &  $1.22\cdot 10^{-3}$  &  $9.58\cdot 10^{-4}$ \\
\hline       
\end{tabular}
\end{table}

\begin{figure}[!ht]
\centering
\includegraphics[width=0.45\textwidth] {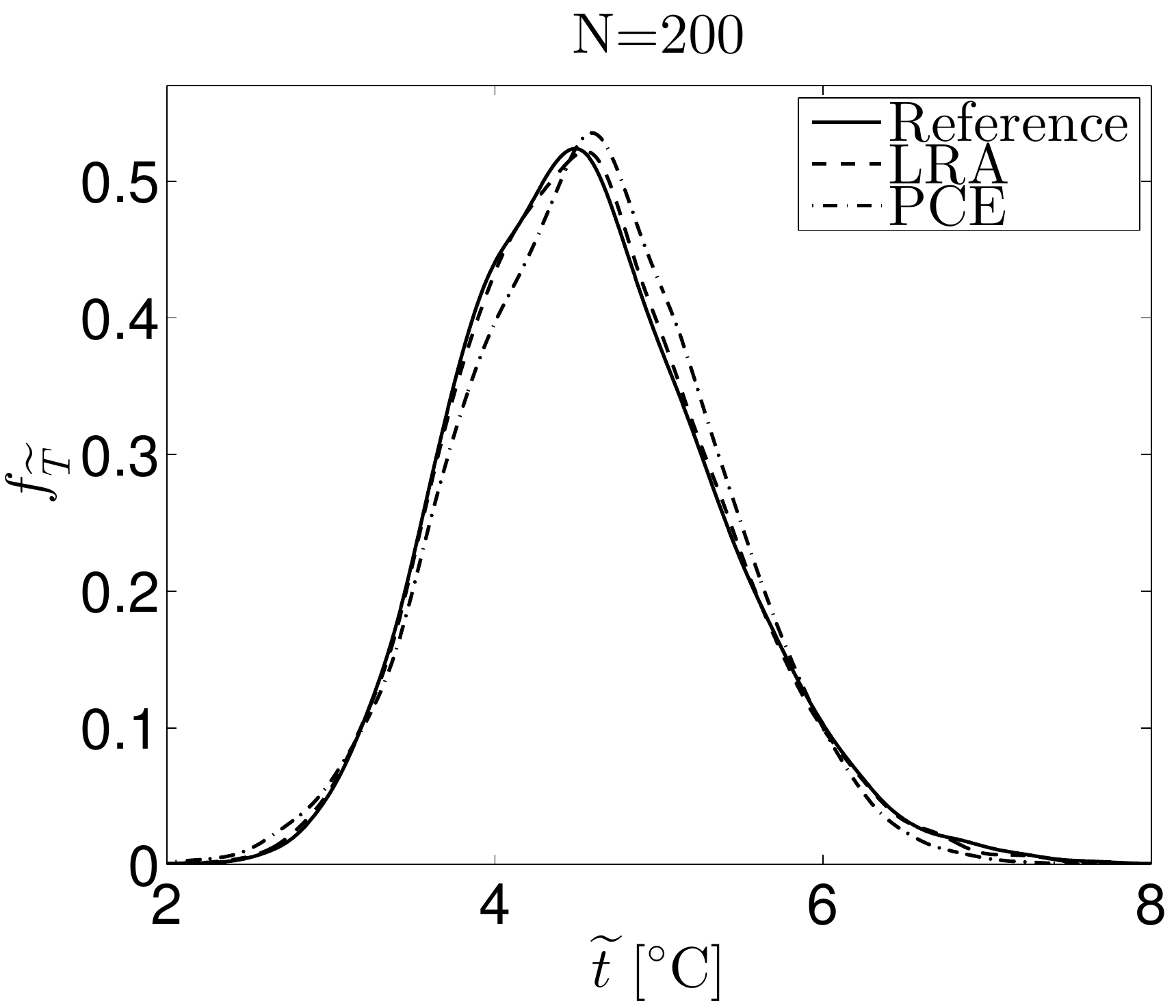}
\includegraphics[width=0.45\textwidth] {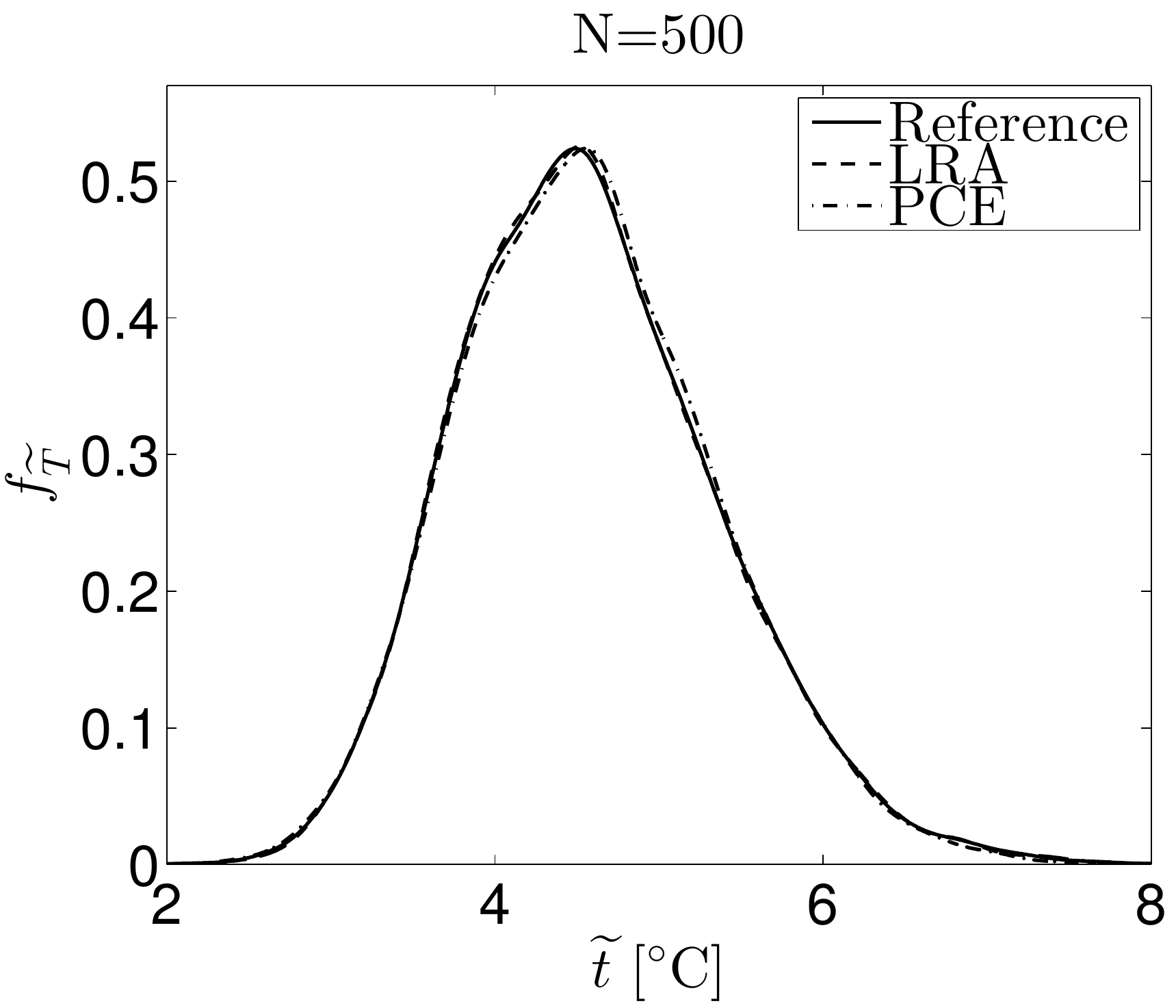}
\caption{Heat-conduction problem: Probability density function of the response (normal scale).}
\label{fig:RF_KDE}
\end{figure}

\begin{figure}[!ht]
\centering
\includegraphics[width=0.45\textwidth] {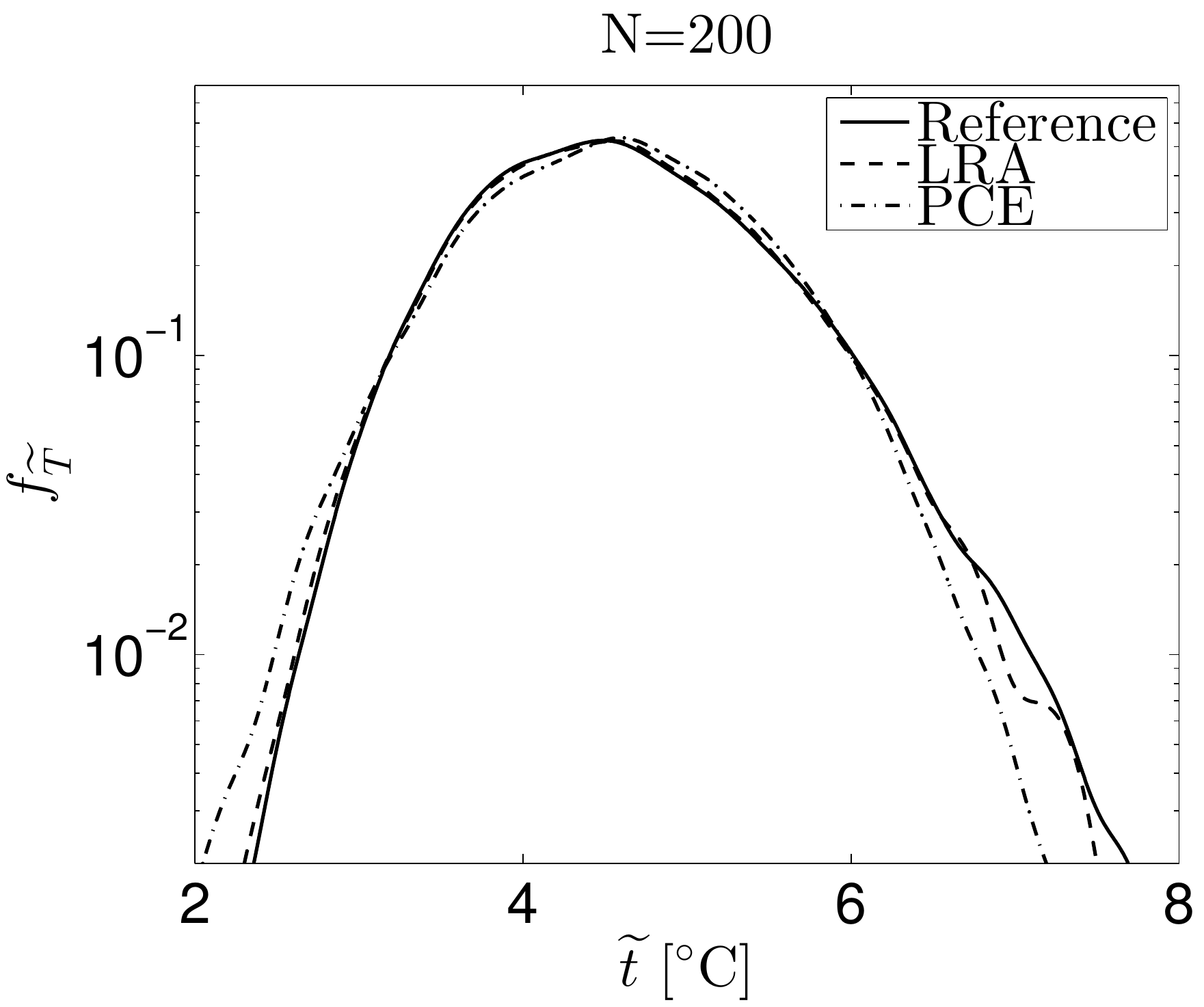}
\includegraphics[width=0.45\textwidth] {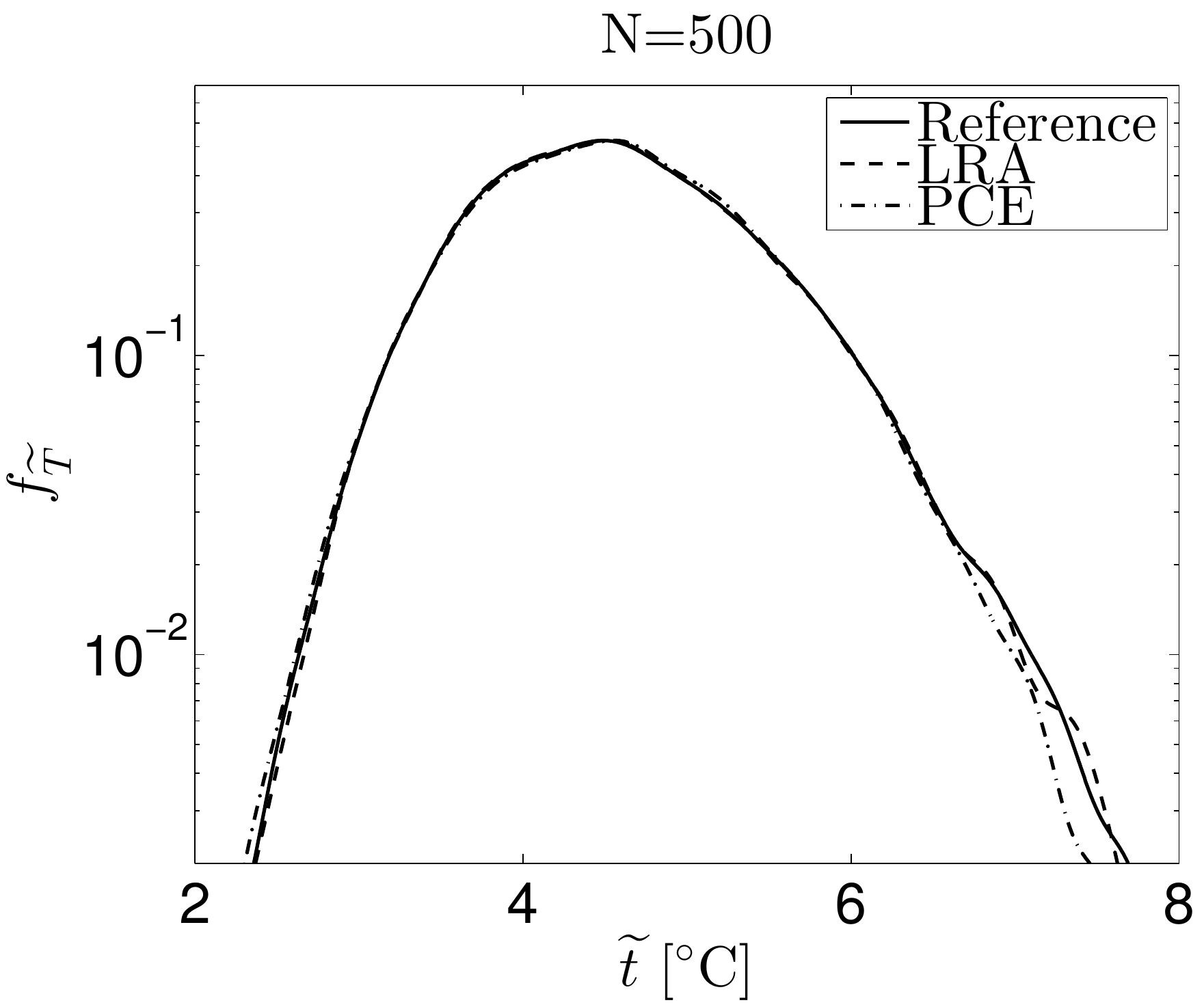}
\caption{Heat-conduction problem: Probability density function of the response (log-scale).}
\label{fig:RF_KDElog}
\end{figure}

We first compare the KDEs of the response PDF $f_{\widetilde{T}}$ obtained with the LRA and sparse PCE meta-models with that obtained with the actual model, which is considered the reference solution for $f_{\widetilde{T}}$. All aforementioned KDEs are based on the evaluation of the different models at a MCS sample of $n=10^4$ points in the input space. In Figure~\ref{fig:RF_KDE}, we depict the KDEs for the cases with $N=200$ and $N=500$; by using a logarithmic scale in~Figure \ref{fig:RF_KDElog}, we emphasize the behavior at the tails of the PDF. With only $N=200$, the LRA approach yields a good approximation of the reference PDF in both the normal and the logarithmic scales, which is clearly superior to the PCE approximation. For $N=500$, the PCE estimate becomes fairly accurate, except for the upper tail of the PDF. We underline that by using the meta-models, we can easily sample larger sets of responses and thus obtain KDEs with smooth tails; for instance, the current implementation of PCE and LRA allows sampling $10^6$ values in only a few seconds with a standard desktop. On the other hand, obtaining such large sets of responses by using the actual finite-element model requires prohibitively high computational times (a single evaluation takes approximately $16~\rm sec$). We herein consider the same set comprising $n=10^4$ points for the evaluation of all KDEs for the sake of comparison.

\begin{figure}[!ht]
\centering
\includegraphics[width=0.45\textwidth] {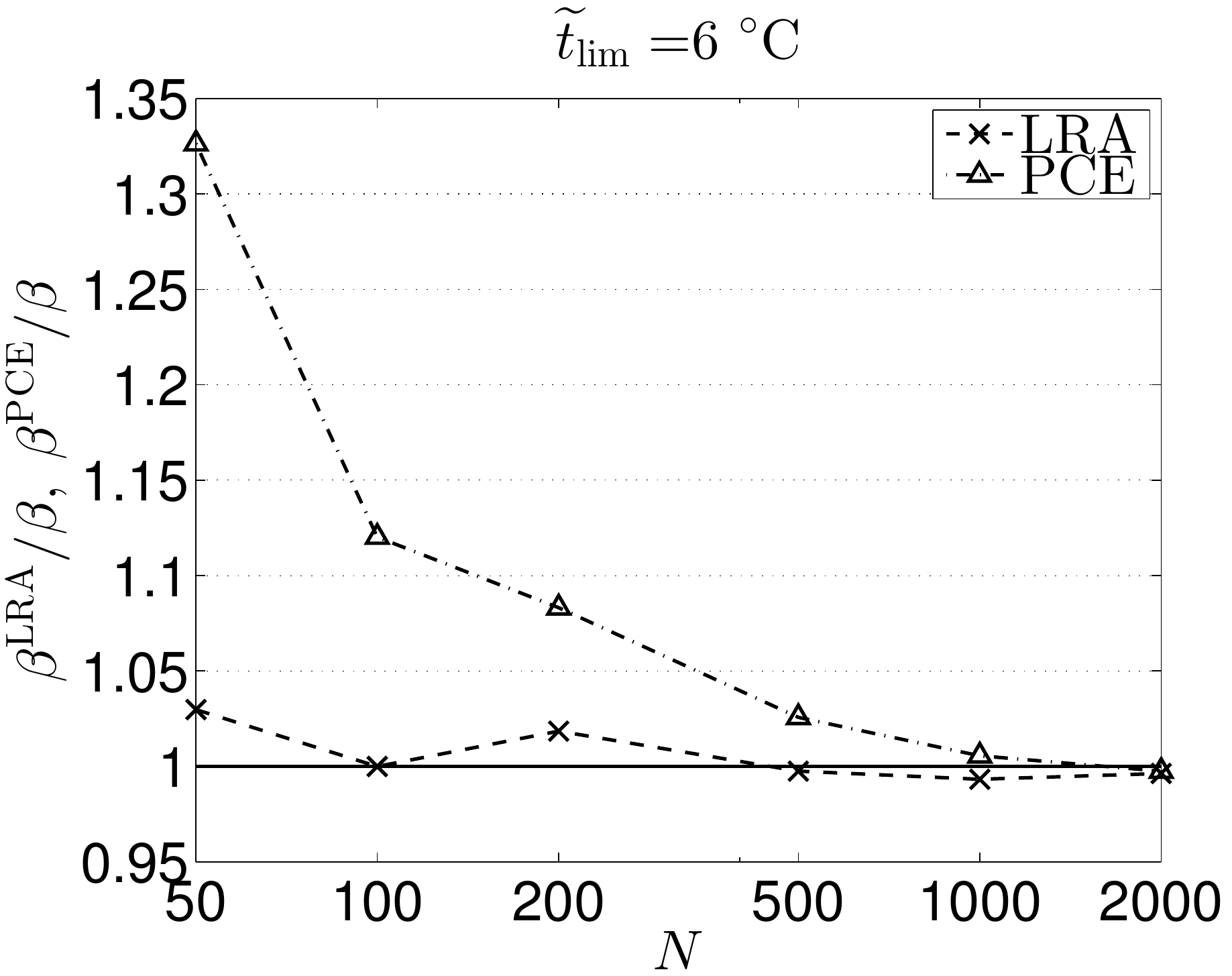}
\includegraphics[width=0.45\textwidth] {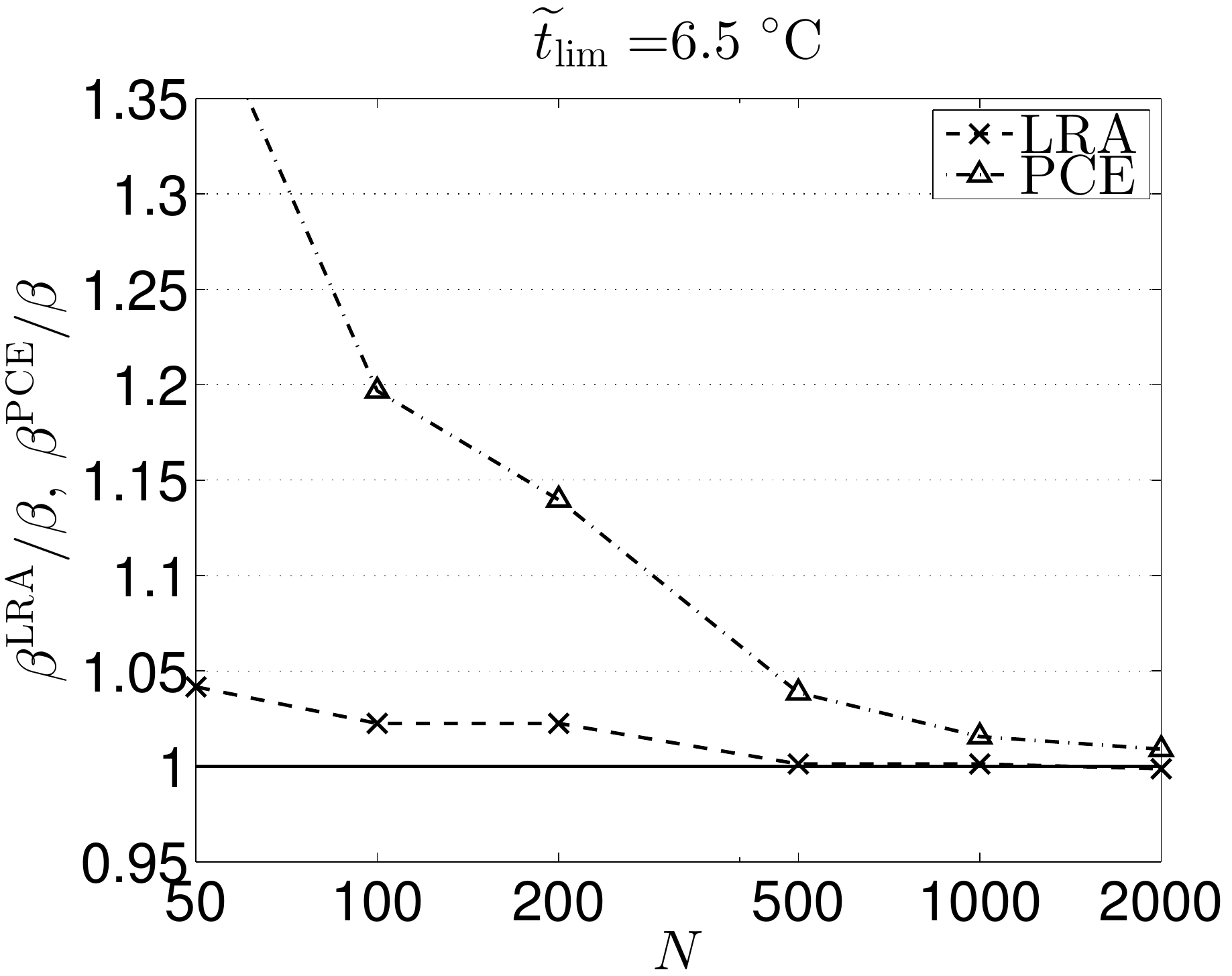}
\caption{Heat-conduction problem: Ratios of meta-model-based to reference reliability indices.}
\label{fig:RF_beta}
\end{figure}

Next, we assess the accuracy of the LRA and sparse PCE meta-models in estimating tail probabilities of the form $P_f=\Pp(\widetilde{t}_{\rm lim}-\widetilde{T}\leq0)=\Pp(\widetilde{T}\geq\widetilde{t}_{\rm lim})$ for the thresholds $\widetilde{t}_{\rm lim}\in\{6.0, 6.5\} ~\rm^{\circ} C$.  The reference failure probabilities are obtained with a MCS approach using $n=10^4$ evaluations of the actual model, which leads to $P_f=4.53\cdot 10^{-2}$ ($\beta=1.69$) for $\widetilde{t}_{\rm lim}=6.0~\rm^{\circ} C$ and $P_f=1.43\cdot 10^{-2}$ ($\beta=2.19$) for $\widetilde{t}_{\rm lim}=6.5~\rm^{\circ} C$; the CoVs of these estimates are $<0.10$.  The same input sample is used to estimate the failure probabilities with the two types of meta-models. Figure~\ref{fig:RF_beta} shows the ratios of the reliability indices $\beta^{\rm LRA}$ and $\beta^{\rm PCE}$, based on the LRA and sparse PCE meta-models, to the reference reliability index $\beta$ versus the ED size. It is remarkable that with an ED of size as small as $N=50$, $\beta^{\rm LRA}$ approximates $\beta$ with a relative error $<5\%$ for both thresholds, whereas for $N\geq500$, this error becomes nearly zero. Obviously, the convergence of $\beta^{\rm PCE}$ to $\beta$ with increasing $N$ is much slower. The values of the LRA- and PCE-based reliability indices and respective failure probabilities for the cases with $N=500$ and $N=2,000$ are listed in the Appendix.

\begin{figure}[!ht]
\centering
\includegraphics[width=0.45\textwidth] {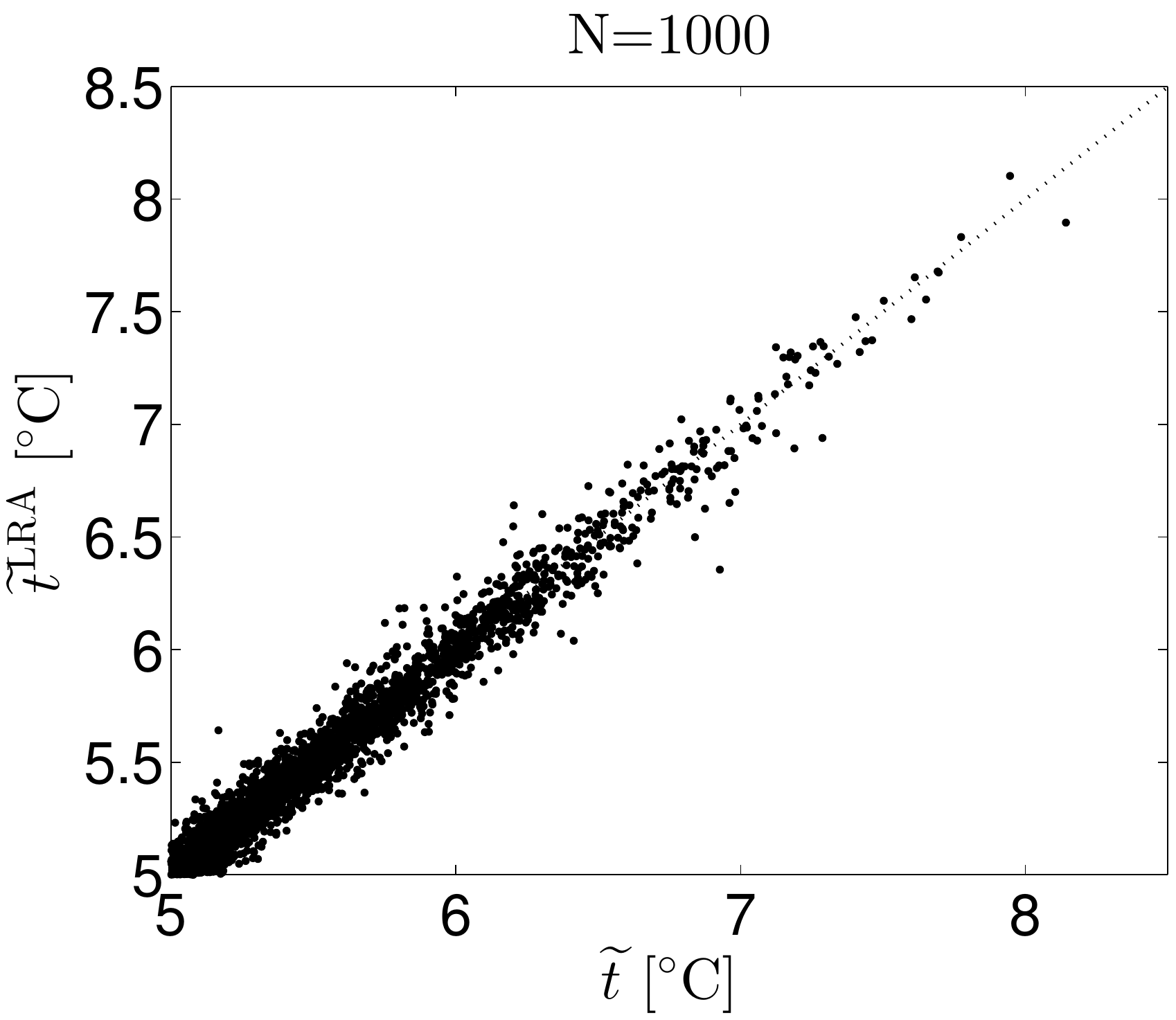}
\includegraphics[width=0.45\textwidth] {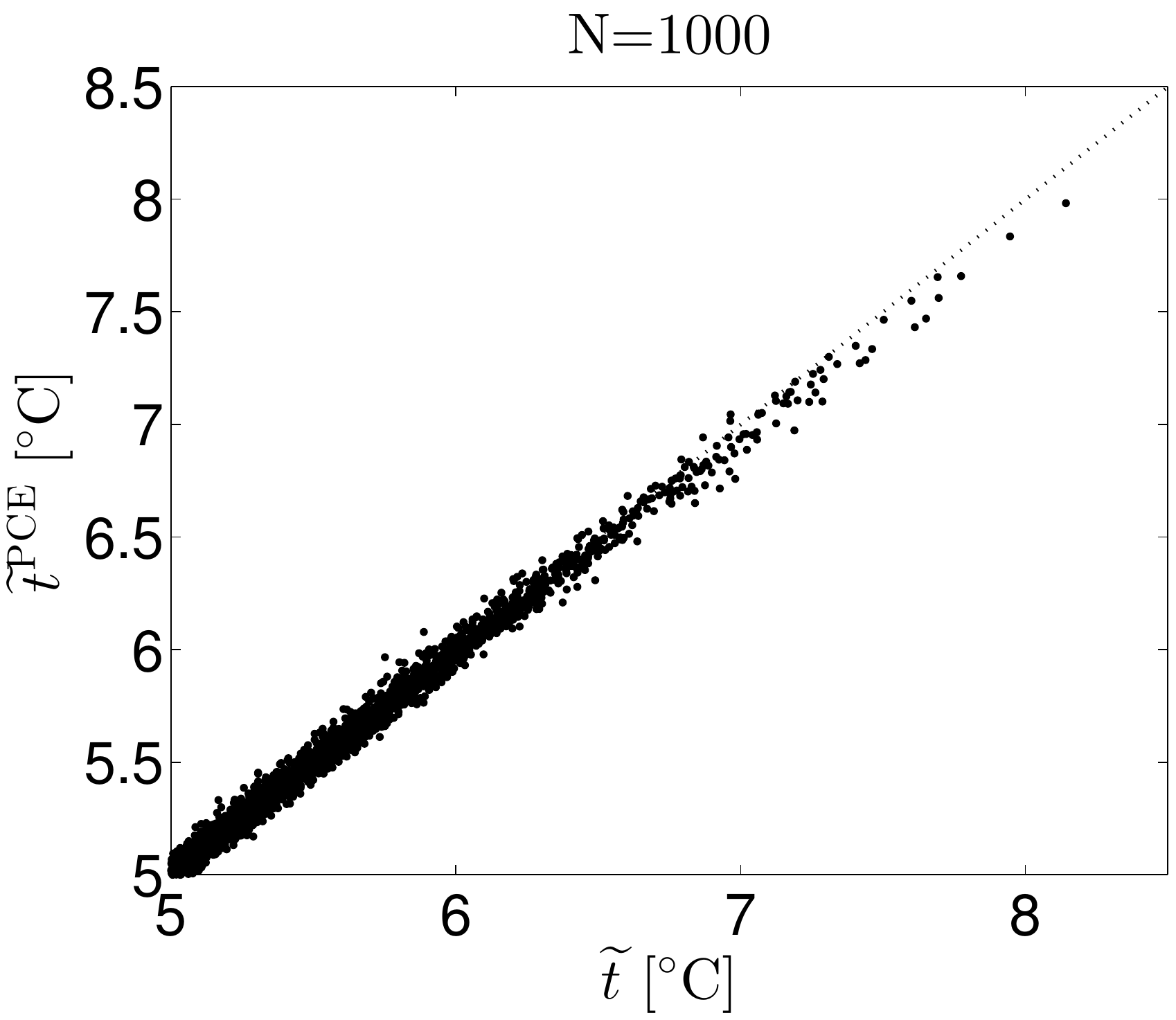}
\caption{Heat-conduction problem: Meta-model versus actual model responses at a subset of the validation set corresponding to the upper tail of the response distribution.}
\label{fig:RF_t_that}
\end{figure}

The above analysis demonstrates that although sparse PCE are characterized by smaller generalization errors for $N\geq 500$, the LRA-based estimates of the tail probabilities remain superior. This is because, contrary to PCE, the LRA responses tend to be unbiased at the tails, even in cases when they exhibit a larger dispersion around the actual responses than the PCE ones. This is illustrated in Figure~\ref{fig:RF_t_that} for the case with $N=1,000$. This figure depicts the responses of the LRA and sparse PCE meta-models, denoted by $\widetilde{t}^{\rm LRA}$ and $\widetilde{t}^{\rm PCE}$ respectively, versus the actual model responses, denoted by $\widetilde{t}$, at the points $\ve x$ of the validation set satisfying the condition $\widetilde{t}=\cm(\ve x)\geq 5.0~\rm^{\circ} C$ ($27.7$-th upper percentile). The LRA responses are characterized by an overall larger dispersion around the actual model responses, which leads to a larger generalization error. However, the LRA responses tend to be unbiased, whereas PCE systematically underestimate the actual model responses at the upper tail. In this case, the conditional generalization errors evaluated at $\cx_{\rm val}^{\rm C}=\{\ve x \in \cx_{\rm val}: \widetilde{t}=\cm(\ve x)\geq \widetilde{t}_{\rm lim}\}$ for $\widetilde{t}_{\rm lim}\in\{6.0, 6.5\} ~\rm^{\circ} C$ (the same thresholds considered in the reliability analysis) remain larger for LRA. By accounting only for the absolute differences between the estimated responses from the actual ones and disregarding their signs, the generalization errors do not reflect herein the superior performance of LRA over PCE in the estimation of tail probabilities.

\begin{figure}[!ht]
\centering
\includegraphics[width=0.45\textwidth] {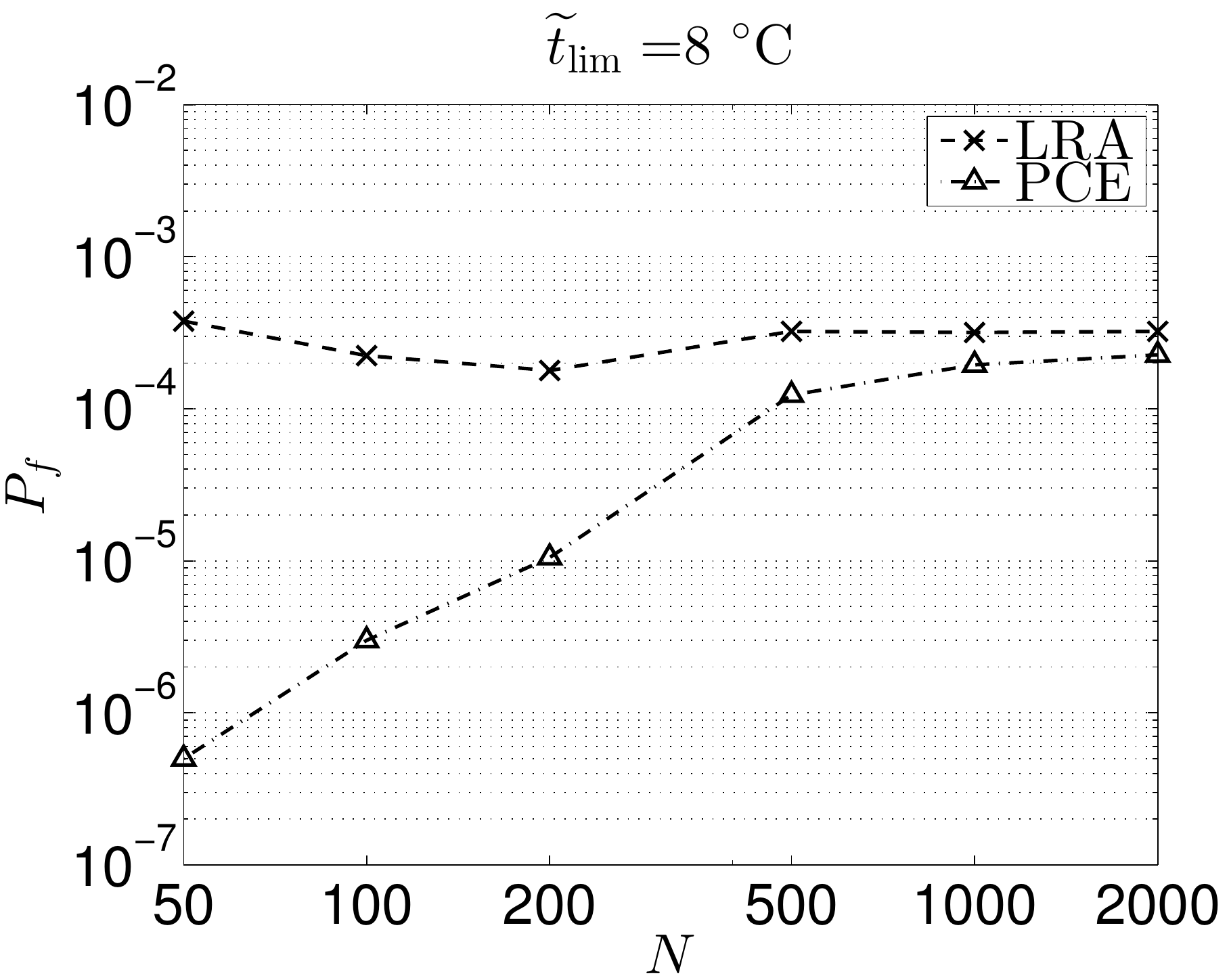}
\includegraphics[width=0.45\textwidth] {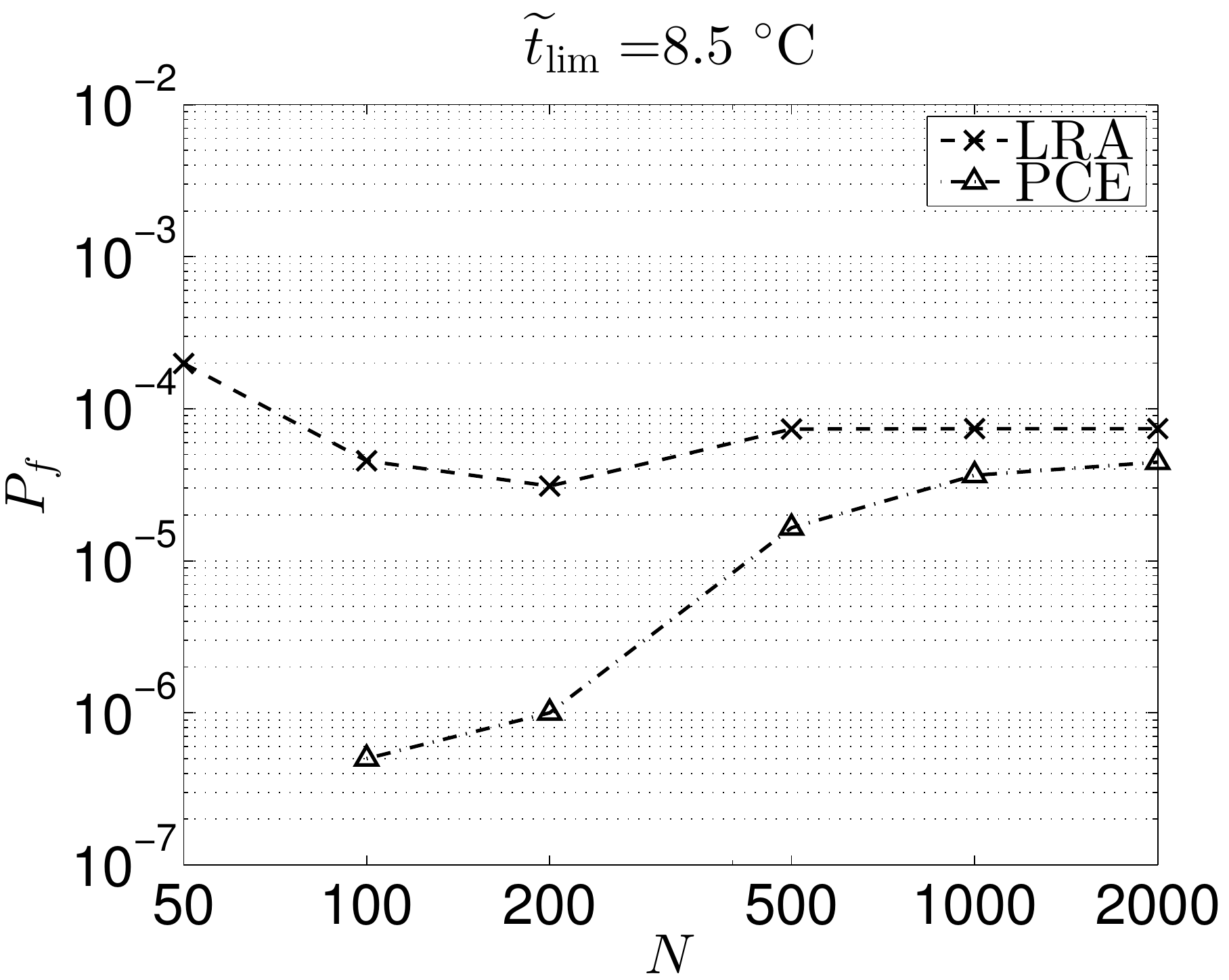}
\caption{Heat-conduction problem: Failure probabilities.}
\label{fig:RF_Pf}
\end{figure}

As mentioned earlier, by using the meta-models, we can easily sample larger sets of responses and thus, estimate lower failure probabilities than the above. In Figure~\ref{fig:RF_Pf}, we plot the LRA- and PCE-based estimates of $P_f$ versus the ED size for $\widetilde{t}_{\rm lim}\in\{8.0, 8.5\} ~\rm^{\circ} C$. A MCS sample of size $n=2\cdot10^6$ is used in order to achieve a CoV$<0.10$. For both temperature thresholds, the LRA estimates practically reach convergence at $N=500$; the corresponding failure probabilities are $P_f=3.22\cdot10^{-4}$ for $\widetilde{t}_{\rm lim}=8.0~\rm^{\circ} C$ and $P_f=7.35\cdot10^{-5}$ for $\widetilde{t}_{\rm lim}=8.5~\rm^{\circ} C$. The PCE estimate appears to converge to a similar value with increasing ED size, but at a much slower rate. It is remarkable that an ED of size as small as $N=50$ is sufficient to obtain a reasonable preliminary estimate of the failure probability with LRA even for the higher temperature threshold, whereas at least $N=500$ points are required to obtain such an estimate with sparse PCE. The values of the LRA- and PCE-based failure probabilities and respective reliability indices for the cases with $N=500$ and $N=2,000$ are listed in the Appendix. We emphasize that for the considered thresholds, reference solutions that rely on the evaluation of the actual finite-element model cannot be obtained at an affordable computational time. Note that because of the high dimensionality of the problem, the typically low-cost FORM and SORM approaches are herein inefficient.
 
{  \subsection{Frame displacement}

In the last example, we consider the frame structure shown in Figure~\ref{fig:frame}, also studied in \cite{LiuPL91,BlatmanPEM2010}. We conduct reliability analysis with respect to the horizontal displacement $u$ at the top right corner of the top floor under the depicted horizontal loads acting at the floor levels. The random input comprises the load values $P_1$, $P_2$ and $P_3$, the Young's moduli of the column and beam elements, respectively denoted by $E_C$ and $E_B$, the moments of inertia of the column and beam elements, respectively denoted by $I_{B_i}$ and $I_{C_i}$, $i=1 \enum 4$, and the cross-sectional areas of the column and beam elements, respectively denoted by $I_{B_i}$ and $I_{C_i}$, $i=1 \enum 4$. The distributions of the aforementioned variables are listed in Table~\ref{tab:frame_input}. Contrary to the previous examples where the input variables were independent, the input of the herein considered model has a dependence structure described by means of a Gaussian copula (for further information on the modeling of probabilistic dependence with copulas, the interested reader is referred to \cite{Nelsen2006}). The non-zero elements of the associated linear correlation matrix are defined as follows:
the correlation coefficient  between the two Young's moduli is $\rho_{E_C,E_B}=0.90$; the correlation coefficient between the cross-sectional area $A_i$ and the moment of inertia $I_i$ of a certain element $i$ is $\rho_{A_i,I_i}=0.95$;
the correlation coefficient between the geometric properties of two distinct elements $i$ and $j$ are $\rho_{A_i,I_j}=\rho_{I_i,I_j}=\rho_{A_i,A_j}=0.13$.
In the original example, the above values represent the corresponding linear correlation coefficients in the standard normal space; however, Blatman and Sudret \cite{BlatmanPEM2010} note that the differences between the two are insignificant. For a given realization of the input random vector, the frame displacement is computed with an in-house finite-element analysis code developed in the Matlab environment.

\begin{figure}[!ht]
\centering
\includegraphics[trim = 0mm 0mm 0mm 0mm, width=0.7\textwidth]{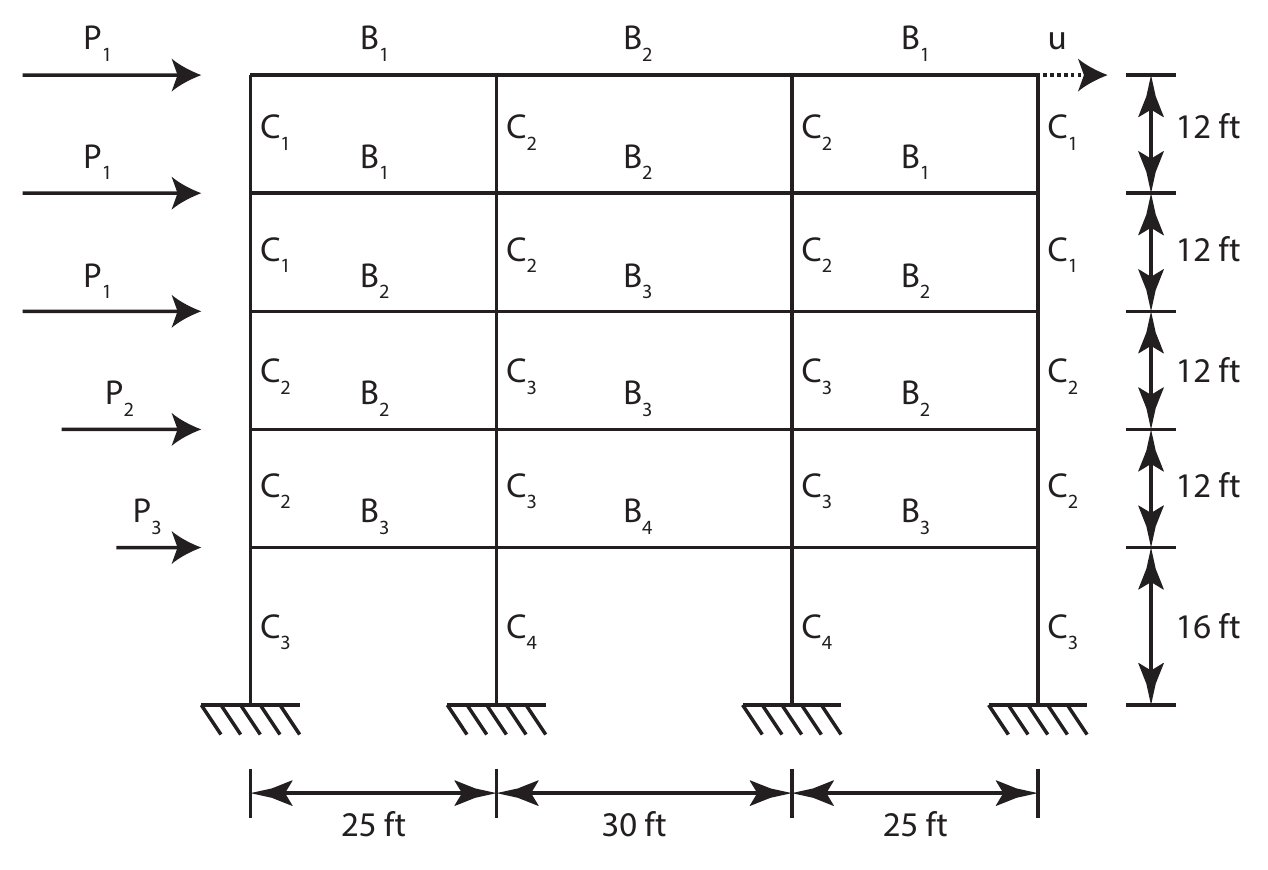}
\caption{Frame structure.}
\label{fig:frame}
\end{figure}

\begin{table} [!ht]
\centering
\caption{Frame-displacement problem: Distributions of input random variables.}
\label{tab:frame_input}
\begin{tabular}{c c c c}
\hline Variable & Distribution & Mean & Standard deviation \\
\hline
$P_1~[\rm KN]$ & Lognormal & $133.45$ & $40.04$ \\          
$P_2~[\rm KN]$ & Lognormal & $88.97$ & $35.59$ \\           
$P_3~[\rm KN]$ & Lognormal & $71.17$ & $28.47$ \\
$E_C~[\rm KN/m^2]$ & Truncated Gaussian over $[0,\infty)$ & $2.3796\cdot 10^{7}$ & $1.9152\cdot 10^{6}$ \\   
$E_B~[\rm KN/m^2]$ & Truncated Gaussian over $[0,\infty)$ & $2.1738\cdot 10^{7}$ & $1.9152\cdot 10^{6}$ \\
$I_{C_1}~[\rm m^4]$ & Truncated Gaussian over $[0,\infty)$ & $8.1344\cdot 10^{-3}$ & $1.0834\cdot 10^{-3}$ \\
$I_{C_2}~[\rm m^4]$ & Truncated Gaussian over $[0,\infty)$ & $1.1509\cdot 10^{-2}$ & $1.2980\cdot 10^{-3}$ \\
$I_{C_3}~[\rm m^4]$ & Truncated Gaussian over $[0,\infty)$ & $2.1375\cdot 10^{-2}$ & $2.5961e-03$ \\
$I_{C_4}~[\rm m^4]$ & Truncated Gaussian over $[0,\infty)$ & $2.5961\cdot 10^{-2}$ &$3.0288\cdot 10^{-3}$ \\
$I_{B_1}~[\rm m^4]$ & Truncated Gaussian over $[0,\infty)$ & $1.0811\cdot 10^{-2}$ & $2.5961\cdot 10^{-3}$ \\
$I_{B_2}~[\rm m^4]$ & Truncated Gaussian over $[0,\infty)$ & $1.4105\cdot 10^{-2}$ & $3.4615\cdot 10^{-3}$ \\
$I_{B_3}~[\rm m^4]$ & Truncated Gaussian over $[0,\infty)$ & $2.3279\cdot 10^{-2}$ & $5.6249\cdot 10^{-3}$ \\
$I_{B_4}~[\rm m^4]$ & Truncated Gaussian over $[0,\infty)$ & $2.5961\cdot 10^{-2}$ & $6.4902\cdot 10^{-3}$ \\
$A_{C_1}~[\rm m^2]$ & Truncated Gaussian over $[0,\infty)$ & $3.1256\cdot 10^{-1}$ & $5.5815\cdot 10^{-2}$ \\
$A_{C_2}~[\rm m^2]$ & Truncated Gaussian over $[0,\infty)$ & $3.7210\cdot 10^{-1}$ & $7.4420\cdot 10^{-2}$ \\
$A_{C_3}~[\rm m^2]$ & Truncated Gaussian over $[0,\infty)$ & $5.0606\cdot 10^{-1}$ & $9.3025\cdot 10^{-2}$ \\
$A_{C_4}~[\rm m^2]$ & Truncated Gaussian over $[0,\infty)$ & $5.5815\cdot 10^{-1}$ & $1.1163\cdot 10^{-1}$ \\
$A_{B_1}~[\rm m^2]$ & Truncated Gaussian over $[0,\infty)$ & $2.5302\cdot 10^{-1}$ & $9.3025\cdot 10^{-2}$ \\
$A_{B_2}~[\rm m^2]$ & Truncated Gaussian over $[0,\infty)$ & $2.9117\cdot 10^{-1}$ & $1.0232\cdot 10^{-1}$ \\
$A_{B_3}~[\rm m^2]$ & Truncated Gaussian over $[0,\infty)$ & $3.7303\cdot 10^{-1}$ & $1.2093\cdot 10^{-1}$ \\
$A_{B_4}~[\rm m^2]$ & Truncated Gaussian over $[0,\infty)$ & $4.1860\cdot 10^{-1}$ & $1.9537\cdot 10^{-1}$ \\
\hline    
\end{tabular}
\end{table}

We develop LRA and sparse PCE meta-models of $X=\{P_1,P_2,P_3,E_C,E_B,I_{C_1}\enum I_{B_4},A_{C_1}\enum A_{B_4}\}$  using two EDs of size $N=500$ and $N=1,000$. For both types of meta-models, we use Hermite polynomials to build the basis functions, after an isoprobabilistic transformation of the input variables to independent standard normal variables. In the LRA algorithm, we define the stopping criterion in the correction step by setting $\Imax=50$ and $\Derrmin= 10^{-6}$.  Parameters and error estimates of the LRA and PCE meta-models are listed in Tables~{\ref{tab:frame_LRA}} and {\ref{tab:frame_PCE}}, respectively. The generalization errors $\errG$ are estimated using a validation set of size $\nval=10^6$ sampled with MCS. For $N=500$, the two types of meta-models exhibit similar generalization errors, while for $N=1,000$, the PCE error is slightly smaller. Note that the generalization errors are approximated fairly well by the corresponding ED-based error estimates.

\begin{table} [!ht]
\centering
\caption{Frame-displacement problem: Parameters and error estimates of LRA meta-models.}
\label{tab:frame_LRA}
\begin{tabular}{c c c c c}
\hline
$N$ & $R$ & $p$ & $\widehat{err}_{\rm CV3}$ & $\widehat{err}_G$ \\
\hline
500  & 1 & 3  &  $2.93\cdot10^{-3}$ & $3.35\cdot10^{-3}$ \\
1,000 & 1 & 3  &  $2.73\cdot10^{-3}$ & $3.19\cdot10^{-3}$ \\
\hline       
\end{tabular}
\end{table}

\begin{table} [!ht]
\centering
\caption{Frame-displacement problem: Parameters and error estimates of PCE meta-models.}
\label{tab:frame_PCE}
\begin{tabular}{c c c c c}
\hline
$N$ & $q$ & $p^t$ & $\widehat{err}_{\rm LOO}^*$ & $\widehat{err}_G$ \\
\hline
500  & 0.50 & 7  &  $ 1.67\cdot10^{-3}$ & $ 3.06\cdot10^{-3}$ \\
1,000 & 1    & 3  &  $ 9.37\cdot10^{-4}$ & $ 1.42\cdot10^{-3}$ \\
\hline       
\end{tabular}
\end{table}

In Figure~\ref{fig:frame_Pf}, we assess the accuracy of the LRA and sparse PCE meta-models in estimating the failure probability $P_f=\Pp(u_{\rm lim}-U\leq0)=\Pp(U\geq u_{\rm lim})$, with the displacement threshold $u_{\rm lim}$ varying in the range $[3, 8]$~cm. All failure probabilities are herein computed with the IS technique. The reference values $P_f$ are obtained by employing IS in conjunction with the actual finite-element model; the LRA and PCE values are computed with exactly the same algorithm but using the respective meta-models in lieu of the original model. As in the truss-deflection problem, the auxiliary PDF in IS is defined in terms of the design point indicated by a previous FORM analysis. Samples of size $N_{\rm IS}=1,000$ are then sequentially added until the coefficient of variation of the estimated probability becomes smaller than $0.01$. Values of the so-obtained reference failure probability vary in the range $[ 8.72\cdot 10^{-2}, 7.38\cdot 10^{-6}]$; for the largest displacement thresholds, these tend to be underestimated when the actual model is replaced by the meta-models. The LRA estimates are characterized by higher accuracy than the PCE ones and remain within the order of magnitude of the reference value, even for the smallest ED and the largest displacement threshold considered. Note again the superior performance of LRA in the prediction of extreme responses, despite the smaller generalization error of PCE. Figure~\ref{fig:frame_beta} shows the corresponding ratios of the reliability indices $\beta^{\rm LRA}$ and $\beta^{\rm PCE}$, based on the LRA and sparse PCE meta-models, to the reliability index $\beta$, obtained in terms of the reference $P_f$. The latter varies in the range $[1.36, 4.33]$. When the LRA approach is employed, the relative error in $\beta$ remains smaller than $5\%$ for $N=500$ and smaller than $4\%$ for $N=1,000$. The PCE errors are slightly higher; for the largest displacement threshold, they exceed $10\%$ and $5\%$ when $N=500$ and $N=1,000$, respectively. The values of the failure probabilities and corresponding reliability indices depicted in Figures~\ref{fig:frame_Pf} and \ref{fig:frame_beta} are listed in the Appendix.

\begin{figure}[!ht]
\centering
\includegraphics[width=0.45\textwidth] {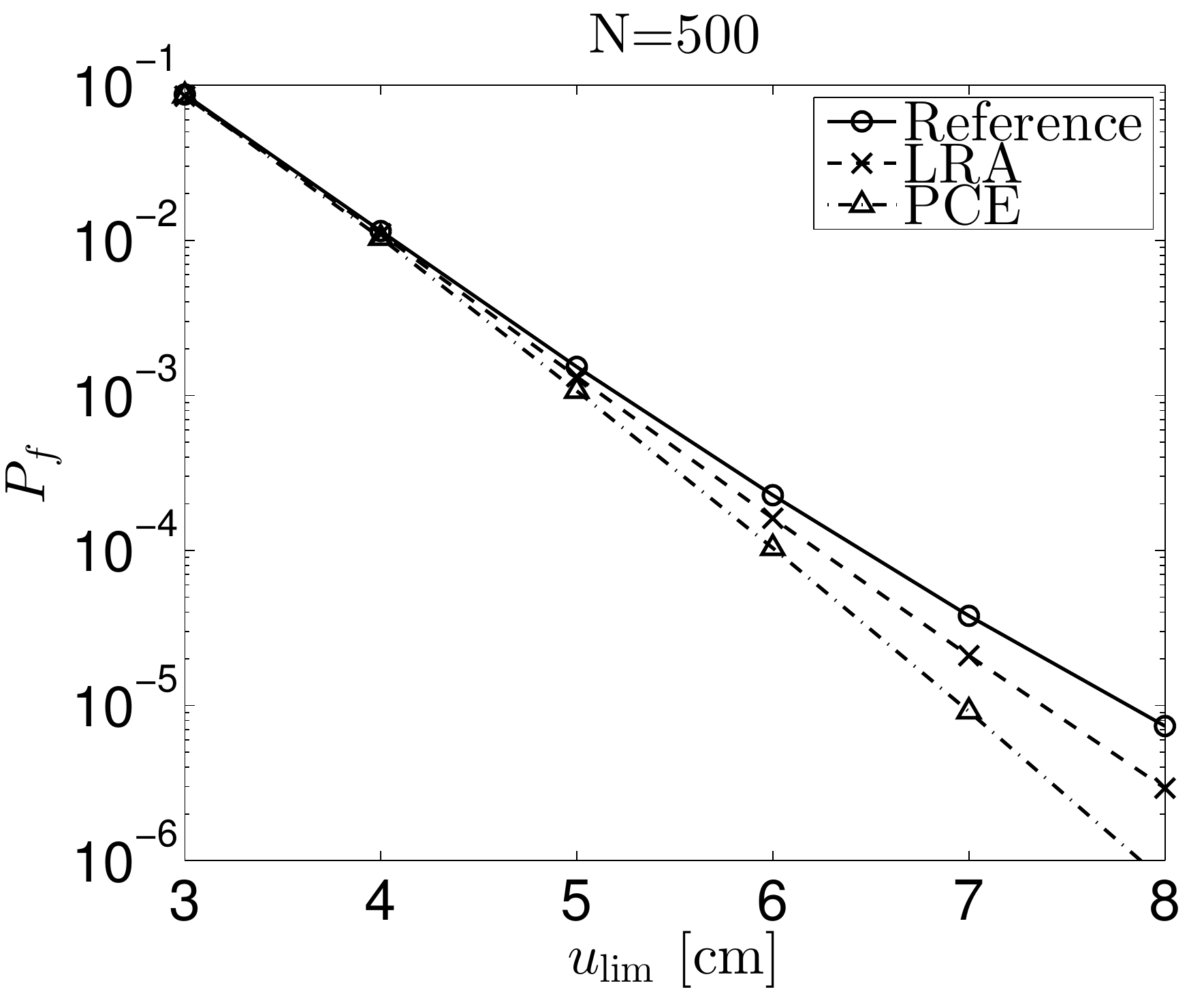}
\includegraphics[width=0.45\textwidth] {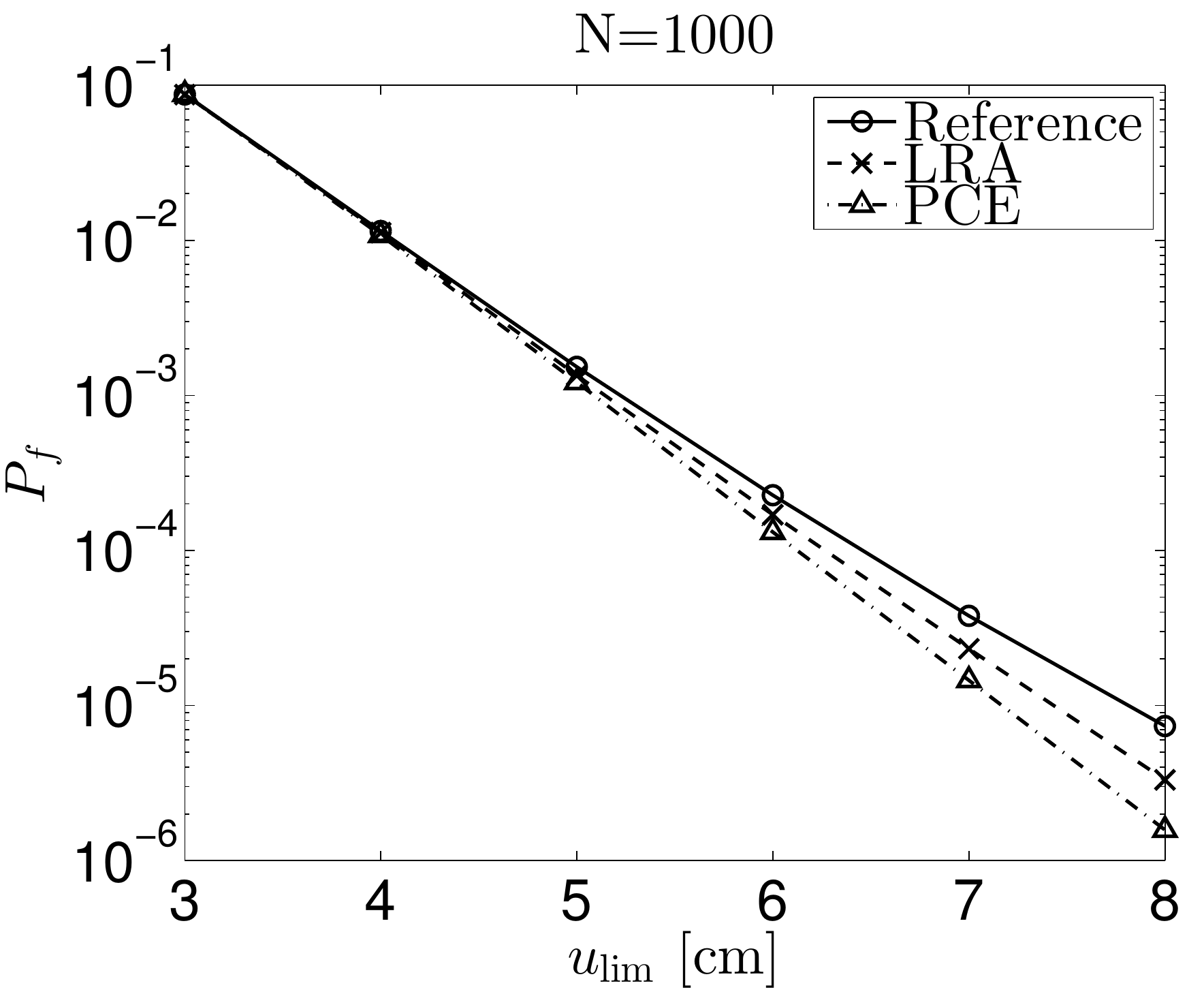}
\caption{Frame-displacement problem: Failure probabilities.}
\label{fig:frame_Pf}
\end{figure}

\begin{figure}[!ht]
\centering
\includegraphics[width=0.45\textwidth] {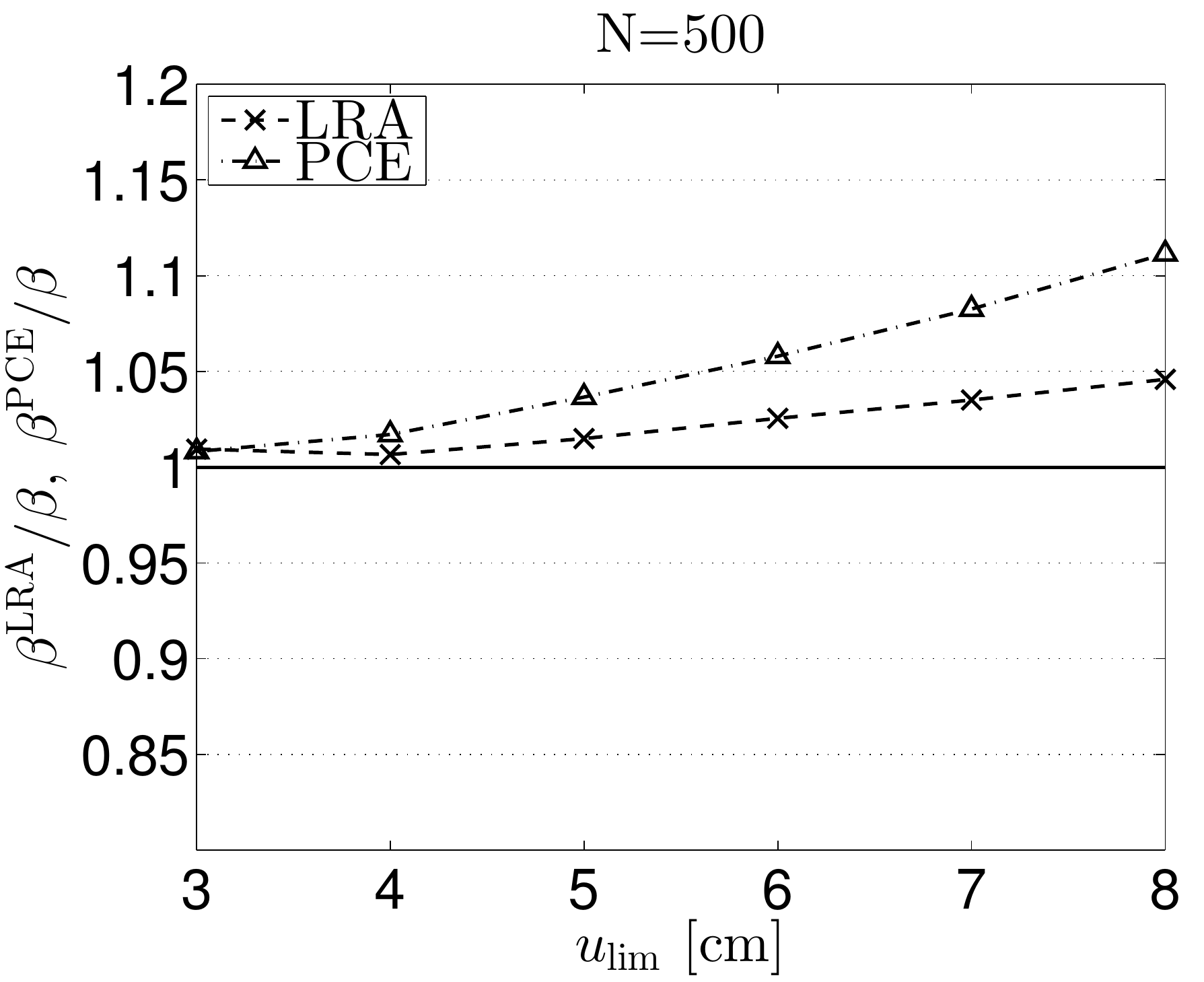}
\includegraphics[width=0.45\textwidth] {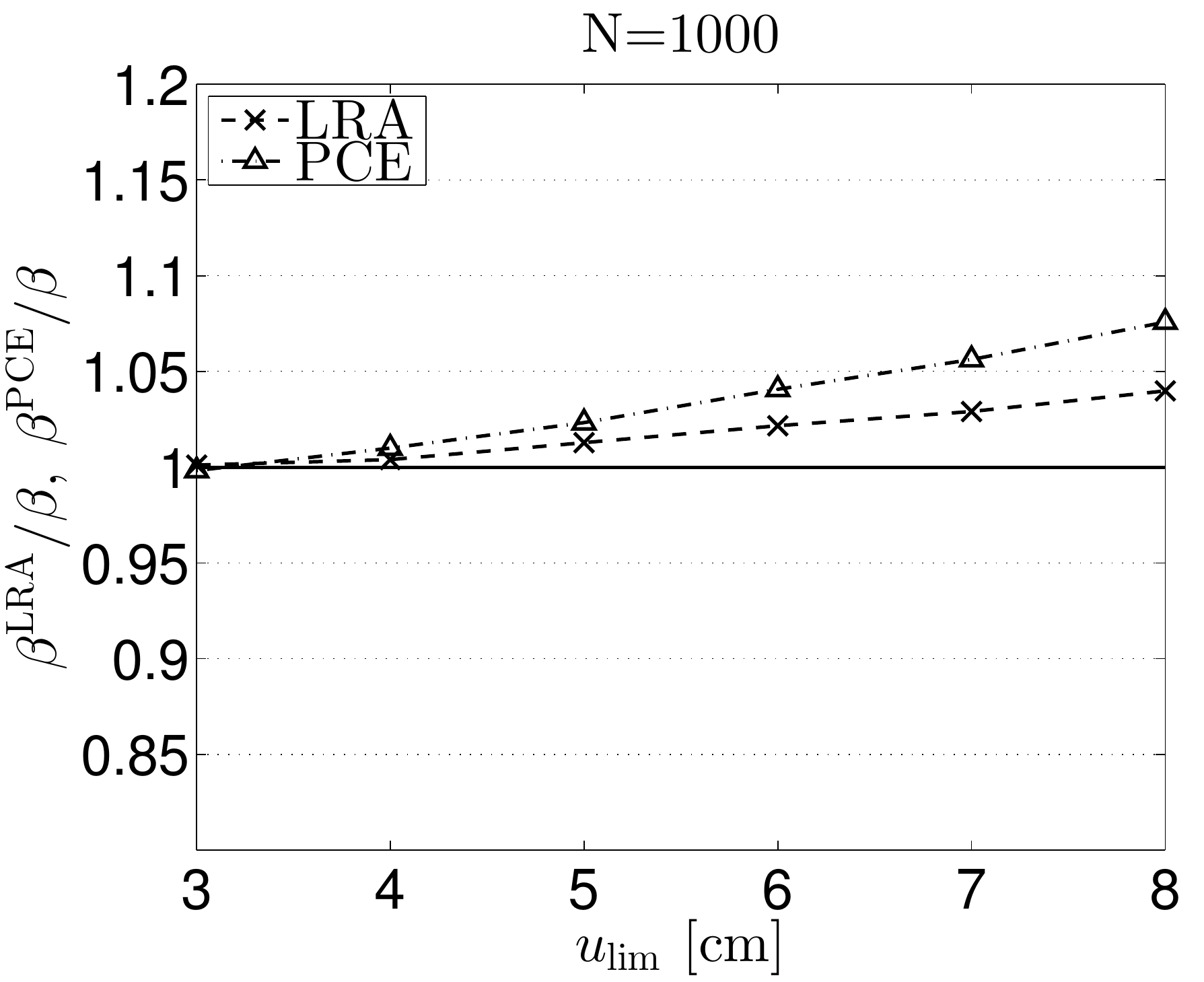}
\caption{Frame-displacement problem: Ratios of meta-model-based to reference reliability indices.}
\label{fig:frame_beta}
\end{figure}

To highlight the computational gain achieved by using a meta-modeling approach, in Table~\ref{tab:frame_eval}, we list the number of evaluations of the actual model (for each threshold) required to compute the reference failure probability. The given numbers include the model evaluations used to obtain the FORM estimate; as in the truss-deflection problem, the FORM analysis for each threshold, except for the smallest, is initiated from the design point corresponding to the previous threshold. The total computational cost of the reference solution comprises the sum of the listed model evaluations for all considered thresholds. We underline that this cost can increase fast with decreasing target coefficient of variation (note the orders-of-magnitude larger number of model evaluations in the present example, requiring CoV$<0.01$, as compared to that in the truss-deflection problem, requiring CoV$<0.10$). Conversely, when the analysis relies on the meta-models, IS is essentially costless, independently of the target coefficient of variation, and the computational effort is confined to the evaluations of the actual model at the ED. As seen above, LRA herein provides an estimate of a reliability index higher than $4$ with an error smaller than $5\%$ based on an ED as small as $N=500$.}

\begin{table} [!ht]
\centering
\caption{Frame-displacement problem: Number of model evaluations required in the computation of the reference failure probabilities.}
\label{tab:frame_eval}
\begin{tabular}{c c c}
\hline $u_{\rm lim}$ [cm] & IS \\
\hline
3 & 18,340 \\              
4 & 29,116 \\              
5 & 41,162 \\              
6 & 58,185 \\              
7 & 79,231 \\              
8 & 132,300 \\ 
\hline       
\end{tabular}
\end{table}

\section{CONCLUSIONS}

Reliability analysis faces challenges in cases when the systems under consideration are represented by complex high-dimensional computational models. In this paper, we demonstrate that meta-models belonging to the class of canonical low-rank approximations (LRA) can provide an accurate representation of the probability density function (PDF) of the model response at the tails, thus leading to efficient estimation of the small exceedence probabilities required in reliability analysis. By replacing a complex model by a meta-model that possesses similar statistical properties, evaluation of a response quantity of interest becomes essentially costless from a computational viewpoint. The LRA approach can be particularly efficient in high-dimensional problems because: (i) the number of unknowns grows only linearly with the input dimension and (ii) their construction relies on a series of least-square minimization problems of small size that is independent of the input dimension.

{  In this paper, canonical LRA developed with polynomial bases are of interest because of the simplicity and versatility characterizing the use of polynomial functions. The formulation and construction of such LRA meta-models in a non-intrusive manner is detailed. Furthermore, the links between canonical LRA and the popular meta-modeling technique of polynomial chaos expansions (PCE) are explained. Canonical LRA are confronted to sparse PCE in reliability applications involving a rank-one model (for which an analytical solution is available) and three finite-element models pertinent to structural mechanics and heat conduction. In all aforementioned applications, canonical LRA are found to outperform sparse PCE for cases when the size of the experimental design is relatively small with respect to the input dimension. By providing unbiased representations of the model responses at the tails, canonical LRA provide superior estimates of small exceedence probabilities compared to sparse PCE, even in cases when the latter exhibit smaller generalization errors. In the examined applications, failure probabilities of the order of $10^{-5}$ are predicted with sufficient accuracy by canonical LRA based on $5 M-25 M$ evaluations of the original model, where $M$ denotes the dimension of the random input. The LRA approach is also found to outperform methods particularly targeted to reliability analysis in terms of the required number of model evaluations. We underline that contrary to such methods, LRA provide a full probabilistic description of the model response, which can be used to estimate any statistical measure of interest beyond the probabilities of exceedence.

Having introduced canonical LRA in reliability analysis and demonstrated its strong potential for dealing with high dimensionality, we underline the need for further studies that will establish the efficacy of the approach in diverse reliability applications. The construction and use of tensor approximation of different formats, beyond the herein considered canonical formulation, is an active research topic in the field of uncertainty quantification. We hope that the present study will motivate further investigations into the capacities of such meta-models to accurately predict extreme responses of high-dimensional models, thus opening new paths to the risk assessment of complex systems.}

\section*{References}
\label{References}
\bibliographystyle{chicago}
\bibliography{bib}

\begin{thebibliography}{}

\bibitem[\protect\citeauthoryear{Abramowitz and Stegun}{Abramowitz and
  Stegun}{1970}]{Abramowitz}
Abramowitz, M. and I.~Stegun (1970).
\newblock {\em Handbook of mathematical functions}.
\newblock Dover Publications, Inc.

\bibitem[\protect\citeauthoryear{Acar, Camtepe, and Yener}{Acar
  et~al.}{2006}]{Acar2006collective}
Acar, E., S.~A. Camtepe, and B.~Yener (2006).
\newblock Collective sampling and analysis of high order tensors for chatroom
  communications.
\newblock In {\em Intelligence and security informatics}, pp.\  213--224.
  Springer.

\bibitem[\protect\citeauthoryear{Acharjee and Zabaras}{Acharjee and
  Zabaras}{2006}]{Acharjee2006}
Acharjee, S. and N.~Zabaras (2006).
\newblock Uncertainty propagation in finite deformations -- {A} spectral
  stochastic {Lagrangian} approach.
\newblock {\em Comput. Methods Appl. Mech. Engrg.\/}~{\em 195}, 2289--2312.

\bibitem[\protect\citeauthoryear{Allen}{Allen}{1971}]{Allen1971}
Allen, D. (1971).
\newblock {\em {The prediction sum of squares as a criterion for selecting
  predictor variables}}.
\newblock Number~23. Technical report, Dept. of Statistics, University of
  Kentucky.

\bibitem[\protect\citeauthoryear{Ammar, Mokdad, Chinesta, and Keunings}{Ammar
  et~al.}{2006}]{Ammar2006new}
Ammar, A., B.~Mokdad, F.~Chinesta, and R.~Keunings (2006).
\newblock A new family of solvers for some classes of multidimensional partial
  differential equations encountered in kinetic theory modeling of complex
  fluids.
\newblock {\em J. Non-Newton. Fluid\/}~{\em 139\/}(3), 153--176.

\bibitem[\protect\citeauthoryear{Andersen and Rayens}{Andersen and
  Rayens}{2004}]{Andersen2004structure}
Andersen, A.~H. and W.~S. Rayens (2004).
\newblock Structure-seeking multilinear methods for the analysis of f{MRI}
  data.
\newblock {\em NeuroImage\/}~{\em 22\/}(2), 728--739.

\bibitem[\protect\citeauthoryear{Appellof and Davidson}{Appellof and
  Davidson}{1981}]{Appellof1981strategies}
Appellof, C.~J. and E.~Davidson (1981).
\newblock Strategies for analyzing data from video fluorometric monitoring of
  liquid chromatographic effluents.
\newblock {\em Anal. Chem.\/}~{\em 53\/}(13), 2053--2056.

\bibitem[\protect\citeauthoryear{Arlot and Celisse}{Arlot and
  Celisse}{2010}]{Arlot2010}
Arlot, S. and A.~Celisse (2010).
\newblock A survey of cross-validation procedures for model selection.
\newblock {\em Stat. Surv.\/}~{\em 4}, 40--79.

\bibitem[\protect\citeauthoryear{Au and Beck}{Au and Beck}{2003}]{Au2003b}
Au, S. and J.~Beck (2003).
\newblock Important sampling in high dimensions.
\newblock {\em Structural Safety\/}~{\em 25}, 139--163.

\bibitem[\protect\citeauthoryear{Balesdent, Morio, and Marzat}{Balesdent
  et~al.}{2013}]{Balesdent2013}
Balesdent, M., J.~Morio, and J.~Marzat (2013).
\newblock Kriging-based adaptive importance sampling algorithms for rare event
  estimation.
\newblock {\em Structural Safety\/}~{\em 44}, 1--10.

\bibitem[\protect\citeauthoryear{Beylkin, Garcke, and Mohlenkamp}{Beylkin
  et~al.}{2009}]{Beylkin2009multivariate}
Beylkin, G., J.~Garcke, and M.~J. Mohlenkamp (2009).
\newblock Multivariate regression and machine learning with sums of separable
  functions.
\newblock {\em SIAM J. Sci. Comput.\/}~{\em 31\/}(3), 1840--1857.

\bibitem[\protect\citeauthoryear{Blatman and Sudret}{Blatman and
  Sudret}{2010}]{BlatmanPEM2010}
Blatman, G. and B.~Sudret (2010).
\newblock An adaptive algorithm to build up sparse polynomial chaos expansions
  for stochastic finite element analysis.
\newblock {\em Prob. Eng. Mech.\/}~{\em 25\/}(2), 183--197.

\bibitem[\protect\citeauthoryear{Blatman and Sudret}{Blatman and
  Sudret}{2011}]{BlatmanJCP2011}
Blatman, G. and B.~Sudret (2011).
\newblock Adaptive sparse polynomial chaos expansion based on least angle
  regression.
\newblock {\em J. Comput. Phys.\/}~{\em 230}, 2345--2367.

\bibitem[\protect\citeauthoryear{Breitung}{Breitung}{1989}]{Breitung1989}
Breitung, K. (1989).
\newblock Asymptotic approximations for probability integrals.
\newblock {\em Prob. Eng. Mech.\/}~{\em 4\/}(4), 187--190.

\bibitem[\protect\citeauthoryear{Bro}{Bro}{1997}]{Bro1997parafac}
Bro, R. (1997).
\newblock {PARAFAC}. tutorial and applications.
\newblock {\em Chemometr. Intell. Lab.\/}~{\em 38\/}(2), 149--171.

\bibitem[\protect\citeauthoryear{Carroll and Chang}{Carroll and
  Chang}{1970}]{Carroll1970a}
Carroll, J.~D. and J.-J. Chang (1970).
\newblock Analysis of individual differences in multidimensional scaling via an
  n-way generalization of ``{E}ckart-{Y}oung'' decomposition.
\newblock {\em Psychometrika\/}~{\em 35\/}(3), 283--319.

\bibitem[\protect\citeauthoryear{Chapelle, Vapnik, and Bengio}{Chapelle
  et~al.}{2002}]{Chapelle2002}
Chapelle, O., V.~Vapnik, and Y.~Bengio (2002).
\newblock {Model selection for small sample regression}.
\newblock {\em Mach. Learn.\/}~{\em 48\/}(1), 9--23.

\bibitem[\protect\citeauthoryear{Chevreuil, Lebrun, Nouy, and Rai}{Chevreuil
  et~al.}{2013}]{Chevreuil2013least}
Chevreuil, M., R.~Lebrun, A.~Nouy, and P.~Rai (2013).
\newblock A least-squares method for sparse low rank approximation of
  multivariate functions.
\newblock {\em arXiv preprint arXiv:1305.0030\/}.

\bibitem[\protect\citeauthoryear{Chevreuil, Rai, and Nouy}{Chevreuil
  et~al.}{2013}]{Chevreuil2013}
Chevreuil, M., P.~Rai, and A.~Nouy (2013).
\newblock Sampling based tensor approximation method for uncertainty
  propagation.
\newblock In {\em Proc. 11th Int. Conf. Struct. Safety and Reliability
  {(ICOSSAR2013)}, New York}.

\bibitem[\protect\citeauthoryear{De~Lathauwer and Castaing}{De~Lathauwer and
  Castaing}{2007}]{DeLathauwer2007tensor}
De~Lathauwer, L. and J.~Castaing (2007).
\newblock Tensor-based techniques for the blind separation of {DS}--{CDMA}
  signals.
\newblock {\em Signal Process.\/}~{\em 87\/}(2), 322--336.

\bibitem[\protect\citeauthoryear{Deman, Konakli, Sudret, Kerrou, Perrochet, and
  Benabderrahmane}{Deman et~al.}{2016}]{DemanKonakli2016}
Deman, G., K.~Konakli, B.~Sudret, J.~Kerrou, P.~Perrochet, and
  H.~Benabderrahmane (2016).
\newblock Using sparse polynomial chaos expansions for the global sensitivity
  analysis of groundwater lifetime expectancy in a multi-layered
  hydrogeological model.
\newblock {\em Reliab. Eng. Sys. Safety\/}~{\em 147}, 156--169.

\bibitem[\protect\citeauthoryear{{Der Kiureghian} and {de Stefano}}{{Der
  Kiureghian} and {de Stefano}}{1991}]{DerKiureghian1991}
{Der Kiureghian}, A. and M.~{de Stefano} (1991).
\newblock Efficient algorithms for second order reliability analysis.
\newblock {\em J. Eng. Mech.\/}~{\em 117\/}(12), 2906--2923.

\bibitem[\protect\citeauthoryear{Doostan, Validi, and Iaccarino}{Doostan
  et~al.}{2013}]{Doostan2013}
Doostan, A., A.~Validi, and G.~Iaccarino (2013).
\newblock Non-intrusive low-rank separated approximation of high-dimensional
  stochastic models.
\newblock {\em Comput. Method. Appl. M.\/}~{\em 263}, 42--55.

\bibitem[\protect\citeauthoryear{Dubourg, Sudret, and Deheeger}{Dubourg
  et~al.}{2013}]{Dubourg2013}
Dubourg, V., B.~Sudret, and F.~Deheeger (2013).
\newblock Metamodel-based importance sampling for structural reliability
  analysis.
\newblock {\em Prob. Eng. Mech.\/}~{\em 33}, 47--57.

\bibitem[\protect\citeauthoryear{Efron, Hastie, Johnstone, and
  Tibshirani}{Efron et~al.}{2004}]{Efron2004}
Efron, B., T.~Hastie, I.~Johnstone, and R.~Tibshirani (2004).
\newblock Least angle regression.
\newblock {\em Ann. Stat.\/}~{\em 32}, 407--499.

\bibitem[\protect\citeauthoryear{Felippa and Ohayon}{Felippa and
  Ohayon}{1990}]{Felippa1990mixed}
Felippa, C. and R.~Ohayon (1990).
\newblock Mixed variational formulation of finite element analysis of
  acoustoelastic/slosh fluid-structure interaction.
\newblock {\em J. Fluid. Struct.\/}~{\em 4\/}(1), 35--57.

\bibitem[\protect\citeauthoryear{Furukawa, Kawasaki, Ikeuchi, and
  Sakauchi}{Furukawa et~al.}{2002}]{Furukawa2002appearance}
Furukawa, R., H.~Kawasaki, K.~Ikeuchi, and M.~Sakauchi (2002).
\newblock Appearance based object modeling using texture database: Acquisition
  compression and rendering.
\newblock In {\em Rendering Techniques}, pp.\  257--266.

\bibitem[\protect\citeauthoryear{Geuzaine and Remacle}{Geuzaine and
  Remacle}{2009}]{Geuzaine2009gmsh}
Geuzaine, C. and J.-F. Remacle (2009).
\newblock Gmsh: A 3-d finite element mesh generator with built-in pre-and
  post-processing facilities.
\newblock {\em Int. J. Numer. Meth. Eng.\/}~{\em 79\/}(11), 1309--1331.

\bibitem[\protect\citeauthoryear{Grasedyck, Kressner, and Tobler}{Grasedyck
  et~al.}{2013}]{Grasedyck2013}
Grasedyck, L., D.~Kressner, and C.~Tobler (2013).
\newblock A literature survey of low-rank tensor approximation techniques.
\newblock {\em arXiv preprint arXiv:1302.7121\/}.

\bibitem[\protect\citeauthoryear{Hackbusch}{Hackbusch}{2012}]{Hackbusch2012}
Hackbusch, W. (2012).
\newblock {\em Tensor spaces and numerical tensor calculus}, Volume~42.
\newblock Springer Science \& Business Media.

\bibitem[\protect\citeauthoryear{Hadigol, Doostan, Matthies, and
  Niekamp}{Hadigol et~al.}{2014}]{Hadigol2014}
Hadigol, M., A.~Doostan, H.~G. Matthies, and R.~Niekamp (2014).
\newblock Partitioned treatment of uncertainty in coupled domain problems: A
  separated representation approach.
\newblock {\em Comput. Method. Appl. M.\/}~{\em 274}, 103--124.

\bibitem[\protect\citeauthoryear{Harshman}{Harshman}{1970}]{Harshman1970foundations}
Harshman, R.~A. (1970).
\newblock {\em Foundations of the {PARAFAC} procedure: Models and conditions
  for an ``explanatory'' multi-modal factor analysis}.
\newblock University of California at Los Angeles Los Angeles.

\bibitem[\protect\citeauthoryear{Hasofer and Lind}{Hasofer and
  Lind}{1974}]{Hasofer1974}
Hasofer, A.~M. and N.~C. Lind (1974).
\newblock Exact and invariant second-moment code format.
\newblock {\em J. Eng. Mech.\/}~{\em 100\/}(1), 111--121.

\bibitem[\protect\citeauthoryear{Hitchcock}{Hitchcock}{1927}]{Hitchcock1927}
Hitchcock, F. (1927).
\newblock The expression of a tensor or a polyadic as a sum of products.
\newblock {\em J. Math. Phys. Camb.\/}~{\em 6}, 164--189.

\bibitem[\protect\citeauthoryear{Jones, Doostan, and Born}{Jones
  et~al.}{2013}]{Jones2013}
Jones, B.~A., A.~Doostan, and G.~H. Born (2013).
\newblock Nonlinear propagation of orbit uncertainty using non-intrusive
  polynomial chaos.
\newblock {\em J. Guid. Control Dyn.\/}~{\em 36\/}(2), 430--444.

\bibitem[\protect\citeauthoryear{Khoromskij and Schwab}{Khoromskij and
  Schwab}{2011}]{Khoromskij2011}
Khoromskij, B.~N. and C.~Schwab (2011).
\newblock Tensor-structured {G}alerkin approximation of parametric and
  stochastic elliptic {PDE}s.
\newblock {\em {SIAM} J. Sci. Comput.\/}~{\em 33\/}(1), 364--385.

\bibitem[\protect\citeauthoryear{Kolda and Bader}{Kolda and
  Bader}{2009}]{Kolda2009}
Kolda, T.~G. and B.~W. Bader (2009).
\newblock Tensor decompositions and applications.
\newblock {\em SIAM Rev.\/}~{\em 51\/}(3), 455--500.

\bibitem[\protect\citeauthoryear{Konakli and Sudret}{Konakli and
  Sudret}{2015a}]{Konakli2015arxiv}
Konakli, K. and B.~Sudret (2015a).
\newblock Low-rank tensor approximations versus polynomial chaos expansions for
  meta-modeling in high-dimensional spaces.
\newblock {\em arXiv preprint arXiv:1511.07492\/}.

\bibitem[\protect\citeauthoryear{Konakli and Sudret}{Konakli and
  Sudret}{2015b}]{Konakli2015UNCECOMP}
Konakli, K. and B.~Sudret (2015b).
\newblock Uncertainty quantification in high-dimensional spaces with low-rank
  tensor approximations.
\newblock In {\em Proc. 1st Int. Conf. on Uncertainty Quantification in Comput.
  Sci. and Eng., {(UNCECOMP)}, Crete island, Greece}.

\bibitem[\protect\citeauthoryear{Lebrun and Dutfoy}{Lebrun and
  Dutfoy}{2009a}]{Lebrun2009b}
Lebrun, R. and A.~Dutfoy (2009a).
\newblock A generalization of the {N}ataf transformation to distributions with
  elliptical copula.
\newblock {\em Prob. Eng. Mech.\/}~{\em 24\/}(2), 172--178.

\bibitem[\protect\citeauthoryear{Lebrun and Dutfoy}{Lebrun and
  Dutfoy}{2009b}]{Lebrun2009a}
Lebrun, R. and A.~Dutfoy (2009b).
\newblock An innovating analysis of the {N}ataf transformation from the copula
  viewpoint.
\newblock {\em Prob. Eng. Mech.\/}~{\em 24\/}(3), 312--320.

\bibitem[\protect\citeauthoryear{Li and Der~Kiureghian}{Li and
  Der~Kiureghian}{1993}]{Li1993optimal}
Li, C.-C. and A.~Der~Kiureghian (1993).
\newblock Optimal discretization of random fields.
\newblock {\em J. Eng. Mech.\/}~{\em 119\/}(6), 1136--1154.

\bibitem[\protect\citeauthoryear{Li, Bect, and Vazquez}{Li
  et~al.}{2012}]{Li2012bayesian}
Li, L., J.~Bect, and E.~Vazquez (2012).
\newblock Bayesian subset simulation: a kriging-based subset simulation
  algorithm for the estimation of small probabilities of failure.
\newblock {\em arXiv preprint arXiv:1207.1963\/}.

\bibitem[\protect\citeauthoryear{Liu and {Der Kiureghian}}{Liu and {Der
  Kiureghian}}{1991}]{LiuPL91}
Liu, P.-L. and A.~{Der Kiureghian} (1991).
\newblock Optimization algorithms for structural reliability.
\newblock {\em Structural Safety\/}~{\em 9}, 161--177.

\bibitem[\protect\citeauthoryear{Marelli, Sch\"obi, and Sudret}{Marelli
  et~al.}{2015}]{UQdoc_09_107}
Marelli, S., R.~Sch\"obi, and B.~Sudret (2015).
\newblock Uqlab user manual -- reliability analysis.
\newblock Technical report, Chair of Risk, Safety \& Uncertainty
  Quantification, ETH Zurich.
\newblock Report \# UQLab-V0.9-107.

\bibitem[\protect\citeauthoryear{Marelli and Sudret}{Marelli and
  Sudret}{2014}]{MarelliICVRAM2014}
Marelli, S. and B.~Sudret (2014).
\newblock {UQLab}: a framework for uncertainty quantification in {MATLAB}.
\newblock In {\em Proc. 2nd Int. Conf. on Vulnerability, Risk Analysis and
  Management {(ICVRAM2014)}, Liverpool, United Kingdom}.

\bibitem[\protect\citeauthoryear{Marelli and Sudret}{Marelli and
  Sudret}{2015}]{UQdoc_09_104}
Marelli, S. and B.~Sudret (2015).
\newblock Uqlab user manual -- polynomial chaos expansions.
\newblock Technical report, Chair of Risk, Safety \& Uncertainty
  Quantification, ETH Zurich.
\newblock Report \# UQLab-V0.9-104.

\bibitem[\protect\citeauthoryear{Melchers}{Melchers}{1989}]{Melchers1989}
Melchers, R. (1989).
\newblock Importance sampling in structural systems.
\newblock {\em Structural Safety\/}~{\em 6}, 3--10.

\bibitem[\protect\citeauthoryear{Mocks}{Mocks}{1988}]{Mocks1988topographic}
Mocks, J. (1988).
\newblock Topographic components model for event-related potentials and some
  biophysical considerations.
\newblock {\em IEEE T. Bio-Med. Eng.\/}~{\em 35\/}(6), 482--484.

\bibitem[\protect\citeauthoryear{Morio, Balesdent, Jacquemart, and
  Verg{\'e}}{Morio et~al.}{2014}]{Morio2014}
Morio, J., M.~Balesdent, D.~Jacquemart, and C.~Verg{\'e} (2014).
\newblock A survey of rare event simulation methods for static input--output
  models.
\newblock {\em Simul. Model. Pract. Th.\/}~{\em 49}, 287--304.

\bibitem[\protect\citeauthoryear{Najm, Debusschere, Marzouk, Widmer, and
  Le~Ma{\^\i}tre}{Najm et~al.}{2009}]{Najm2009}
Najm, H.~N., B.~J. Debusschere, Y.~M. Marzouk, S.~Widmer, and O.~Le~Ma{\^\i}tre
  (2009).
\newblock Uncertainty quantification in chemical systems.
\newblock {\em Int. J. Numer. Meth. Engng.\/}~{\em 80\/}(6-7), 789--814.

\bibitem[\protect\citeauthoryear{Nelsen}{Nelsen}{2006}]{Nelsen2006}
Nelsen, R. (2006).
\newblock {\em An introduction to copulas\/} (2nd ed.), Volume 139 of {\em
  Lecture {Notes} in {Statistics}}.
\newblock Springer-Verlag, New York.

\bibitem[\protect\citeauthoryear{Niederreiter}{Niederreiter}{1992}]{Niederreiter1992}
Niederreiter, H. (1992).
\newblock {\em Random number generation and quasi-{M}onte Carlo methods}.
\newblock Society for {I}ndustrial and {A}pplied {M}athematics, {P}hiladelphia,
  PA, USA.

\bibitem[\protect\citeauthoryear{Nouy}{Nouy}{2010}]{Nouy2010b}
Nouy, A. (2010).
\newblock Proper generalized decompositions and separated representations for
  the numerical solution of high dimensional stochastic problems.
\newblock {\em Archives of Computational Methods in Engineering\/}~{\em 17},
  403--434.

\bibitem[\protect\citeauthoryear{Rackwitz and Fiessler}{Rackwitz and
  Fiessler}{1978}]{Rackwitz1978}
Rackwitz, R. and B.~Fiessler (1978).
\newblock Structural reliability under combined load sequences.
\newblock {\em Computers \& Structures\/}~{\em 9}, 489--494.

\bibitem[\protect\citeauthoryear{Rai}{Rai}{2014}]{RaiThesis}
Rai, P. (2014).
\newblock {\em Sparse Low Rank Approximation of Multivariate Functions --
  {A}pplications in uncertainty quantification}.
\newblock Ph.\ D. thesis, Engineering Sciences [physics]. Ecole Centrale
  Nantes.

\bibitem[\protect\citeauthoryear{Shashua and Levin}{Shashua and
  Levin}{2001}]{Shashua2001linear}
Shashua, A. and A.~Levin (2001).
\newblock Linear image coding for regression and classification using the
  tensor-rank principle.
\newblock In {\em Proc. 2001 IEEE Computer Society Conference on Computer
  Vision and Pattern Recognition {(CVPR)}}, Volume~1, pp.\  I--42. IEEE.

\bibitem[\protect\citeauthoryear{Sidiropoulos, Bro, and Giannakis}{Sidiropoulos
  et~al.}{2000}]{Sidiropoulos2000parallel}
Sidiropoulos, N.~D., R.~Bro, and G.~B. Giannakis (2000).
\newblock Parallel factor analysis in sensor array processing.
\newblock {\em IEEE T. Signal Proces.\/}~{\em 48\/}(8), 2377--2388.

\bibitem[\protect\citeauthoryear{Soize and Ghanem}{Soize and
  Ghanem}{2004}]{Soize2004}
Soize, C. and R.~Ghanem (2004).
\newblock Physical systems with random uncertainties: chaos representations
  with arbitrary probability measure.
\newblock {\em SIAM J. Sci. Comput.\/}~{\em 26\/}(2), 395--410.

\bibitem[\protect\citeauthoryear{Sudret}{Sudret}{2007}]{SudretHDR}
Sudret, B. (2007).
\newblock {\em Uncertainty propagation and sensitivity analysis in mechanical
  models -- Contributions to structural reliability and stochastic spectral
  methods}.
\newblock Universit\'e Blaise Pascal, Clermont-Ferrand, France.
\newblock Habilitation \`a diriger des recherches, 173 pages.

\bibitem[\protect\citeauthoryear{Sudret and Der~Kiureghian}{Sudret and
  Der~Kiureghian}{2000}]{Sudret2000stochastic}
Sudret, B. and A.~Der~Kiureghian (2000).
\newblock {\em Stochastic finite element methods and reliability: a
  state-of-the-art report}.
\newblock Department of Civil and Environmental Engineering, University of
  California.

\bibitem[\protect\citeauthoryear{Validi}{Validi}{2014}]{Validi2014}
Validi, A. (2014).
\newblock Low-rank separated representation surrogates of high-dimensional
  stochastic functions: Application in {B}ayesian inference.
\newblock {\em J. Comput. Phys.\/}~{\em 260}, 37--53.

\bibitem[\protect\citeauthoryear{Viana, Haftka, and Steffen~Jr}{Viana
  et~al.}{2009}]{Viana2009}
Viana, F.~A., R.~T. Haftka, and V.~Steffen~Jr (2009).
\newblock Multiple surrogates: how cross-validation errors can help us to
  obtain the best predictor.
\newblock {\em Struct. Multidiscip. O.\/}~{\em 39\/}(4), 439--457.

\bibitem[\protect\citeauthoryear{Xiu and Karniadakis}{Xiu and
  Karniadakis}{2003}]{Xiu2003}
Xiu, D. and G.~Karniadakis (2003).
\newblock Modelling uncertainty in steady state diffusion problems via
  generalized polynomial chaos.
\newblock {\em Comput. Methods Appl. Mech. Engrg.\/}~{\em 191\/}(43),
  4927--4948.

\bibitem[\protect\citeauthoryear{Xiu and Karniadakis}{Xiu and
  Karniadakis}{2002}]{Xiu2002wiener}
Xiu, D. and G.~E. Karniadakis (2002).
\newblock The {W}iener--{A}skey polynomial chaos for stochastic differential
  equations.
\newblock {\em SIAM J. Sci. Comput.\/}~{\em 24\/}(2), 619--644.

\end{thebibliography}

\section*{Appendix}

\begin{table}[!ht]        
\centering
\caption*{Table A1: Beam-deflection problem: Failure probabilities (see Figure~\ref{fig:beam_Pf}).}            
\begin{tabular}{c c c c c c}
\hline
 \multirow{2}{4.5em}{$u_{\rm lim}$ [mm]} & Analytical & \multicolumn{2}{c} {LRA} & \multicolumn{2}{c} {PCE} \\
 {} & {} & $N = 30$ & $N = 50$ & $N = 30$ & $N = 50$ \\
\hline                                
4 & $6.60\cdot10^{-2}$ & $6.56\cdot10^{-2}$ & $6.57\cdot10^{-2}$ & $5.67\cdot10^{-2}$ & $6.37\cdot10^{-2}$ \\
5 & $1.19\cdot10^{-2}$ & $1.15\cdot10^{-2}$ & $1.18\cdot10^{-2}$ & $5.94\cdot10^{-3}$ & $9.23\cdot10^{-3}$ \\
6 & $2.00\cdot10^{-3}$ & $1.81\cdot10^{-3}$ & $1.96\cdot10^{-3}$ & $4.00\cdot10^{-4}$ & $1.00\cdot10^{-3}$ \\
7 & $3.37\cdot10^{-4}$ & $2.80\cdot10^{-4}$ & $3.25\cdot10^{-4}$ & $2.07\cdot10^{-5}$ & $9.08\cdot10^{-5}$ \\
8 & $5.86\cdot10^{-5}$ & $4.57\cdot10^{-5}$ & $5.57\cdot10^{-5}$ & $5.00\cdot10^{-7}$ & $6.70\cdot10^{-6}$ \\
9 & $1.07\cdot10^{-5}$ & $7.90\cdot10^{-6}$ & $1.01\cdot10^{-5}$ & - & $3.00\cdot10^{-7}$ \\
\hline
\end{tabular} 
\end{table}

\begin{table}[!ht]        
\centering 
\caption*{Table A2: Beam-deflection problem: Reliability indices (see Figure~\ref{fig:beam_beta}).}            
\begin{tabular}{c c c c c c}
\hline
 \multirow{2}{4.5em}{$u_{\rm lim}$ [mm]} & Analytical & \multicolumn{2}{c} {LRA} & \multicolumn{2}{c} {PCE} \\
 {} & {} & $N = 30$ & $N = 50$ & $N = 30$ & $N = 50$ \\
\hline                              
4 & 1.51 & 1.51 & 1.51 & 1.58 & 1.52 \\
5 & 2.26 & 2.27 & 2.26 & 2.52 & 2.36 \\
6 & 2.88 & 2.91 & 2.88 & 3.35 & 3.09 \\
7 & 3.40 & 3.45 & 3.41 & 4.10 & 3.74 \\
8 & 3.85 & 3.91 & 3.86 & 4.89 & 4.35 \\
9 & 4.25 & 4.32 & 4.26 & - & 4.99 \\
\hline
\end{tabular} 
\end{table}

\begin{table}[!ht]        
\centering
\caption*{Table A3: Truss-deflection problem: Failure probabilities (see Figure~\ref{fig:truss_Pf}).}            
\begin{tabular}{c c c c c c c}
\hline
\multirow{2}{4.5em}{$u_{\rm lim}$ [cm]} & \multicolumn{2}{c} {Reference}  & \multicolumn{2}{c} {LRA} & \multicolumn{2}{c} {PCE} \\
 {} & {SORM} & { IS} & $N = 50$ & $N = 100$ & $N = 50$ & $N = 100$ \\
\hline                                
10 & $4.47\cdot10^{-2}$ & $4.13\cdot10^{-2}$ & $4.35\cdot10^{-2}$ & $4.23\cdot10^{-2}$ & $3.17\cdot10^{-2}$ & $3.83\cdot10^{-2}$ \\
11 & $9.28\cdot10^{-3}$ & $9.86\cdot10^{-3}$ & $8.88\cdot10^{-3}$ & $8.47\cdot10^{-3}$ & $3.66\cdot10^{-3}$ & $6.45\cdot10^{-3}$ \\
12 & $1.63\cdot10^{-3}$ & $1.34\cdot10^{-3}$ & $1.52\cdot10^{-3}$ & $1.41\cdot10^{-3}$ & $2.58\cdot10^{-4}$ & $8.19\cdot10^{-4}$ \\
13 & $2.57\cdot10^{-4}$ & $2.14\cdot10^{-4}$ & $2.28\cdot10^{-4}$ & $2.09\cdot10^{-4}$ & $1.20\cdot10^{-5}$ & $8.03\cdot10^{-5}$ \\
14 & $3.77\cdot10^{-5}$ & $3.46\cdot10^{-5}$ & $3.23\cdot10^{-5}$ & $2.73\cdot10^{-5}$ & $4.00\cdot10^{-7}$ & $6.03\cdot10^{-6}$ \\
15 & $5.31\cdot10^{-6}$ & $3.90\cdot10^{-6}$ & $4.73\cdot10^{-6}$ & $3.10\cdot10^{-6}$ & - & $4.33\cdot10^{-7}$ \\
\hline
\end{tabular} 
\end{table}

\begin{table}[!ht]        
\centering 
\caption*{Table A4: Truss-deflection problem: Reliability indices (see Figure~\ref{fig:truss_beta}).}            
\begin{tabular}{c c c c c c c}
\hline
\multirow{2}{4.5em}{$u_{\rm lim}$ [cm]} & \multicolumn{2}{c} {Reference}  & \multicolumn{2}{c} {LRA} & \multicolumn{2}{c} {PCE} \\
 {} & {SORM} & { IS} & $N = 50$ & $N = 100$ & $N = 50$ & $N = 100$ \\
\hline                              
10 & 1.70 & 1.74 & 1.71 & 1.72 & 1.86 & 1.77 \\
11 & 2.35 & 2.33 & 2.37 & 2.39 & 2.68 & 2.49 \\
12 & 2.94 & 3.00 & 2.96 & 2.99 & 3.47 & 3.15 \\
13 & 3.47 & 3.52 & 3.51 & 3.53 & 4.22 & 3.77 \\
14 & 3.96 & 3.98 & 4.00 & 4.04 & 4.94 & 4.38 \\
15 & 4.40 & 4.47 & 4.43 & 4.52 & - & 4.92 \\ 
\hline
\end{tabular} 
\end{table}

\begin{table}[!ht]        
\centering
\caption*{Table A5: Heat-conduction problem: Failure probabilities (see Figures~\ref{fig:RF_beta} and \ref{fig:RF_Pf}).}            
\begin{tabular}{c c c c c c}
\hline
 \multirow{2}{4em}{$\widetilde{t}_{\rm lim}~[\rm^{\circ} C]$} & Reference & \multicolumn{2}{c} {LRA} & \multicolumn{2}{c} {PCE} \\
 {} & {} & $N = 500$ & $N = 2,000$ & $N = 500$ & $N = 2,000$ \\
\hline                                
6 & $4.53\cdot10^{-2}$ & $4.57\cdot10^{-2}$ & $4.59\cdot10^{-2}$ & $4.13\cdot10^{-2}$ & $4.57\cdot10^{-2}$ \\  
6.5 & $1.43\cdot10^{-2}$ & $1.42\cdot10^{-2}$ & $1.44\cdot10^{-2}$ & $1.15\cdot10^{-2}$ & $1.36\cdot10^{-2}$ \\
8 & - & $3.22\cdot10^{-4}$ & $3.23\cdot10^{-4}$ & $1.23\cdot10^{-4}$ & $2.27\cdot10^{-4}$ \\  
8.5 & - & $7.35\cdot10^{-5}$ & $7.40\cdot10^{-5}$ & $1.65\cdot10^{-5}$ & $4.45\cdot10^{-5}$ \\
\hline
\end{tabular} 
\end{table}

\begin{table}[!ht]        
\centering
\caption*{Table A6: Heat-conduction problem: Reliability indices (see Figures~\ref{fig:RF_beta} and \ref{fig:RF_Pf}).}            
\begin{tabular}{c c c c c c}
\hline
 \multirow{2}{4em}{$\widetilde{t}_{\rm lim}~[\rm^{\circ} C]$} & Reference & \multicolumn{2}{c} {LRA} & \multicolumn{2}{c} {PCE} \\
{} & {} & $N = 500$ & $N = 2,000$ & $N = 500$ & $N = 2,000$ \\
\hline                                
6 & 1.69 & 1.69 & 1.69 & 1.74 & 1.69 \\  
6.5 & 2.19 & 2.19 & 2.19 & 2.27 & 2.21 \\
8 & - & 3.41 & 3.41 & 3.67 & 3.51 \\  
8.5 & - & 3.80 & 3.79 & 4.15 & 3.92 \\
\hline
\end{tabular} 
\end{table}

\begin{table}[!ht]        
\centering
\caption*{Table A7: Frame displacement: Failure probabilities (see Figure~\ref{fig:frame_Pf}).}            
\begin{tabular}{c c c c c c}
\hline
 \multirow{2}{2em}{$u_{\rm lim}$ [cm]} & Reference & \multicolumn{2}{c} {LRA} & \multicolumn{2}{c} {PCE} \\
{} & {} & $N = 500$ & $N = 1,000$ & $N = 500$ & $N = 1,000$ \\
\hline                                
3 & $8.72\cdot10^{-2}$ & $8.51\cdot10^{-2}$ & $8.69\cdot10^{-2}$ & $8.54\cdot10^{-2}$ & $8.75\cdot10^{-2}$ \\
4 & $1.15\cdot10^{-2}$ & $1.10\cdot10^{-2}$ & $1.12\cdot10^{-2}$ & $1.04\cdot10^{-2}$ & $1.08\cdot10^{-2}$ \\
5 & $1.53\cdot10^{-3}$ & $1.32\cdot10^{-3}$ & $1.35\cdot10^{-3}$ & $1.07\cdot10^{-3}$ & $1.22\cdot10^{-3}$ \\
6 & $2.27\cdot10^{-4}$ & $1.61\cdot10^{-4}$ & $1.70\cdot10^{-4}$ & $1.04\cdot10^{-4}$ & $1.31\cdot10^{-4}$ \\
7 & $3.78\cdot10^{-5}$ & $2.10\cdot10^{-5}$ & $2.32\cdot10^{-5}$ & $9.16\cdot10^{-6}$ & $1.45\cdot10^{-5}$ \\
8 & $7.38\cdot10^{-6}$ & $2.94\cdot10^{-6}$ & $3.32\cdot10^{-6}$ & $7.36\cdot10^{-7}$ & $1.58\cdot10^{-6}$ \\
\hline
\end{tabular} 
\end{table}

\begin{table}[!ht]        
\centering 
\caption*{Table A8: Frame displacement: Reliability indices (see Figure~\ref{fig:frame_beta}).}            
\begin{tabular}{c c c c c c}
\hline
 \multirow{2}{2em}{$u_{\rm lim}$ [cm]} & Reference & \multicolumn{2}{c} {LRA} & \multicolumn{2}{c} {PCE} \\
{} & {} & $N = 500$ & $N = 1,000$ & $N = 500$ & $N = 1,000$ \\
\hline                              
3 & 1.36 & 1.37 & 1.36 & 1.37 & 1.36 \\
4 & 2.27 & 2.29 & 2.28 & 2.31 & 2.30 \\
5 & 2.96 & 3.01 & 3.00 & 3.07 & 3.03 \\
6 & 3.51 & 3.60 & 3.58 & 3.71 & 3.65 \\
7 & 3.96 & 4.10 & 4.07 & 4.28 & 4.18 \\
8 & 4.33 & 4.53 & 4.51 & 4.82 & 4.66 \\
\hline
\end{tabular} 
\end{table}

\end{document}